# In search for an alternative to the computer metaphor of the mind and brain

**Damian G. Kelty-Stephen**[1,*], **Paul E. Cisek**[2], **Benjamin De Bari**[3,4], **James Dixon**[3,4], **Luis H. Favela**[5,6], **Fred Hasselman**[7], **Fred Keijzer**[8], **Vicente Raja**[9], **Jeffrey B. Wagman**[10], **Brandon J. Thomas**[11], **and Madhur Mangalam**[12,*]

[1]*Department of Psychology, State University of New York at New Paltz, New Paltz, NY 12561, USA*

[2]*Department of Neuroscience, University of Montréal, Montréal, QC H3C 5J9, Canada*

[3]*Center for the Ecological Study of Perception and Action, Storrs, CT 06269, USA*

[4]*Department of Psychological Sciences, University of Connecticut, Storrs, CT 02629, USA*

[5]*Department of Philosophy, University of Central Florida, Orlando, FL 32816, USA*

[5]*Cognitive Sciences Program, University of Central Florida, Orlando, FL 32816, USA*

[7]*Department of Philosophy, University of Groningen, 9712 CP Groningen, Netherlands*

[8]*Behavioural Science Institute, Radboud University, 6525 XZ Nijmegen, Netherlands*

[9]*Rotman Institute of Philosophy, Western University, London, ON N6A 3K7, Canada*

[10]*Department of Psychology, Illinois State University, Normal, IL 61790, USA*

[11]*Department of Psychology, University of Wisconsin, Whitewater, WI 53190, USA*

[12]*Department of Physical Therapy, Movement and Rehabilitation Sciences, Northeastern University, Boston, MA 02115, USA*

*Correspondence: keltysd@newpaltz.edu (D. G. Kelty-Stephen); m.mangalam@northeastern.edu (M. Mangalam).

ORCIDs: 0000-0001-7332-8486 (D. G. Kelty-Stephen); 0000-0001-7060-7287 (P. E. Cisek); 0000-0002-7894-7849 (B. De Bari); 0000-0002-6678-3621 (J. Dixon); 0000-0002-6434-959X (L. H. Favela); 0000-0003-1384-8361 (F. Hasselman); 0000-0001-7362-1770 (F. Keijzer); 0000-0001-5780-9052 (V. Raja); 0000-0001-5361-3802 (J. B. Wagman); 0000-0002-7471-2778 (B. J. Thomas); 0000-0001-6369-0414 (M. Mangalam).

**Abstract**

The brain-as-computer metaphor has anchored the professed computational nature of mind, wresting it down from the intangible logic of Platonic philosophy to a material basis for empirical science. However, as with many long-lasting metaphors in science, the computer metaphor has been explored and stretched long enough to reveal its boundaries. These boundaries highlight widening gaps in our understanding of the brain's role in an organism's goal-directed, intelligent behaviors and thoughts. In search of a more appropriate metaphor that reflects the potentially noncomputable functions of mind and brain, eight author groups answer the following questions: (1) What do we understand by the computer metaphor of the brain and cognition? (2) What are some of the limitations of this computer metaphor? (3) What metaphor should replace the computational metaphor? (4) What findings support alternative metaphors? Despite agreeing about feeling the strain of the strictures of computer metaphors, the authors suggest an exciting diversity of possible metaphoric options for future research into the mind and brain.

**Keywords:** behavioral neuroscience; cascade dynamics; cognitive neuroscience; complexity; dissipative system; dynamical system; ecological psychology; fractal; perception and action; representation; resonance

**Glossary**

**Knitting together the mind, brain, and behavior with Turing's cascade instability**

**Avalanches:** A series of bursts of activity [in neural networks] that can be described by a power law in terms of size distribution.

**Cascade:** A physical process characterized by a blending of information or structure built at multiple scales. For instance, when events spread from one scale to another, e.g., cellular to genetic, or cellular to whole-tissue, and then to organ- and to whole-organism scale, what we have is a cascade of effects.

**Intermittency:** An uneven distribution of events across time or space, characterized by the unpredictable, sudden appearance or disappearance of stability—or by the unpredictable, sudden transitions among stable states.

**Multifractality:** A generalization of a fractal system in which one fractal dimension is not enough to describe its dynamics; instead, a continuous spectrum of exponents (the so-called singularity spectrum) is needed.

**Power law:** A power law is a functional relationship between two quantities in which a change in one quantity causes a corresponding change in the other quantity, regardless of their original sizes: one quantity changes as a power of another.

**The brain is a control system**

**Control system:** A physical system that regulates the state of its input variables by providing appropriate output commands.

**Dorsal stream:** A visual pathway in the primate brain from the occipital lobe and pulvinar to the parietal lobe, in which neurons are sensitive to spatial information and involved in movement control.

**Hypothalamus:** A brain region that regulates the autonomic nervous system, body temperature, hunger, thirst, rest, and other bodily functions.

**Telencephalon:** A region of the forebrain that includes the cerebral cortex, cerebral nuclei, hippocampus, and many other structures implicated in advanced primate behavior.

**Ventral stream:** A visual pathway in the primate brain from the occipital lobe to the temporal lobe, in which neurons are sensitive to information about the identity of objects in the world.

**Virtual machine:** The simulation of a computer system implemented using software running on a different physical computer system.

**Dissipative structures as an alternative to the machine metaphor of the mind and brain**

**Dissipative Structure:** Self-organized forms (morphological or dynamical) derived from non-equilibrium conditions.

**Dynamical stability:** A time-invariant system state (e.g., fixed point, limit cycle) maintained by irreversible dissipative processes. If the system is perturbed, it will relax back to the same state. In some systems, multiple dynamics may be possible, in which case large perturbations may drive the system to converge on a different state.

**Far-from-equilibrium:** Systems are sustained out of equilibrium if there are fluxes of energy and matter driven by an imbalance of the distribution of energy within the system (e.g., a thermal gradient), called a thermodynamic force. Far-from-equilibrium refers to the regimes where the magnitude of thermodynamic forces on the system is sufficiently large to drive the emergence of dissipative structures or turbulent dynamics and in which force-flux relations are nonlinear. The near-equilibrium regime is one in which linear equations describe force-flux relations.

**Fractionability:** A system is fractionable if it can be decomposed into discrete parts, and those parts have isolable functions.

**Self-organization:** The emergence of novel ordered forms (e.g., morphological or dynamical) due to processes within a system's boundaries, not owing to prescribed instructions from another agent.

**Complexity: Understanding brains and minds on their own terms**

**Complexity science:** The interdisciplinary study of complex systems, which, in general, are phenomena composed of many interacting parts that give rise to irreducible order at particular spatial and temporal scales.

**Emergence:** When a scale of activity or organization is more than the sum of its parts. A concept that refers to an ordinary feature of nature, has an enormous literature across disciplines, and little agreement on its exact meaning.

**Fractal:** Activities or structures that are self-similar; also referred to as scale-free or scale-invariant. Self-similar patterns are the repetition of whole or part of an entity or activity at various spatial or temporal scales. Perfect self-similarity means the global structure is maintained at every scale of observation (e.g., Koch snowflake, Mandelbrot set, and Sierpinski triangle). Statistical self-similarity means only particular features are repeated at different scales and to certain degrees (e.g., bronchial tubes, coastlines, and tree branches).

**Nonlinearity:** Disproportional or nonadditive relationships among variables, particularly exponential and multiplicative interactions. Systems constituted by nonlinearly interacting variables quickly become computationally challenging to analyze, are no more predictable than statistical probability, and can give rise to phase transitions.

**Self-organization:** Patterns of activity or structural organization without direct intervention or instruction from a central controller, outside source, or preprogramming. Typically results from simple rules or processes, is spontaneous, and reduces degrees of freedom.

**Universality:** Recurring patterns of activity and structural organization in nature by way of vastly different substrates and contexts. *Universality classes* are mathematical models of widespread classes of systems exhibiting qualities largely independently of the dynamical details.

**Radical embodied computation: Emergence of meaning through the reproduction of similarity by analogy**

**Adaptive behavior:** Any interaction with the internal or external environment of a system that appears to be (partially) coordinated by a unique history of previously experienced interactions and/or by semantic information about the internal or external environment of the system.

**Embodied computation:** The processes responsible for maintaining or evolving the complexity of the internal structure of physical systems that are alive. Radical Embodied Computation posits that the massive redundancy of reality is sufficient for the adaptive coordination of behavior by evolved agents.

**Information**: A measurable quantity that resolves uncertainty about the configuration of an information source.

**Physical information**: A physical system can represent an amount of information which is associated with the degrees of freedom it has available to manifest its current state and behavior. When changes in the internal state configuration of a system concern changes to the available degrees of freedom, this changes the physical information represented by the system. Therefore, when a system self-organizes from one stable state into another, this can be described as information processing or natural computation.

**Self-replication:** One of the hallmarks of complexity is the (approximate) reproduction of dynamic patterns, often across different spatial or temporal scales. In the context of living systems, self-replication continuously occurs at the scale of periodic neuronal oscillations, reproduction of complex molecules, reproduction of cells, all the way up to the scale of the approximate replication of the individual organism through sexual reproduction. The emergence of self-replicating systems has been suggested to be an unavoidable consequence of a universe in which energy, matter, and information have to be optimally dissipated as entropy.

**Semantic information**: If physical information sources are systematically associated through their internal configuration, the mutual information they represent can be described as meaningful, or, semantic information. Generally, the redundancies between information sources are exposed or translated by an analogy, a code, which itself can be a physical information source. The code together with the information sources constitute the physical embodiment of meaning. Adaptive behavior can be redefined as translating the semantic information represented by the redundancies between the configuration of the body and the structure of the environment.

**Tunnel vision, tunnel action, tunnel mind: Just get out**

**Biogenic approach:** A methodological proposal formulated by Pamela Lyon: The cognitive sciences can better start from general biological principles, such as continuity, control, interaction, and normativity, among others. A biogenic approach is contrasted with an *anthropogenic approach* that starts with humans as the central exemplar and works from the human case to other organisms.

**Brain-body dualism:** Given the notion of a mind-brain, the brain becomes conceptually dissociated from the body with which it is physically fully intertwined and a functional unit. The mind-brain amounts to a conceptual chimera that combines its biological constitution and evolution with mind-based concepts centered on reason, rationality, and logic. A new form of dualism that cuts straight through the body, unclear and messy.

**Conceptual problems:** Solving conceptual problems is as important to scientific progress as solving empirical ones, said Laudan in his classic "Progress and its problems." Such conceptual problems usually relate to how empirical findings are interpreted and what might consist of possible solutions. However, the problems can also become intertwined with extra-scientific beliefs—e.g., coming from metaphysics, logic, ethics, and theology—that in this way can impact our scientific conceptions of a set of phenomena.

**Mind-brain:** The brain is treated as a physical instantiation of the mind. Mental properties and characteristics are subsequently used to provide a global view of the brain's functionality, which strongly influences the agenda of cognitive neuroscience.

**Skin brain thesis:** A proposal that the origin of nervous systems was (initially) driven by the way it enabled multicellular contraction-based motility for early animals. Horizontal neural transmission is considered central to coordinating the spread-out activity of many separate contractile cells.

**Vertical and horizontal neural transmission:** Vertical transmission refers to neural activity traveling from sensors to effectors, potentially via a central nervous system. Horizontal transmission refers to neural activity (such as provided by a nerve net) that travels across an effector surface or an array of effectors to coordinate such macroscopic effectors.

**Resonances in the brain**

**Ecological psychology:** Theory of perception and action first developed by James J. Gibson and Eleanor J. Gibson. Ecological psychology proposes ideas like direct perception, perception-action loops, and affordances. It stands in complete contrast to information-processing theories of perception and action.

**Ecological resonance:** The process by which brain dynamics become constrained by the (ecological) perceptual information that is available at, and is already constraining, the organism-environment scale.

**Informational invariant:** Main elements of perceptual information within the ecological psychology. When organism moves around in an environment, most of the structural patterns of the energy arrays available at its sensory receptors change. However, some of those patterns remain unchanged or invariant. These invariant patterns specify permanent properties of the environment the organism is within and, therefore, constitute perceptual information.

**Resonance:** As a physical phenomenon, resonance is described as the increase of the amplitude in the oscillation of a system when another system influences it at a frequency equal or close to its natural frequency.

**Tau:** A variable of perceptual information (i.e., an informational invariant) described in the ecological literature. It is defined as the inverse of the relative rate of closure of a given gap. Such a gap can be defined in different dimensions: geometrical gap, energy gap, etc. The units of tau are seconds.

**The brain as a fractal antenna**

**Affordances:** The opportunities for behavior that result from the fit between animal and the environment.

**Animal-environment system:** The symmetrical and reciprocal relationship between animals and the environment constitutes the fundamental unit of analysis in the Ecological approach to perception and action.

**Degrees of freedom problem:** The fact that in performing any given movement of the body requires coordinating a seemingly overwhelming number of individual components Formal system: an abstract and idealized system in which symbols are manipulated and combined according to rules.

**Ecological approach to perception and action:** An approach to understanding the everyday performance of goal-directed behavior in which relationships between animal and environment lawfully structure energy patterns such that they are informative about those relations to the animal.

**Everyday behavior problem:** The concern that a description of perception as resulting in mental experience is ultimately challenged to explain how such mental experience could lead to a process resulting in the successful performance of everyday behaviors.

**Fractal antenna:** The fact that performing any given movement of the body requires coordinating a seemingly overwhelming number of individual components.

**Formal system:** an abstract and idealized system in which symbols are manipulated and combined according to rules.

**Grounding problem:** The concern over how symbols acquire meaning or are related to what they symbolize, especially in the context of computational approaches to cognition.

**Turing machine:** A model of computation in which a machine reads and writes symbols in sequence along with an infinite according to a finite set of rules.

# 1. Introduction—Madhur Mangalam[1] and Damian G. Kelty-Stephen[2]

Viewing and modeling the brain as a computer has become second nature—but it was not always this way. In 1958, John von Neumann released the influential book *The Computer and the Brain*. In this ambitious attempt, he brought together what was known about the machine-like qualities of the brain and what such machines that are not brains might one day accomplish. Since then, the computer metaphor has been used to explain the computational powers of the mind through its origins in a computational brain. This move reflects a physicalist attempt to view the mind as a computer by putting hierarchically organized computational powers onto the most likely supporting anatomical organ. As a result, the computational mind has been anchored by the brain-as-computer metaphor, which has wrested it down from the abstract ideals of Platonic philosophy to a material testbed for empirical science. Undoubtedly, the computer metaphor has been a valuable and fruitful metatheoretical framework for understanding the mind and behavior. However, as with many long-lasting metaphors in science, the computer metaphor has been explored and stretched long enough to reveal its boundaries. A systematic reinterpretation of "computation" in the context of mind and brain entails the development of new metaphors that describe the sometimes noncomputable processes are perhaps the only ways to rescue the scientific study of the mind and behavior from the shortcomings of the computer metaphor.

The current work is a collection of candidate metaphors to describe the mind and brain, and that could potentially replace the computer metaphor. So far, these metaphors have been developed and expanded on the outskirts of psychology and neuroscience, and while they have quite a long way to go in displacing the computer metaphor, psychologists and neuroscientists are continually learning more about them and their usefulness in explaining behavior and cognition. Moreover, these metaphors either generate or evaluate novel theoretical foundations for describing the structures and processes that support perception, action, and cognition in ways that the computer metaphor does not.

## 1.1. The computer metaphor of the mind and brain

The invention of the telegraph in the early 20th century, followed by the telephone, influenced initial conceptions of the brain, but soon it became evident that the numerous "switches" in the nervous system do not function as they do in these technologies (Cobb, 2020). Biological discoveries quickly matched or outstripped these technological paradigms (e.g., the canonical Hodgkin-Huxley model; Hodgkin and Huxley, 1952). As a result, other metaphors were soon required to explain what the brain does and how it accomplishes it. Since von Neumann published *The Computer and the Brain* (Von Neumann, 1958), the computer metaphor has dominated our theories about what the brain is (a computer), what it does (information processing), and how it accomplishes it (neural encoding). Three coincidental developments in the mid-twentieth century accelerated the adoption of the computer metaphor. First, Behaviorism, the then-dominant metatheoretical perspective in psychology, failed to explain how people understand and develop language (Chomsky, 1959), creating an immediate need for a new metatheoretical model for psychology. Second, the development of Communication Theory (Shannon, 1948) provided a method for measuring the amount of information flowing through a system. Third, the invention of digital computers offered psychologists a metaphor for psychological constructs and a concrete approach to investigating brain functioning. Ulric Neisser, one of the founders of cognitive psychology, went so far as to claim that the "task of … trying to understand human cognition is analogous to that of … trying to discover how a computer has been programmed" (Neisser, 1967, p. 6) Neisser justified this approach by claiming that a computer program is a "recipe for selecting, storing, recovering, combining, outputting and generally manipulating [information]" (Neisser, 1967, p. 8).

---

1   Correspondence: m.mangalam@northeastern.edu (M. Mangalam).
2   Correspondence: keltystd@newpaltz.edu (D. G. Kellty-Stephen).

The ongoing proposal has been that a digital computer that is as "intelligent" as the human brain can be built if we can just decipher the algorithms used by the brain, code them, and put them into action (Wood, 2019). This overarching computational view has remained the dominant metaphor among contemporary psychologists and neuroscientists (Arbib, 1975; Boden, 1988; Churchland and Sejnowski, 1994; Pinker, 1997a; Wolpert and Ghahramani, 2000), and, of course, among researchers in artificial intelligence as well (Cox and Dean, 2014; Floreano and Mattiussi, 2008; Hassabis et al., 2017). Computationalism is so broadly applied to the study of the mind and brain that the fields of computer science and neuroscience have become almost indistinguishable (Lillicrap et al., 2020; Lindsay, 2021; Marblestone et al., 2016; Mollon et al., 2022; Richards et al., 2019; Saxe et al., 2021).

Computers operate on symbolic representations, they store and retrieve representations, process them, and have physical memories with physical addresses of these representations, guided by algorithms that depend on specific physical structures. Humans can certainly use symbolic representations and sometimes they use algorithms to do it (e.g., while using an actual computer, paper, or other recording device). Still, attempts to explain human intelligence by referring to an anatomical organ as an entity that "computes" is likely a case of circular reasoning. Aside from the presumption that these symbolic functions can be explained by situating them in particular anatomical tissue, the computer metaphor has grown into a delimiter of the questions we can ask and the proper answers that can be given. For instance, the underlying assumption has been that brain function can be inferred by perturbing particular brain components and seeing impacts on functioning of the total system (Jonas and Kording, 2017). Such a research program may be insightful and uncover evidence about the brain's workings. Yet, no empirical evidence can ever have the power to correct what could be an unduly constrictive assumption: the idea that function is best understood as the output of components (i.e., distinct and separate enough to be deleted or perturbed) will only ever be able to motivate the discovery of new evidence of distinct, independent component activity (Van Orden et al., 2001). This assumption leaves unconsidered the possibility that the brain function or computational ability does not rest on the independent contributions of distinct and separable components. Hence, the computer metaphor risks foreclosing on any interest in the interactivity and flexibility of the brain, which may have implications for the adaptability of behavior and function.

There is nothing specifically negative about using a metaphor as a scientific or pedagogical device. Metaphors are useful for successful science, especially when attempting to understand the unknown in terms of something known. There is a long history of metaphors in the brain sciences (Cobb, 2020; Smith, 1993). A strength in this history has been the willingness to engage with new metaphors when old ones have shown to be untenable. A remarkable idiosyncrasy of this history has been the tendency to base each new metaphor on the newest machine technology. René Descartes considered the brain to be a hydraulic pump propelling the spirits of the nervous system through the body (Smith, 1999), the psychoanalyst Sigmund Freud envisioned the brain as a steam engine (Bock von Wülfingen, 2013), and the neuroscientist Karl Pribram discussed the brain as a holographic storage device (Pribram, 1982). These metaphors afforded a cutting edge theoretical lens through which to plan new experiments and reinterpret experimental findings. However, metaphors can be harmful when they pose limits upon what and how we can think. Consequently, even our seemingly most fundamental ideas can come to be challenged, such as the usefulness of understanding brains and nervous systems in terms of neuronal networks representing the outside world (e.g., Anderson, 2014; Chemero, 2009). The increasing dissatisfaction in the scientific community with understanding the mind and brain via contemporary dominant technology might encourage healthy skepticism about our current pattern of metaphorizing the brain.

What our history of metaphor tells us is, first, that the psychology and neuroscience communities see the brain as the central component in highly intelligent systems and, second, that none of the current metaphors of the mind and brain are likely to outlast the next technological innovation. An important question here is: shall we continue to appeal to technological fads as the source of our next metaphor for the brain, or should the pattern be broken to consider other options? Proponents of the computer metaphor might argue that "no one believes" in strong computational functionalism—that is, no one believes that all mental states and events reflect computational states of the brain. Instead, they believe in weak computational functionalism, which holds that the brain's computational organization is reflected by the mind, which is realized by neural network structure and functions (Putnam, 1988).

At this point, at least some of the computer metaphor's proponents are prepared to rein in the scope of the metaphor. Richards and Lillicrap (2022) have recently argued that the brain-computer metaphor debate is "useless"—a matter of semantics because brains are either computers or not, depending on one's definitions. Reducing the computer metaphor to mere semantics raises questions about how it ought to be taken seriously in scientific investigations and explanations. If we and you adhere to different definitions of "computer," then it becomes unclear what the scientific use of a computer metaphor is. Alan Turing (1936) famously provided proof of the possibility of developing universal machines capable of computing all computable procedures. To define a computer in other terms would raise challenging questions about whose "computer" applies to what target of inquiry. Turing was the beginning of this inquiry not the end.

Meanwhile, a minority of psychologists and neuroscientists are ready to let the computer metaphor go and to dispense with even the semantic form of the metaphor. Researchers in this group have sometimes asserted that the mind and brain defy lawful explanation given the understanding of physical laws at present, requiring new, yet-to-be-discovered laws and principles to comprehend them (Chalmers, 1996). Such alternatives are typically not based on empirical data, contradicting the physicalist working hypotheses, nor do they propose a different metaphor that might guide us toward the new laws. But without a different metaphor and a systematic, valid organization of exceptions or violations of the dominant metaphor, these disavowals of computer metaphors leave it unclear what to do. Short of waiting for the new laws to fall from the nearest apple tree, it is difficult to know what to do in the meantime.

## 1.2. In search for an alternative to the computer metaphor of the mind and brain

We believe that the tide may be turning. Hungry for a metaphor that generalizes beyond word interpretation, scientists are bringing evidence to bear on the questions of what to do besides or beyond the computer metaphor. While the computer remains a crowning technological achievement, many scientists are ready to try out new metaphors. Our format for this article is intended to provide the reader with an overview of the limitations of the computer metaphor and several candidate metaphors that are ready to replace it. Eight author groups respond to the following four questions and point to future directions for research prompted by these metaphors.

**Question 1:** *What do we understand by the computer metaphor of the mind and brain?*

**Question 2:** *What are some of the limitations of this computer metaphor?*

**Question 3:** *What metaphor should replace the computational metaphor?*

**Question 4:** *What empirical findings support this alternative metaphor?*

The resulting contributions represent a range of contemporary responses from the primary disciplines in behavior and brain sciences, including complexity science (De Bari, Dixon, Favela, Hasselman, Kelty-Stephen, Mangalam), experimental psychology (De Bari, Dixon, Favela, Kelty-Stephen, Mangalam, Thomas, Wagman), movement science (Kelty-Stephen, Mangalam), neuroscience (Cisek, Mangalam), philosophy of cognitive sciences (Favela, Keijzer, Raja), and statistical physics (Dixon, Kelty-Stephen, Mangalam). While we see a consensus that pushes against the strictures of computer metaphors across these sciences, we also see a variety that suggests an exciting diversity of possible metaphoric options. Conversely, showcasing this diversity in a single avenue might provoke the behavior and brain sciences towards uncovering those fundamental facts about the brain that the computer metaphor keeps beyond our grasp.

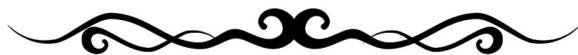

## 2. Knitting together the mind, brain, and behavior with Turing's cascade instability—Damian G. Kelty-Stephen[3] and Madhur Mangalam[4]

We see the computer metaphor as a holdover from premodern scientific traditions hoping to anchor the mind's computational ability in a material anatomical part. Despite having prompted decades of valuable empirical insights, the computer metaphor has likely outgrown its usefulness. Brains are context-sensitive and capable of adapting to novelty, eschewing the locality of meaning necessary to computation. Rather than proposing a new metaphor for the mind and brain, we think that Turing's old idea of cascading instability is a metaphor worth reconsidering. It recommends a power-law-driven geometrical framework that might knit together the mind, brain, and behavior in context.

### 2.1. What do we understand by the computer metaphor of the mind and brain?

The computer metaphor of the mind and brain has evolved historically from the rationalist tradition that the mind's uniqueness lies in its capacity to use logical inference (Fodor, 1981). Because physicalist science has aimed for centuries to localize various physiological functions in independent anatomic structures (e.g., Harvey identified that the heart is the anatomical structure responsible for blood circulation; Ribatti, 2009), a long-standing, relentless goal has been to sort out the proper anatomical anchoring for the logical processes. A fundamental obstacle in achieving this goal has been the abstract quality of logical thought. Platonic traditions have left the Western world with the impression of thought as somehow removed from the tangible details—even overshadowing the Aristotelian tradition that even mind is material. However, the prevailing tradition has been, in effect, that whatever material substrate can support the logical reasoning must itself carry some logical structure to it, thanks to our scientific culture that reasons from the principle of similarity, i.e., similar effects should arise from similar causes (Hume, 2020).

---

3   Correspondence: keltystd@newpaltz.edu (D. G. Kelty-Stephen).
4   Correspondence: m.mangalam@northeastern.edu (M. Mangalam).

The computational aspect of the mind thus led psychological sciences to seek a computational aspect of organic tissues. The brain has long topped the list of anatomical supports for the mind. Before the advent of neuroimaging, ancient cultures could see plain deficits in mental functioning following injuries to the head and, specifically, the brain. Then again, the brain did not always appear as a computer. For all his insights in discerning a difference between efferent and afferent nerves, even the Greek physician and surgeon Claudius Galen promoted the retrospectively laughable idea that the brain was more of a stomach to digest incoming stimuli (Galen, 2019). This perspective ignored the then-laughable idea that the cortical surface was essential to cognition and emphasized the ventricles filled with cerebrospinal fluid. Holdovers of this now-strange idea survived well into premodern science, even populating some pages of Leonardo Da Vinci's otherwise visionary notebooks (Gross, 1999).

The computer may not have entered the explicit discourse until the modern age of stored-program computing. Notwithstanding, the computer metaphor began to take root centuries before scholars like Descartes legitimized the mind-body dualism as an acceptable premise from which to investigate subjective experiences (Fodor, 1981). Descartes did not have an IBM PC on his desk, but he did grow up in a culture that had become fascinated with early-stage robots, i.e., automata that artful engineers gave a biological and sometimes humanoid form—Descartes's referred to automata in his writing, and the culture was so full of imagination about automata that myths grew about Descartes potentially having crafted automata of his own and perhaps even in the image of his daughter (Kang, 2017). Descartes would have held out for the mind to be something other than mechanical-biological tissue, but he viewed the operation of the mind as sufficiently elegant and rational to organize and manipulate the logical operation of machines. So, without calling the brain a computer, Descartes sowed the seed of expectation that the brain and its structures should have a logical operation that made them ready instruments for the mental process to play upon.

Computer technology caught up to the Cartesian dream in the early 20th century with Alan Turing's (**Fig. 1**) intriguing discourse on mechanical intelligence (Hodges, 1983; Turing, 1950a). Turing leaned heavily into creative speculations about how to build minds or intelligence with computing technology, and these speculations always seemed to stop short of the conclusion that the human mind or the brain was explicitly and exclusively a computer (Sprevak, 2017). Turing was always keen to point out that, just because the computer-like rules were not in plain evidence, it was always possible that the rules could later become evident with further investigation. However, he found neurobiology tedious and irrelevant to the arguments he was developing about computation and the role of the logical process relative to physical supports (Hodges, 1983). Turing's later work in morphology clarifies a long-brewing suspicion that the physical supports could be of at least comparable importance as the logical process (Kelso, 1995; Turing, 1952). Specifically, Turing never saw the grounds to lock the brain up into a computer

metaphor—for him, there was always the opportunity that nonlinear and perhaps random cascading processes through the physical embodiment was crucial for generating the computational powers. In short, Turing saw that the illogical physics of biological form might have been the unwitting engineer of the sometimes computational human mind.

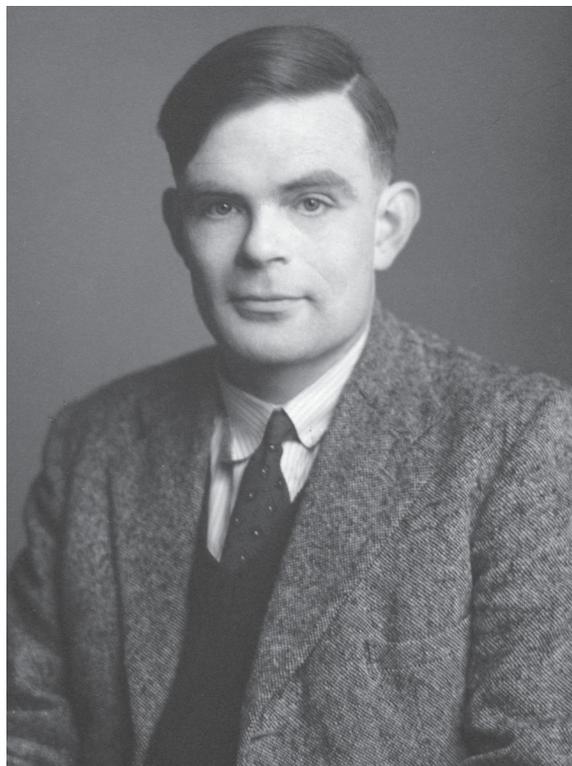

**Fig. 1.** Alan Turing (1912– 1954). Copyright © The Royal Society.

The crucial step towards the computer metaphor was McCulloch and Pitts's recognition of Turing-style computability as a good match for the brain—the binary language of 0s and 1s made neat analogy with the binary of neural dynamics: all-or-nothing action potentials (McCulloch and Pitts, 1943). The computer metaphor suggested physicalism a neat anatomical support, leading the behavioral and brain sciences to generally foreclose on Turing's curiosity about whether illogical, unstable cascades might sometimes produce computation. Instead, the modern computer metaphor has drawn from Turing's earlier work using Bayesian statistics to hack coded messages with predictive logic. The branching neurons coevolved together in networks spanning cortical layers across the brain now appear to us as a hierarchical Bayesian network of logic (de Lange et al., 2018). Indeed, some modern-day neuroscientists are prepared to design/inform psychiatric diagnoses and interventions based on how psychopathology might result from creaking, collapsing Bayesian hierarchies (Petzschner et al., 2021, 2017).

The computer metaphor is so predominant that one might easily never know anyone thought any different. We also do not think it has been such a bad metaphor. One of us has confessed in print to having loved the Cartesian promise of mental clockwork impressed upon neural clockwork (Kelty-Stephen and Dixon,

2013). However, we see a couple of crucial signs of how the computer metaphor is not living up to its promise. For instance, there is a long history of comparative biology that cannot quite settle the question of brain size. This research often starts from the premise that humans have the greatest command of mental computational power, and for decades it has been stuck on the persistent, daunting question of why many animals with bigger brains than we have cannot manage our feats of logical reasoning (Stout, 2018). The research then spins its wheel a bit with qualifications and attempts to revisit the question from different angles, e.g., by disentangling global size from specific parts' size, disentangling different ways to consider proportional or relative size metrics. None of these attempts to find the better, clearer meaning of what a more "intelligent" brain entails are bad, but the discourse quickly becomes slippery and hard to keep transparent. The only potentially clear thread in the discourse is a persistent desire to show that the human brain is best suited to our vaunted intellectual powers (Barton and Montgomery, 2019).

Furthermore, no matter the undisputed elegance of hierarchical Bayesian networks (Berezutskaya et al., 2020; Daube et al., 2019), these models fail to yield the full diversity of the higher cognitive functions centrally symptomatic to the computational capacities we seek to explain. Yes, we are thinking specifically of language, that exceptional case of rule-based manipulation of arbitrary symbols that distinguishes human intelligence from other comers. Chomsky (1980) could make the compelling case that language is a computational behavior without claiming to name any specific brain or genetic structure with the requisite computational hardware. Indeed, it was the computational structure of language that gave Chomsky the certainty that this only-computational portrayal requires innate knowledge (Bickhard, 2001; Fodor, 1983, 1975; Pinker, 1997b). We authors do not agree that language must be innate, but we agree that anything strictly computational must be; the arbitrariness of the symbol and the rule does not admit any learning. If computation is in fact about the local operation of rules directly on symbols with local meanings (Fodor, 2000a; Pattee, 1977, 2013), then no amount of experience will be sufficient to steer the linguistic behavior towards its typically developing human form in as short a time. This point is as true for language as it is for protein folding, and in their own department, organic chemists grappled the same issue that many amino acids engaged only in local interactions would never randomly discover functional protein forms (Levinthal, 1969). In any event, shuttling localized meaning from one part to another of a linguistic interaction is all that Bayesian networks can accomplish. The computational commitment of local meaning to individual symbols is simply not enough to explain the dynamic, context-sensitive interplay of factors (i.e., motoric, acoustic, neural, social, cultural) spreading from one scale to another (e.g., from single-neuron to brain region to social group, from last week's gossip to tomorrow's exam) in language behaviors (Fusaroli et al., 2014; Olmstead et al., 2021; Skipper et al., 2017).

For all we know about the brain, the evolution of language remains a mystery (Hauser et al., 2014), and the computer metaphor may foreclose any new insights by enforcing the need for fixity of local meaning and arbitrary symbols. Stored-program computers have no apparent natural heritage except wholesale innate installation and wiring. We can build the arbitrary symbols in, but that is only because we organisms can do computation—to say we can only do that because we have computers in our head just begs the question. Computers must have come from somewhere and arisen at some point in the past, but their symbols and rules are arbitrary—not by accident but by design. To say the brain is entirely a computer does not just commit all meaning to local tokens and local interactions but also seals off the chance for new structures. It is further to assume that the brain is also arbitrary and thus disconnected from any natural law in the evolutionary past before computers. Organisms might build modern computers because they have computing brains, but where would the "computer" brains have come from? It is hard to address this question if computers are always arbitrary. If computers have truly arbitrary structures, no natural law could guarantee their emergence. Then again, if our brain has evolved according to natural principles (e.g., of selection), they could not be arbitrary and are thus unlike every other computer we humans have designed.

For all that the computer metaphor of the brain and cognition affords us, it reflects an aspiration to find an organic, material home for an idealized notion of what the mind is doing—the mind is ultimately more flexible than just a computing process. It carries more profound imprints of illogical, unstable cascades, and the brain is no different. Computation is a thing that minds can do, a thing that organisms poised at abacuses or drawing geometrical proofs with pencil and paper can do as well. However, without even denying that the brain must support that organism in computation, we see the computer metaphor as a reasonable approximation that leaves out the flexibility to depart from logical validity. The 0s and 1s that McCulloch and Pitts (1943) used to portray action potentials are a coarse-graining of a neuron's nonlinear dynamical properties. The brain is a fluidic nonlinear dynamical system that engages in drastic transitions at the crossings of activity thresholds, and that adapts its local exchanges from neuron to neuron sensitively to the bodily, task, and environmental contexts. Not only do we authors fail to see a computer in the brain, but we also fail to see the computer metaphor doing adequate justice to the illogical, unstable cascades that brains engage in bringing to the support of the mind.

## What are some of the limitations of the computer metaphor?

The fundamental limitation of the computer metaphor follows from the central requirement of computation that meaning must be locally encoded into the symbols shuttling meaning from input to output. This limitation is not an accident. Instead, it is the sheer genius of the computer metaphor. Computation is good at what it does precisely because arbitrary symbols can carry meaning from premise to conclusion in a way that is perfectly readable or transferable by any system equipped with the same symbols and rules. 1 + 1 = 2 no matter whether it is raining outside or where the program is enacted. We are aware of connectionist attempts to drop below the symbolic level and allow individual tokens to carry only part of the meaning, but we share Fodor and Pylyshyn's (1988) impatience with this strategy because it presumes to make computation no longer computation. Specifically, if the proposed sub-meaning of subsymbols is somehow not also local to the subsymbol, then it is no longer a computer but a fluid system that needs constant supervision to ensure that the nodes are somehow not locally free to absorb meaningless stimulation and somehow they do something reliable.

However, with meaning localized to the symbol to be operated upon, the limitations accrue as the inability of the computer to discover new symbols (or subsymbols), to correctly identify new (or any) proper features of the world to which the symbols (or subsymbols) refer, or to adaptively sense the vast (i.e., non-local) contextual frames that might change the local meanings of the symbols. These are the symbol grounding problem (Harnad, 2007; Searle, 1982, 1980), the problem of projectable predicates or abduction (Goodman, 1983; Misak, 1992), and the frame problem (Dennett, 2006; Fodor, 2000a), respectively—all of which have been discussed in such, to our minds, harrowing detail that we resist detailing them here. The upshot of all of these issues is that, as Fodor (1983, 1975) suggested and as Pinker (1997) readily agreed, the computer metaphor for the mind and so the brain requires immense amounts of innate knowledge. Additionally, this innate knowledge supports an array of special-purpose computational modules that are encapsulated and so impenetrable to learning or experience. So, in effect, the computer metaphor guarantees that the brain should fare very poorly with novelty or context-sensitive awareness.

Of course, the plain fact is that the brain is quite good at dealing with novelty and context-sensitive awareness (Ghazizadeh et al., 2020; Hunter and Daw, 2021). The context-sensitivity of the brain is so good that bees with much smaller brains than us can solve spatial-reasoning problems like the traveling-salesperson problem (Lihoreau et al., 2010). The traveling salesperson problem is centrally the problem of efficiency: finding the shortest, least costly path through multiple locations without visiting the same location twice. Specifically, for computing terms, the traveling-salesperson problem is called "NP-complete," meaning that it requires processing time from the computer that explodes exponentially with the number of locations the salesperson must visit. The computing time grows so fast with the number of locations because computers lacking background knowledge about the spatial layout have to sift through all possible orderings of locations before identifying the shortest path. Thus, the only way computers can efficiently solve this problem is with built-in background knowledge. The beneficent programmer can give their computer algorithmic hints as to which sequences to try first, and the programmer aims to optimize the algorithm to need minimal hints. Meanwhile, modern computation attempts to optimize the traveling-salesperson problem are currently invoking optimization procedures that mimic the activity of bees and ants (Gao, 2020). So, small-brained animals are helping the computers solve cumbersome problems—the traveling salesman problem being just one example.

Indeed, the computer metaphor may have inverted the facts: the computer is slowly trying to be more like a brain. What makes it curious is that if it is a computer, the brain is very leaky, noisy. The brain is full of delays thanks to synapses and contextual pressures on neural transmission (Sabatini and Regehr, 1996; Xu et al., 2012). So, for all of the challenge of limiting the immense computing time to get the traveling-salesperson types of jobs done, computers are processing electrical relays at speeds far beyond the brain's capacity to spread electricity (Luo, 2020). So, the slow system performs much better than the computer with no synapses and has to do more work even if it can work much faster. Curiously, the computer engineers are gearing up to build computers that have synapses (Choi et al., 2020; van de Burgt et al., 2017; Wei et al., 2021), which would allow all the flexibility that we biological systems with neurons enjoy, but then the computers will in effect sacrifice speed—for tasks that are already done more efficiently by biological systems.

Of course, it is not news that the brain is unlike any other computer in evidence. Daniel Dennett even said as much in an Edge.org conversation almost ten years ago (*https://www.edge.org/conversation/daniel_c_dennett-the-normal-well-tempered-mind*; see also Levin and Dennett, *https://aeon.co/essays/how-to-understand-cells-tissues-and-organisms-as-agents-with-agendas*). However, we think that a metaphor is in a resemblance that will suggest new research directions. The metaphor is always a more accessible thing we understand better than the system we wish to understand, and we gaze at the metaphor's internal mechanism precisely because we expect a similar form in the less-accessible, lesser-known system (Klein, 2021). We can entirely understand Turing's (1950) and McCulloch and Pitts's (1943) points when coining the explicit computer metaphor, and we find the intervening decades of evidence very informative. But much of that evidence pointed away from the computer metaphor. Now, we are ready to look for a metaphor with a slightly firmer resemblance now that we know much more than the all-or-nothing dichotomy of action potentials.

## What metaphor should replace the computer metaphor?

We do prefer another metaphor rather than no metaphor. We do not think that "the data" ever speak for themselves. This point applies as much to computers as it applies to how we use metaphors, computer metaphors or otherwise. The data are all the computers know. The computers are not the ones interpreting the data, however. Meaning requires an interpreter to organize the data into an interpretable form. This organization requires contextualizing continuous variation (e.g., the data; or any flow) by applying constraints (e.g., units and geometrical frameworks). Constraints are necessary for meaning to ever arise from the dynamics of any kind, and this necessity follows meaning-making down to the simplest stages of biological uses of information like protein folding (Pattee, 2013).

    Metaphors do not, we expect, ever go out of style for building a meaningful scientific explanation. Indeed, we might suggest that, for real and metaphorical computers alike, the computer is most beneficial in its capacity for supporting our metaphor building. In a sense, we use computers or computation as metaphor machines: symbols and their rule-based operations are the models we use to distill a set of events to "bare-bones"—consistently reducing somehow to get a more efficient grasp on control or communication, ensuring what George E.P. Box famously reminded us, that "all models are wrong, but some models are useful." Of course, what we can grasp and wield physically does not need a model (e.g., we can pick up a book no matter whether we understand the text inside), and simply hefting a computer does not mean we know how to use the models. We just find it regrettable how persistently scientists have hewn to the computer metaphor, in effect using their office computers to make models of a "computer" that is going to, as a brain, to do much less but then again much more than what a computer can do. This point demands some emphasis: the computer metaphor is a reduction like all models before it, but the persistence of the computer metaphor depends on a faith that the metaphorical computer for the mind/brain might do things that computers cannot do. The computer metaphor is thus an overstatement of what a computer is and not the more-valuable reduction of the events we expect.

    Not every model is a computer. Hence, metaphors capable of generating scientifically valid hypotheses about theory are not limited to logical machines like computers. Indeed, biological and artificial sciences alike have a yawning portfolio of models that are not logical but somewhat tangible, living, and embodied (Bravo et al., 2021; Matthews and Vosshall, 2020; Michael et al., 2021; Nielsen, 2019; Vergara et al., 2017). Much like the computer metaphor, these wilder models carry our best predictions forward into the future. Notwithstanding, unlike computer metaphors, they have a nuanced texture, a set of tissues, and material capacities that can gain traction in the world outside the silicon microprocessor and bring new variables and new relationships to our attention.

    The alternative metaphor need not be new either—we prefer an old one. Indeed, Turing also gave us the other metaphor: the cascade. In both strains of his work, when pondering intelligent computers (Turing, 1950a) or when pondering biological morphogenesis (Dawes, 2016; Turing, 1952), Turing was focused on the same metaphor, just phrased in different ways. Turing (1950) opined that the creativity of the mind could be compared to the critical amplification of a nuclear fission pile (**Figs. 2, 3, and 4**). Turing (1952) opined that the large-scale creativity of bodily forms could be modeled as the random collisions of organic molecules under boundary conditions set by genetics. The two opinions are effectively the same: they both acknowledge that minds and bodies alike follow from the percolations of small-scale fluctuations giving rise to coherent forms at a large scale and then to large-scale patterns doubling back to rein in the small-scale dynamics.

A hierarchical configuration of events nesting at multiple scales achieves adaptive, context-sensitive behavior through a balance of noise and order. Dynamical fluctuations collide with constraints at multiple scales, pitching the hierarchy intermittently from one globally stable mode to another. In a sense, this description is just a tweak on the hierarchical Bayesian logic network some see in the brain currently (Colombo and Seriès, 2012; Friston, 2012; Kersten et al., 2004; Knill and Pouget, 2004; Körding and Wolpert, 2004). The significant difference is that, whereas its purely logical form makes the whole Bayesian network one single and very ornate constraint, the cascade is fluid at all scales, leaving dynamics to interleave nonlinearly with the constraints across each of those scales. The interactions across scales offer a nonlinearity that pervades the system rather than depending on any nonlinear parametrization at individual levels of the hierarchy (Lovejoy and Schertzer, 2018).

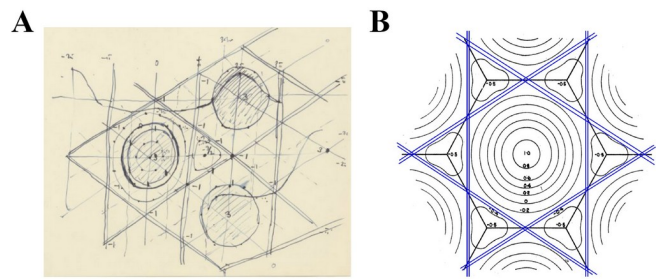

**Fig. 2.** Hexagonal planforms for patterns. (**A**) The original sketch by Turing. Reproduced from AMT/C/27/19a. Copyright © W.R. Owens. (**B**) Reproduced sketches from Pellew and Southwell (1940). The pairs of blue thick solid lines have been added to relate the two figures.

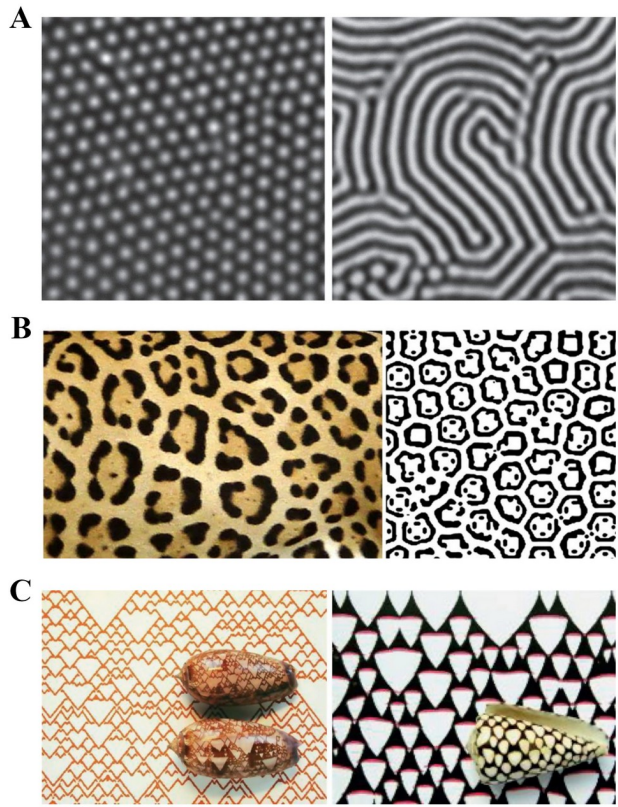

**Fig. 3.** Turing's cascade instability can lead to complex patterns observed in the nature. (**A**) Illustrative generic patterns of an activator–inhibitor scheme. From Ball (2015), courtesy of Jacques Boissonade and Patrick De Kepper, University of Bordeaux, France. (**B**) Naturally-observed "rosette" spots of a jaguar and its analog produced by two coupled activator–inhibitor processes. From Liu et al. (2006), © American Physical Society. (**C**) Naturally-observed patterns on seashells and their analogues produced by theoretical activator–inhibitor systems. From Meinhardt (2009), courtesy of Hans Meinhardt, MPI for Developmental Biology, Tübingen, Germany.

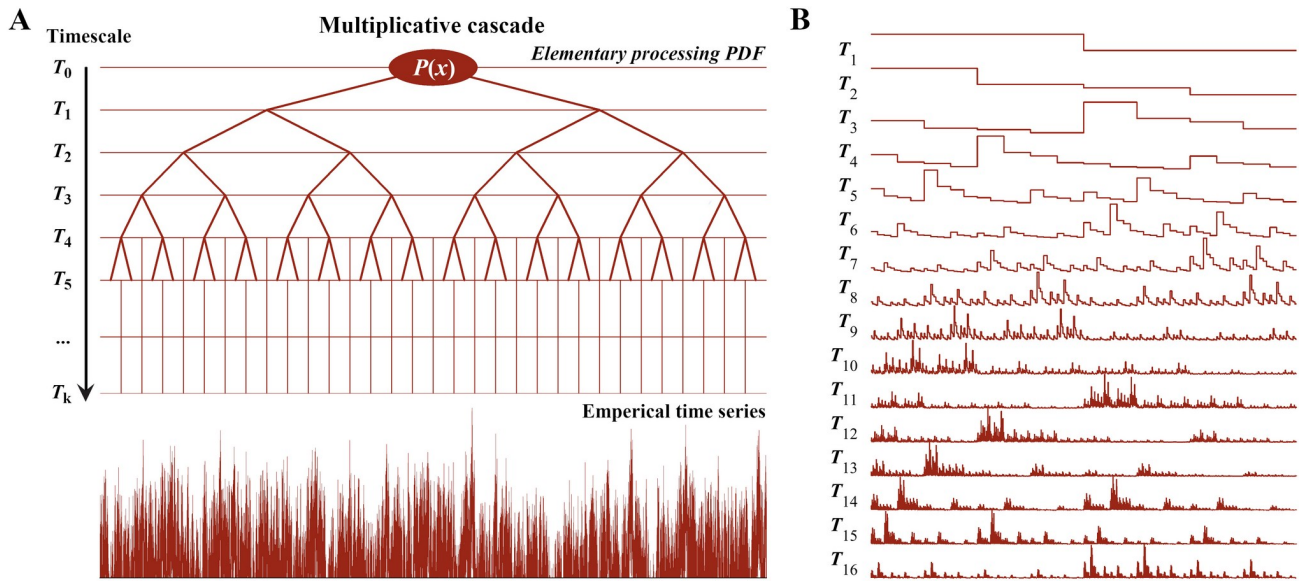

**Fig. 4.** Multiplicative cascade process. (**A**) Schematic of a multiplicative cascade process at increasingly fine binary timescales, driven by a general pdf, $P(x)$, starting with the whole time interval at $T_0$ and increasing the number of intervals by an integer power of two until the number of intervals at time scale $T_k$ is $2^k$. The activity time series at the bottom of the graph is believed to be formed by a multiplicative cascade, with $P(x)$ being a Gaussian density with mean $\mu$ and variance $\sigma^2$. (**B**) A binomial multiplicative cascade as a mathematical concept of how proportions of events can distribute themselves across progressively smaller sample sizes. The bottom-most curve shows the $2^{16}$-samples series representing the cascade after 16 generations. Each value reflects the multiplication of different sequences of the proportions 0.25 and 0.75. The leftmost samples reflect the successive multiplications of predominantly smaller proportions, and the rightmost samples reflect the successive multiplications of predominantly larger proportions. Each curve is normalized by its maximum value and displaced by 1 unit in the vertical for clarity.

**What empirical findings support your preferred alternative metaphor?**

The evidence supporting the cascade metaphor is already in: brains exhibit cascade-like forms most evidently seen as "neuronal avalanches" (Beggs and Plenz, 2003; Dalla Porta and Copelli, 2019; Hahn et al., 2010; Jannesari et al., 2020; Klaus et al., 2011; Mariani et al., 2021; Miller et al., 2019; Pazzini et al., 2021; Plenz and Thiagarajan, 2007; Shew et al., 2011; Wu et al., 2019). Neuronal avalanches epitomize the cascade metaphor and the capacity of cascading processes across multiple scales to generate fluid and flexible structures that keep adapting to ever-changing circumstances (Mariani et al., 2021; Miller et al., 2019). Indeed, they may reflect essential resources in counteracting prediction errors that might otherwise accrue in a leaky computer-like relay of stimulation from sensors to actuators (Kaur et al., 2018). It may not be necessary to predict exactly or always correctly if the brain instead has the resources to weave vast contextual information together with small-scale stimulation (Van Orden et al., 2012).

This cascade metaphor offers researchers eager to sort out the relationship between brain, body, and context is the unifying framework of fractal and, more generally, multifractal geometry. A characteristic feature of cascades—avalanche or not, in the brain or not—is power-law scaling (Lovejoy and Schertzer, 2018). The pervading of cross-scale interactions, very big to very small, entails a scale-invariant structure. Whether or not it looks "self-similar," the cascade is always spreading a similar abstract form across all of its scales. The cascade may roll this way or that, but the action of the cascade may be understood through the evolution and change of power-law distribution functions (Kelty-Stephen et al., 2020). Here is, in fact, the reduction that the cascade metaphor enacts: the cascade metaphor offers to reduce various physiological and neural dynamics into a set of power-laws whose change can signal the response to or growth of constraints at various scales (Furmanek et al., 2020; Kelty-Stephen et al., 2020; Mangalam et al., 2021; Mangalam and Kelty-Stephen, 2021; Wallot and Van Orden, 2012). Helpfully for this purpose, power-law behavior can break down, even at specific scales, and it can also manifest in different forms for different sized effects.

As we build our theories to encompass not just brains but also bodies and contexts, we can find bodies and contexts themselves rich in cascades. The power-law form can become a formal common currency to express how brain, body, and context might behave similarly. There is already a tradition of calling the power-law behavior of cascades "universal," suggesting that under the critical dynamics entailed by power-laws may hold similarly across various systems composed of radically different materials (Lovejoy and Schertzer, 2018). Also, power-law behavior can spread across neighboring systems, and so we might begin to see how the cascade behavior spreads among brain, body, and context. For now, empirical research has examined various dyadic relationships between pairs of these; in all cases, the power-law behavior of one cascade is contagious and can spread to another cascade (Carver et al., 2017; Gutiérrez and Cabrera, 2015; Kelty-Stephen, 2017; Mangalam et al., 2020a, 2020b).

The exact circumstances for this contagion are still to be determined, but this mechanism affords a geometrical framework that allows us to consider how brain, body, and context may all coordinate in a way that is always slower than but often more adaptive than a computer. Furthermore, this metaphor may also pay all the same dues that the computer metaphor did in predicting the human, "brainful" responses: indeed, estimating the power-law behavior in the brain has been shown to predict the human cognitive response (Kardan et al., 2020a, 2020b). Less central to brain sciences, it is also important to note that the power-law behavior of bodily movement is predictive of the human perceptual responses (Kelty-Stephen et al., 2021; Mangalam et al., 2020b, 2020a). Of course, it is always possible that the two classes of predictive relationships are just coincidental. Notwithstanding, the progress and falsifiability of theoretically driven research outside the computer metaphor are by no means as ethereal or opaque as critics can sometimes protest (Wagenmakers et al., 2012). We can develop research beyond the computer metaphor through a cascade metaphor that we can elaborate (or fail to elaborate) within a clear geometrical framework of power-law relationships. But any success or failure must have no bearing on our search for an alternative to the computer metaphor.

The time is finally ripe for addressing the cascades that Turing knew about and had already begun to enlist for building the eventually intelligent computer. The computer metaphor has been of immense value and generated incredible decades of empirical research, but it is past the time when we had the available mathematical operations for addressing what Turing saw but could not put into numbers. Thus, there is no newfangled wisdom, just old wisdom with the mathematics that fit, and the policy of ignoring the constraints on the computer metaphor for the brain thrives only on the grounds that similar appearances require similar causes, i.e., as between a computing mind and a computing machine. However, Hume told us years before perceptual science that appearances can be deceiving. Thus, it is high time for us to pursue the strange possibility that a fluid cascade might generate a computation. Granular, symbolic detail may certainly compose the mosaic of a rule-driven computation, but the details themselves may emerge from a fluid substrate, crystallizing out of continuity (Bernstein, 1967). In any event, the physicalist hopes to anchor human computational ability may need to look no further than the cascade.

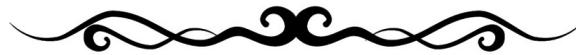

## 3. The brain is a control system—Paul Cisek[5]

All variations of the computer metaphor are based on the premise that the principal function of the brain is to process information and, thus, that brain functions can be subdivided into information processing functions. However, neurophysiological data strongly argues against such subdivisions, rejecting classical concepts of computational modules. Instead, I suggest the brain be viewed from the perspective of the process that produced it—evolution. This leads to the proposal, made many times over the last hundred years, that the brain is a control system. I briefly discuss evolutionary and neurobiological data that supports this view and leads toward a functional architecture for the brain consisting of parallel and nested sensorimotor control loops. I suggest that this architecture is more compatible with the process of brain evolution and comparative data across diverse species, that it better explains a wide range of neurophysiological findings, and that it provides a better conceptual taxonomy for understanding behavior.

### 3.1. What do we understand by the computer metaphor of the mind and brain?

Much of the debate about the computer metaphor results from the fact that it means different things to different people. e.g., some use it as a literal interpretation of the algorithm that the brain uses to implement higher thought, but others use it in a more general sense by describing the brain as a system that processes information. Consequently, some versions of the metaphor are met with criticisms that simply do not apply to other versions, leading to fruitless debates. Here, I will describe some of the ways the computer metaphor has been used, starting with the most specific version and gradually removing some of its assumptions to yield a version that is more general. However, I will ultimately argue that even this more general version of the computer metaphor misses the big picture of what the brain and behavior are really about.

---

5  Correspondence: paul.cisek@umontreal.ca (P. Cisek).

The most specific version of the computer metaphor was espoused by classical cognitive psychology in the middle of the 20th Century, which was strongly influenced by research in artificial intelligence. In that view, the metaphor was taken quite literally, and explanations of mental functions were expressed as computer programs with if-then rules, working memory buffers, explicit storage and retrieval of symbolic beliefs, et cetera. Of course, everyone agreed that biological mechanisms did not resemble such operations, but this was seen as irrelevant to the task of explaining mental life. This reasoning was justified by another metaphor, that of a **virtual machine**, where the hardware of biological tissues was seen as implementing the software of mental functions (Block, 1995). A similar ideas was expressed by Marr's (1982) famous distinctions between the computational, algorithmic, and implementational levels of explanation. This proposed independence of software from hardware was tenaciously defended via strong and persuasive mathematical arguments. However, it should be acknowledged that at least part of the attraction was that it excused its proponents from having to know anything about neuroscience, which at the time was still in its infancy. Thus, psychological theory proceeded largely without the burden of biological constraints.

Nevertheless, despite the prevalence of an explicit attitude that the mental is independent from the biological, proposals of functional architectures that more closely resembled neural architectures were favored over those that did not. In particular, one could relax the assumption that cognition is like a *symbolic* computer program and instead think of it in more general terms of mathematical computations like Boolean logic or linear algebra. This led to network explanations such as those of McCulloch and Pitts (1943) and Rosenblatt (1958), which demonstrated that artificial systems can learn to discriminate patterns given a set of examples. The ability of such systems to learn was so impressive that it was heralded in the media as the start of an AI revolution. That revolution has waxed and waned through several cycles of hype and disappointment, but even the most staunch skeptics must agree that today's AI systems such as multilayer networks are quite impressive at what they do.

But what is it that they do? They take an input pattern and produce an output pattern, effectively implementing a mapping that is learned through a training phase involving example input-output pairings. Often the mapping process is strictly feed-forward, such as in classic three-layer backpropagation networks, and sometimes it includes feedback (recurrent) connections. But regardless of the algorithm under the hood, the computational task that these systems perform is defined as transforming an input pattern into an output pattern–that is, what we can call "information processing." Here, "information" is usually defined in the Shannon and Weaver (1949) sense of the deviation of a pattern from randomness, and "processing" in the Turing (1936) sense of manipulating symbols or patterns. In other words, we can back away from the specific version of the computer metaphor that seeks to express mental operations in terms of computer programs and allow our explanations to instead take the form of neural networks.

Consequently, we can have a computational metaphor that is less literal but still operates within the broader metaphor of information processing, whereby mental functions are seen as processes that transform inputs into outputs. These could range from the kinds of mechanisms one might imagine happen at the retina, transforming a pattern of photons into a pattern of action potentials in the optic tract, to more complex mechanisms that integrate sensory evidence into beliefs and knowledge. They can even extend to behavior as a whole: where sensation is like input, muscle contraction is like output, and the interesting stuff in-between is like computation. Notably, the stage for this proposal was already set long ago by philosophers such as George Berkeley (1685–1753), centuries before electrical computers were available as an existence proof that purely physical systems can implement sophisticated behavior.

Why is the "information processing" metaphor so attractive? I think there are many reasons (Cisek, 1999). For one, it offers a purely physical explanation for mental life, exorcising the pitfalls of dualism. It also provides a mathematical language for describing the phenomena of psychology, thus elevating a hitherto "soft" science into something on par with physics. Finally, it suggests a way of bridging biological and psychological phenomena at various levels: allowing one to start with a largely conceptual bridge at the start of a research program and then progressively fill in the details as knowledge of physiological mechanisms grows. Because its arrival solved so many of the problems facing psychology at the time, to many the computer metaphor became seen as necessary to progress and its critiques as attempts to undermine science (Still and Costall, 1991).

Another reason the computer metaphor was so valuable is that it outlined a research strategy for subdividing behavior and brain mechanisms. Importantly, if we define some behavioral task as an input-output mapping, then as long as we can measure the input and output, we can infer the intervening computation through various methods such as the "system identification" techniques of engineering. If that inference proves challenging for some difficult problem, information processing offers a strategy for subdividing that problem into smaller, presumably more manageable ones. In particular, if one suggests a hypothesis on how the large problem is composed of two steps that involve constructing some intermediate representation, then one has turned the large input-output problem into two smaller problems, one that produces the intermediate representation as its output and a second that then uses it as its input. Most importantly, one now has an explicit and testable *prediction* of what kinds of intermediate representations should be found in the brain through neural recording, functional imaging, or some other empirical technique.

e.g., it seems impossible to explain all behavior with a single mechanistic theory that captures everything from perceptual phenomena to emotional states and the details of motor control. However, we can apply a strategy of "functional decomposition" to subdivide behavior into "perceptual" mechanisms, which use sensory information to build an internal representation of the external world, "cognitive" mechanisms, which use that representation to build knowledge and make decisions, and "action" mechanisms, which implement decisions through muscular commands (**Fig. 5**). Each of these is still a daunting problem, but they too can be subdivided into subproblems. e.g., visual perception could involve distinct mechanisms to separate figure from ground, detect specific behaviorally relevant features, bind them into labeled objects, and other specialized computational modules. Likewise, cognition could include distinct mechanisms for memory storage and retrieval, logical inference processes, and decision-making.

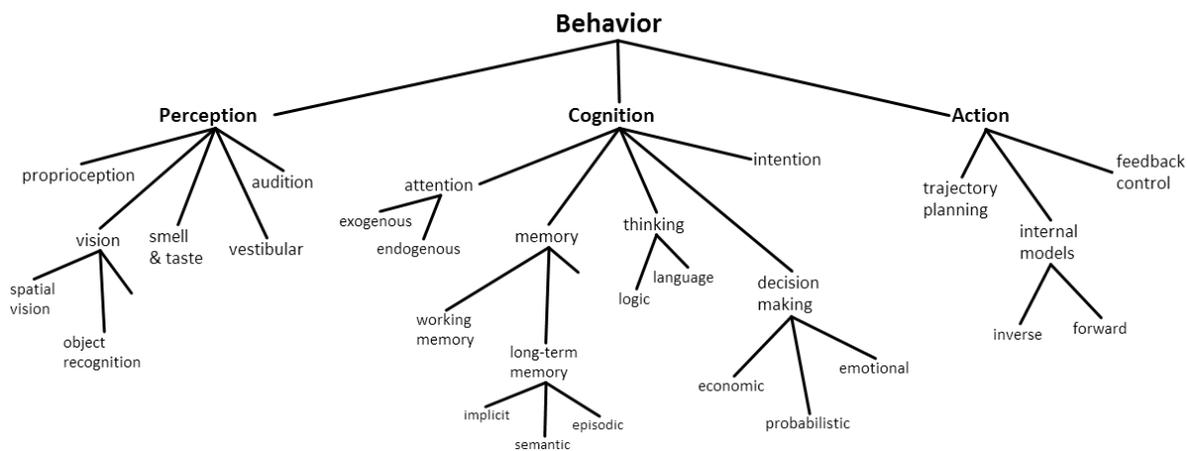

**Fig. 5.** Sketch of the standard conceptual taxonomy in cognitive science / psychology. Reproduced with permission from Cisek (2019).

In summary, the computational metaphor can be formulated at many levels. At the highest and most general level, it is often tacitly assumed that "information processing" is just what the brain does. e.g., Decharms and Zador (2000) write that "The principle function of the central nervous system is to represent and transform information and thereby mediate appropriate decisions and behaviors" (p. 613). The basic assumption that the brain's job is to process representations now pervades neuroscience, as evidenced in the introduction of the major textbook in the field: "The task for the years ahead is to produce a study of mental processes, grounded firmly in empirical neural science, yet still fully concerned with problems of how neural representations and states of mind are generated" (Kandel et al., 2013, p. 4). Indeed, many have taken the attitude that it is a waste of time to argue whether the brain is doing computation and better to just get on with the task at hand—determining what that computation is and how it is implemented. So if that is true, then why do people continue to write articles such as this one?

**3.2. What are some of the limitations of the computer metaphor?**

One challenge to even the most general variants of the computer metaphor comes from the growing body of knowledge in neuroscience. While it is possible to "decode" some relevant behavioral variable from some neural activity pattern in the brain, that activity is so strongly influenced by other variables (including aspects of behavioral context) that the decoding only works in completely unnatural situations and only if one already knows all other relevant variables (Brette, 2019). In other words, the finding that some neural activity covaries with some variable we have defined to describe the external world is trivial—this has to happen in any system that is dynamically coupled with the world. But it does not imply that the functional role of that neural activity is to encode information about that variable in the same way we have defined it. Your heart rate covaries with running speed, but the functional role of the heart is not to encode speed.

Furthermore, the distribution of neural activity across brain circuits does not respect the classical theoretical categories of computational functions (Figure 5). e.g., while introductory textbooks often describe certain parts of the brain as "sensory", or "cognitive", or "attention" regions, in almost all cases further research shows these descriptions to be superficial (Anderson, 2014; Cisek and Kalaska, 2010; Lindquist and Barrett, 2012). e.g., neural responses to visual stimuli can be found in putatively "motor" regions such as premotor and motor cortex in as little as 50 milliseconds, much earlier than in parts of the brain implicated in visual recognition (Ledberg et al., 2007; Schmolesky et al., 1998). And yet, during natural behavior the entire cerebral cortex is dominated by motor-related activity (Musall et al., 2019). The visual system famously diverges into a **dorsal stream** sensitive to spatial locations and a **ventral stream** sensitive to object identity (Ungerleider and Mishkin, 1982) but there's no place where it all comes together to yield a unified internal model of the visual world. Variables related to decisions are found almost everywhere researchers have looked for them, including not only the frontal lobe areas traditionally associated with cognition (Padoa-Schioppa, 2011; Rushworth et al., 2012), but also the parietal cortex (Platt and Glimcher, 1999; Roitman and Shadlen, 2002), frontal eye fields (Schall, 2004), premotor cortex (Pastor-Bernier and Cisek, 2011), basal ganglia (Arimura et al., 2013; Ding and Gold, 2010), and even the primary motor cortex (Thura and Cisek, 2014) and superior colliculus (Basso and May, 2017), two structures just a few synapses away from muscles. Neural activity putatively related to attention has been observed throughout the cortical and subcortical circuits implicated in sensory and motor processes, raising doubt on whether attention is even a coherent concept (Anderson, 2011; Hommel et al., 2019). In general, analyses of brain activation patterns in terms of cognitive concepts such as working memory, semantic judgements, or response inhibition, regularly show a high degree of overlap and distribution (Anderson, 2014). In short, the categories of concepts we inherit from psychology simply do not fit the brain (Buzsáki, 2019; Cisek and Kalaska, 2010; Lindquist and Barrett, 2012).

Given that the assumption of the virtual machine led concepts such as attention, decision-making, working memory, etc., to be deliberately developed in isolation from neurobiological data, it should not be surprising that they do not fit with that data. Furthermore, these concepts were developed almost exclusively to explain higher human thought, which is clearly not the totality of brain function. Even if such a virtual machine metaphor did apply to explaining human thought, why would we expect its concepts to apply to animal groups from which our lineage diverged millions of years before humans existed? Expecting that concepts useful for explaining human thinking should be the appropriate pieces for explaining animal brains is tantamount to a kind of bizarre "inverted phylogeny," as if evolutionary innovations can trickle backwards through time. Obviously, that idea needs to be thoroughly abandoned.

But then what is left? On the one hand, one could retain the strategy of explaining human thought by reference to the computer metaphor and concepts such as working memory, encoding and decoding of memories, and serial processing of perception-cognition-action, but restrict it just to the human brain. After all, the human brain is capable of amazing feats of learning, with each generation adapting to entirely novel domains of interaction, from the written word to smartphones. One could therefore take the position that the computational organization of the human brain is entirely a product of learning, a "software" solution that has little to do with the underlying biological "hardware." However, taking that attitude would imply abandoning data from nonhuman animal studies and discarding all of the insights it offers. It would also imply ignoring the similarities between the apparent functional organization of human and animal brains, and retreating one's thinking to just the places where they differ. Zador (2019) makes a strong case for why this would be a serious mistake. If the functional organization of the brain was entirely a product of learning, then why, given their dramatically different lifestyles, should human brains and monkey brains exhibit such similar activation patterns when they perform similar tasks?

All of these questions lead to a larger one: What if the basic metaphor of the brain as a computer, whether like an explicit symbolic program or a mapping that transforms inputs to outputs, is simply wrong? What would be the alternative?

### 3.3. What metaphor should replace the computer metaphor?

A car is a physical system that processes energy. In the case of an internal combustion engine, the source energy is chemical and it is turned into kinetic energy via a series of small controlled explosions that drive a piston attached to the drive shaft. In the case of an electric motor, the source energy is electrical, and it is turned into kinetic energy using magnetic fields. Regardless of the type of motor, the kinetic energy from the drive shaft undergoes a series of coordinate transformations that turn rotation into translation of the vehicle. Would anyone argue that this description is false? Cars do, in fact, process energy.

However, does energy processing provide a description of cars that leads one to a good understanding of cars? Without going beyond the metaphor of energy processing, how would one explain such things as windshield wipers? How would one understand the difference between cars and other systems that also process energy, such as chloroplasts or nuclear power plants? To understand cars, it is not enough to merely keep repeating the fact that they process energy, but to go beyond that to observe that the purpose of cars is to transport people from place to place. In other words, just because a metaphorical description may not be false does not imply that it is useful as the foundation for understanding. Cars do, of course, process energy, but that is just a means toward a larger purpose–to move people from place to place.

Likewise, brains do indeed process information, but that is just a means toward a larger purpose—to control behavior (Ashby, 1952; Cisek, 1999; Maturana and Varela, 1980; Powers, 1973). In other words, brains are not simply input-output devices, like computers, which take sensory input and produce the proper motor output. Instead, they are more correctly described as **control systems** whose task is to produce motor output that results in the proper input.

Importantly, in the first case (input-to-output) the notion of what is the proper output is ill-posed. Usually, what is considered "proper" is up to the designer of the system–e.g., for a machine vision system the proper output is the correct labeling of objects in the image, based on the rules implicit in the training set. Consequently, the purpose is external to the system, imposed by the goals of the programmer. In contrast, for a behavioral control system the notion of what is the proper input is trivial. Input that specifies being in a state that leads to continued functioning (e.g. fed, warm, and safe) is desirable, whereas input that signals a state of disfunction (e.g. proximity to a predator) is undesirable. The task of behavior is to produce actions that change the input toward one that specifies a desirable state.

Adaptive behavioral control is possible only because certain contingencies between output and input are inherent in the environment. e.g., consuming certain kinds of items improves one's nutrient state. Orienting away from a predator and then running tends to succeed in escaping it. Because these output-input contingencies are consistent and discoverable, they can be exploited to meet the organism's needs by establishing the complementary input-output policy, either through evolution or learning. Thus, the task of behavioral control is not simply to respond to the environment, or to build knowledge about it, but to complement the dynamics inherent in the environment in such a way that the whole organism-environment system tends to move toward desirable states and away from undesirable ones. This is by no means a novel proposal. The idea that the brain is a control system has been repeatedly introduced and re-introduced for over a hundred years of philosophy, from John Dewey in the 19th Century to George Herbert Mead, Maurice Merleau-Ponty, and Andy Clark since then. It was a central tenet of models proposed by engineers like Norbert Wiener, W. Ross Ashby, and Rodney Brooks, psychologists like Jean Piaget, James Gibson, and William Powers, and neuroscientists like Valentino Braitenberg, Karl Friston, and Lisa Feldman Barrett. In particular, Friston and Barrett both emphasize the importance of predictive control, whereby the system does not simply wait for events to perturb it away from desirable states, but continuously anticipates and acts pre-emptively to prevent such perturbations whenever possible (Friston, 2010; Katsumi et al., 2021).

Of course, one could say that a control system is just a special case of an input-output system, one in which the input depends on the output. Indeed that is true. It's a special case, and therefore it offers a more precise and useful description of what a brain is doing. It presents a smaller search space of potential theories, better constraining both scientific and engineering research goals. Just like energy processing is an incomplete and underspecified description of what cars do, mapping inputs to outputs is an incomplete and underspecified description of the function of brains. We can do better.

To summarize, I began this article by starting with the explicit computer metaphor for the brain and stepped back from some of its assumptions toward the more general idea of the brain as an input-output system. But then, that description is too general and not sufficiently constraining to guide further research. Consequently, we need to add a bit more precision—what *kind* of input-output system? The answer, as has been proposed for many decades, is that the brain is a *control system*. Its overall role is to maintain the organism in a desirable state and away from undesirable states, and it accomplishes that role by establishing sensorimotor control policies that complement the motor-sensory contingencies inherent in the world. I don't think that anyone would disagree with this claim, or find it controversial. Much more controversial is my second claim: that accepting the first claim changes everything.

### 3.4. What empirical findings support your preferred alternative metaphor?

One of the first implications of taking control systems as a metaphor for the brain is that it leads to a different way of breaking down the large problem of behavior into smaller sub-problems. To characterize a control system, one must consider the entire causal loop from input to output and around again, without breaking it down into hypothetical serial stages. Given the complexity of the nervous system, with all the sophistication of mechanisms along the way, from sensors to muscles, this might seem like a huge and unwise mistake. And indeed, in many cases, the problem is daunting. However, in many cases explaining behavior as a whole is actually much simpler than explaining any of its putative serial stages.

Consider the example of the "outfielder problem:" How does a baseball outfielder catch a fly ball? One approach to the problem is to break it down into subproblems. First, the sensory information must be used to detect the ball and calculate its trajectory through space. This requires combining the retinal position of the ball with information about the position of the eye in the head and the head on the body, as well as with depth cues such as vergence, to calculate the position of the ball with respect to the body. Next, that position must be sampled for a period of time to determine the ball's trajectory. Next, the trajectory must be extrapolated into the future using knowledge of the parabolic flight of objects moving in gravity, as well as estimates of wind direction and speed, to predict where the ball will land. Finally, a motor plan needs to be produced and executed to bring the player to that location. All of these steps are subject to noise and uncertainty, but these can presumably be dealt with using Bayesian signal integration or other sophisticated techniques. Such an approach is plausible, but it requires highly accurate measurement of multiple variables that are difficult to estimate in a real-world scenario. However, there exists a much simpler alternative strategy. The alternative is to keep your eye on the ball and move forward if your gaze angle (head+eye) shifts downward, move backward if your gaze angle shifts upward, all the while matching the ball's lateral position. As long as you keep your eye on the ball and move to keep your gaze steady, then the ball will fall into a glove brought in front of your face. In fact, that's the strategy used by real players, as shown through studies that perturb vision with virtual reality (Fink et al., 2009). In summary, solving the whole control problem is far easier than solving any of the putative subproblems into which one might be tempted to subdivide it.

There is another good reason to expect that the mechanisms of biological control do not break down into a series of computational stages, and that comes from evolution. Contrary to how it's sometimes treated in popular science, evolution is not nature's method for finding solutions to problems posed by the world. It's not like engineering, where one first defines a problem and then proposes and tests candidate solutions. Evolution does not even identify problems to be solved. Instead, it merely produces variations of an ancestral organism and then, through natural selection, favors those that happen to accomplish something that used to be a problem. Furthermore, in order for a variation to even enter the arena of natural selection, it must first be possible as a modification of the ancestral system. This is a massive source of constraints that makes most variations completely unavailable. The reason is that the genome does not describe a "blueprint" of the body or "connectome" of the brain, but instead specifies a "recipe" for producing the brain and body through a long and convoluted developmental process. Like any recipe, that process consists of a sequence of stages, each of which relies on previous stages and sets up conditions for later ones. Consequently, no mutation, no matter how advantageous, can produce a whole new "module," complete with input and output connections and a full developmental program, which implements some optimal solution to some externally defined problem. Instead, evolution can only elongate or shorten developmental sequences, duplicate systems and then differentially specialize them, and gradually shift tissue growth, cell migration, and axonal projection patterns. All of this dramatically limits the viable organisms that can enter into natural selection, and limits how far two species can diverge away from their common ancestor. Finally, natural selection has no way of evaluating the efficiency of any individual parts of these systems, but only the survival of the entire organism. In short, evolution does not identify computational subproblems and can neither produce nor evaluate their potential solutions. As Hendriks-Jansen (1996) put it, "functional decomposition and natural selection do not mix."

So what is the result? Must we give up on the idea of subdividing brain functions into subproblems and instead try to solve all of behavior all at once? That seems unrealistic, and indeed it is not necessary. A subdivision of neural systems is still possible, but it need not be guided by definitions of putative computational problems that we invent or inherit from psychological traditions. Instead, it can be guided by considering the process that built the brain–again, by considering evolution. The very same developmental constraints that so strongly limit the possibilities for arbitrary modifications to the brain's functional architecture also make that architecture easier to infer. In particular, if we know the phylogenetic relationships between different species of animals and compare what is similar and what is different in their brains, then we can make plausible hypotheses on the evolutionary stages along a given lineage of interest, and reconstruct the sequence of how control systems elongated, differentiated, and specialized along that lineage. I have referred to this approach as "phylogenetic refinement" (Cisek, 2019), but it is really nothing more than an application of the comparative method of biology to theoretical neuroscience.

Following an evolutionary approach, one begins with considering the basic functions of all living things, metabolism and replication, of which the former involves closed-loop control (Cannon, 1939). Control within the organism is what we call "physiology," while control that extends through the environment is what we call "behavior" (Ashby, 1952; Cisek, 1999; Maturana and Varela, 1980; Powers, 1973). Because behavior in a world of growing complexity is an endlessly expanding challenge, the history of evolution of mobile animals is a history of extending that control further and further into the world. Guided by the growing body of comparative data on neural evolution, I have recently proposed how that history of control unfolded along the lineage from the earliest multicellular animals to primates (Cisek, 2022, 2019), drawing upon similar approaches followed by others (Feinberg and Mallatt, 2016; LeDoux, 2020; Murray et al., 2017; Passingham and Wise, 2012; Striedter and Northcutt, 2019).

To summarize, early nervous systems included a high-level controller using hormonal signaling, which in our lineage evolved into the hypothalamus, and a low-level controller using synaptic transmission, which evolved into the midbrain and spinal cord (Arendt et al., 2016). As our ancestors became more mobile, these systems differentiated further, with the **hypothalamus** sprouting a telencephalon responsible for foraging control, while the midbrain controlled the details of visually guided approach and avoidance (Saitoh et al., 2007). With the expanding behavioral repertoire of land animals, sensorimotor control gradually shifted to the **telencephalon**, in particular the "pallium," which gives rise to the cerebral cortex, hippocampus, and many other key structures (Puelles et al., 2013). When mammals retreated into nocturnality during the age of dinosaurs, they regressed midbrain visuomotor control but enhanced cortical olfaction, audition, and somatosensation, and then upon returning to diurnal life as primates re-enlarged their visual systems but now primarily in the cerebral cortex (Kaas et al., 2022). Throughout this long sequence, however, the basic organization was not one of serial processing stages but parallel and nested control loops, competing against each other for execution through increasingly complex modulation and mutual inhibition (Cisek, 2022). Even the primate brain, I would argue, can be understood as such (Cisek, 2007).

One result of following such an evolutionary approach is that it leads to a conceptual taxonomy of functions (**Fig. 6**) that differs in many ways from the classical conceptual taxonomy of psychology. Its elements are not mental faculties such as attention or decision-making, but a hierarchy of control problems. I would claim that this taxonomy fits much more naturally with the growing body of neuroscientific data, mapping to specific circuits and regions as noted in **Fig. 6**. e.g., the divergence of the visual system described above makes perfect sense if the dorsal stream is actually involved in guiding movements (Goodale and Milner, 1992), specifying multiple potential actions (Cisek and Kalaska, 2005; McPeek and Keller, 2002), while the ventral stream collects information for biasing a competition between those potential actions (Cisek, 2007; Cisek and Kalaska, 2010). This also explains the widespread influence of "decision-variables" (Hernández et al., 2010; Pastor-Bernier and Cisek, 2011; Platt and Glimcher, 1999; Roitman and Shadlen, 2002; Romo et al., 2002; Thura et al., 2016; Thura and Cisek, 2014), and "attentional" effects (Boynton, 2005; Hommel et al., 2019; Treue, 2001), as well as the parallel processing observed during even the simplest tasks (Ledberg et al., 2007). A more thorough discussion of relevant neuroscientific data is beyond the scope of the present paper, but I refer the reader to several reviews (Anderson, 2014; Cisek and Kalaska, 2010; Lindquist and Barrett, 2012). In short, if we back off the assumption that the brain must be understood in terms of a series of explicitly defined input-output problems and instead see it as a set of interacting control loops, then all that confusing data begins to make a lot more sense.

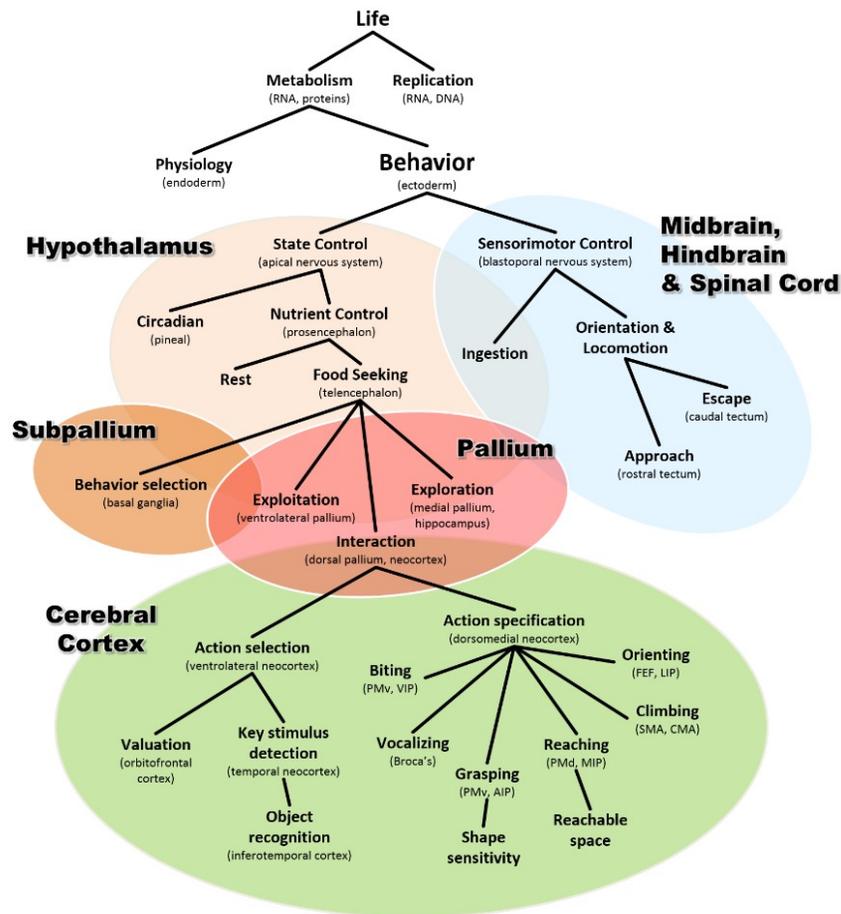

**Fig. 6.** An alternative conceptual taxonomy based on evolution. Here, each functional category is proposed to be a specialization of the one above it, and to correspond to a biological structure that emerged as a specialization within an ancestral structure. Reproduced with permission from Cisek (2019). AIP: anterior intraparietal area; CMA: cingulate motor area; FEF: frontal eye fields; LIP: lateral intraparietal area; PMd: dorsal premotor cortex; PMv: ventral premotor cortex; SMA: supplemental motor area; VIP: ventral intraparietal area.

In summary, in trying to explain the brain we are not obliged to do so in terms of the computer metaphor, input-output mappings, or general "processing" of "information." These concepts are not false per se, but they distract us into fruitless and irrelevant attempts to "decode" the brain or reverse engineer putative functional units that only exist in our imagination. Instead, we can look at the brain in an alternative way–as a control system. The concept of a control system is the basic metaphor in all other domains of biological science, from physiology to animal behavior, and it provides a more natural framework for interpreting neural data. In short, the computer metaphor was useful many decades ago to help psychology get away from dualism, but now it is time to move on.

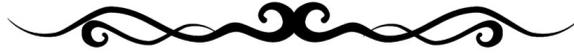

# 4. Dissipative structures as an alternative to the machine metaphor of the mind and brain—Benjamin De Bari[6] and James Dixon[7]

The computer metaphor derives from two parallel histories of thought, the first questioning the operation and activity of minds and brains and the second aiming at finding a minimal model of the mind and brain. These histories informed two separate hypotheses of mind, epistemic constructivism, and the mechanistic hypothesis; we focus on the latter. First, we briefly review how the brain has historically been discussed with machine metaphors and identify five tenets that define a machine. We review findings in neuroscience that motivate that the brain demonstrates exceptions to these tenets and thus ought not to be considered a machine. We offer that an alternative classification may be found in far-from-equilibrium self-organizing systems known as dissipative structures. We review the properties of these systems and suggest that the brain is more like a dissipative structure than a machine. If brains do not fit the mechanistic hypothesis underwriting the computer metaphor, then the cognitive sciences may need to seek alternative metaphors based on the assumption that minds and brains are some other kind of natural systems, namely dissipative structures.

> *The brain being indeed a machine, we must not hope to find its artifice through other ways than those which are used to find the artifice of other machines. It thus remains to do what we would do for any other machine; I mean to dismantle it piece by piece and to consider what these can do separately and together.*
>
> — Nicolaus Steno (Bresadola, 2015)

## 4.1. What do we understand by the computer metaphor of the mind and brain?

While the technology of computational machines is a relatively recent development in human history, the computational metaphor of the mind has roots dating back to antiquity. Two parallel histories of thought have precipitated our contemporary perspectives, each corresponding to a broad question about the mind and brain. The first question concerns *the operation* of minds and brains; what sorts of things do they do? What kinds of processes support intelligent action, guaranteeing stable knowledge, certainty, and understanding in the face of inconstant and imperfect experience? The second question concerns the proper classification of mind and brain as natural systems; what *kind of thing* is a mind or brain? The modern computer metaphor provides answers to both of these questions. Let us briefly tour some landmark ideas in the history of cognitive science that have paved the way.

At least as far back as antiquity, thinkers were engaged with the problem of making epistemic contact: how can an organism (primarily a human) know about

6  Correspondence: .benjamin.de_bari@uconn.edu (B. De Bari).
7  Correspondence: james.dixon@uconn.edu (J. Dixon).

things in the world, and how can the organism use that knowledge? It was taken to be the case that our immediate contact with the world is impoverished relative to the true state of affairs in the world. Plato conceived of this limitation with his "Allegory of the Cave," which likened the mind to a person chained within a cave, facing the back wall, locked away from the world. Events, beings, and objects in the world cast shadows on the cave wall, and these shadows are the only contact with reality that the mind has. Thus, when an organism contacts the world, it is only sensitive to *shadows* cast by objects and events, insufficient information to meaningfully guide behavior. If an organism is to survive and operate effectively in the world, and if the immediate source of information is insufficient to guide behavior, then the organism must *add something* to that stimulation (Lombardo, 2017). This assumption about the fundamental inferential function of perception-action-cognition (PAC) is the basis of the *information processing* perspective on the mind.

A variety of theories of how the PAC system adds to the impoverished stimulation have been proposed, from Plato's recollection of ideal forms to Malebranche's divine intervention, to the Gestaltists' nativist organizational laws (Turvey, 2018). But perhaps the most dominant approach originated with Helmholtz's theory of unconscious inferences (Pastore, 1971). To produce meaningful knowledge of the world, one must extrapolate from the limited sensory information, making inferences (specifically rational, logical, computable ones) about the true properties of the world: given a particular sense datum, what thing-in-the-world is likely to have produced that sensory information? But for such a process to be effective, there must be constraints on the type of inferences we make; rules for linking sensory stimulation to perceptual states related to things-in-the-world. Here we can see a direct connection to computational processes: Information in one form (sense-data—symbols) are processed by a system of rules (inferential processes—logic operations) and transformed into a new system-state (perception—encoded information) that represents the thing-in-the-world. Herein we refer to this enterprise of understanding the mind as a system for enriching stimulation and building a model of the world as *epistemic constructivism*.

This inference-machine metaphor has been extended into a more general computational approach to the mind; it has propagated into nearly every aspect of contemporary psychology. The received wisdom is that the mind is in the business of manufacturing representations of the world, building the world inside the brain and that transformations of these representations comprise cognition. Moreover, the process of representation making and transforming is manifest in an implicit ratiocination, a process of applying logical rules and encoding relationships to transform sensation to perception, ambiguous information to knowledge, physiological states into emotions, and so forth. Or as Hobbes famously offered, "by ratiocination I mean computation" (Hobbes, 1656, as cited in Haugeland, 1985). Cognitive science has built on Turing's insights that such algorithmic transformations of symbols can be mechanically implemented, facilitating the perspective of mind as a computing machine (Turing, 1950b). As computational technology has developed, so too has the theory of this process, from Physical Symbol Systems and mental scripts to connectionist networks, to Bayesian inferences. Nevertheless, the underlying architecture of the argument has remained largely unchanged: the mind and brain are a special-purpose system for generating and manipulating internal representations of the world that can be used to guide behavior.

The discussion above centers on what the mind does or how it operates. However, the computer metaphor is best understood with reference to another implicit hypothesis about what a mind is; that is, to what class of natural systems does the mind belong? The prevailing answer has been that a mind, by virtue of being a product of biological processes (i.e., the brain), is a special kind of *machine*. In the history of cognitive science, Rene Descartes provides a common entry point for discussing mechanical theories of biology; he famously carved up human psychology into the animate incorporeal soul and the inanimate material body. The human body, he offered, was fundamentally mechanical in nature. This perspective was likely influenced in part by the then contemporary inventions of automata, clockwork-style machines with intricate mechanical components that mimicked living systems (Leiber, 1991). Later, in the 19$^{th}$ century when the brain was increasingly understood to participate critically in PAC processes, it too was conceived of with machine-metaphors. e.g., when the electrochemical transduction of neurons was being investigated, analogies were drawn to the telegraph networks of the time (Lenoir, 1994). In the 20$^{th}$ century, when the dynamic nature of interneural communication and connection was identified, analogies were drawn to the switchboards in telephone networks (Bergson, 1950; Kirkland, 2002).

Oddly enough, however, the identification of neural processes with digital computational ones did not arise in quite the same order; rather, conceptions (if impoverished ones) of the activity of neurons informed the design of digital computers. A famous development in the history of cognitive science was McCollough and Pitts' formulation of a model of the brain where neurons were arrays of signal transduction systems that performed formal logical operations dependent on how they were connected (McCulloch and Pitts, 1943). The pair derived this model from their understanding of the organization of nervous systems. The model then went on to revolutionize computational theory, inspiring von Neumann and thus the digital computing legacy we have inherited (Cobb, 2020). Neuroscientists at the time of McCullough and Pitts' writing however noted that the model of nervous activity they schematized was in fact substantially different from how the nervous system is actually organized (Cobb, 2020; Piccinini, 2004). Subsequent attempts to identify logic gates in nervous tissue have had some success, while also demonstrating that their operation is different from McCullough and Pitts' formulations (Dobosiewicz et al., 2019). Nevertheless, the impoverished model of neural organization and activity was incorporated into cognitive science and the notion that the brain is a complex logical computing machine remains foundational in current conceptions of the brain.

These historical developments naturally led to the supposition that the mind and brain are best understood as kinds of machines. This idea, the mechanistic hypothesis, and the hypothesis of epistemic constructivism, jointly underwrite the computer metaphor: The mind and brain are complex machines whose operation consists of encoding and transforming system-states that function as representations. A wealth of discussion has been devoted to questioning the latter hypothesis, primarily derived from the philosophy of American Pragmatism stemming from William James's work (Chemero, 2009), and several camps of thought have been established in opposition. These include Ecological Psychology, as formulated by James Gibson and elaborated on by Turvey, Shaw, and many others (Gibson, 1979; Shaw et al., 1982; Turvey, 2018); Interactivism, as developed by Bickhard (Bickhard, 2009); Haken's Synergetics (Haken, 1981); the dynamical systems stance such as that posed by van Gelder (van Gelder, 1998); and even approaches to robotics such as Brooks's incremental intelligence (Brooks, 1991). Critical distinctions notwithstanding, all these theories are alternatives to the epistemic constructivist perspective on the mental operation. Because such responses have been so thoroughly discussed elsewhere, we do not aim to evaluate this assumption of the computer metaphor and turn our attention instead to the mechanistic hypothesis.

In short, the discussion revolves around whether organisms and, consequently, brains and minds are best described as machines or as systems of a different kind. Borrowing ideas from Robert Rosen (Rosen, 1991), we offer two possible perspectives on natural systems; organisms are either special kinds of machines or they are ontologically distinct (**Fig. 7**). While Rosen's insights motivate this latter perspective, it leaves organisms oddly disconnected from the rest of the physical world in terms of scientific theory. In previous work (Chung et al., 2022; De Bari et al., 2020; Kondepudi et al., 2017), we have expanded on Rosen's proposition by identifying a broader class of natural systems that encompasses organisms and their subsystems (e.g., the mind and brain), but excludes machines. This class, called dissipative structures, has quite different properties from the class of systems to which machines belong, called equilibrium structures. In the subsequent section, we outline the criteria necessary to consider a system as a machine and evaluate whether brains fit those criteria.

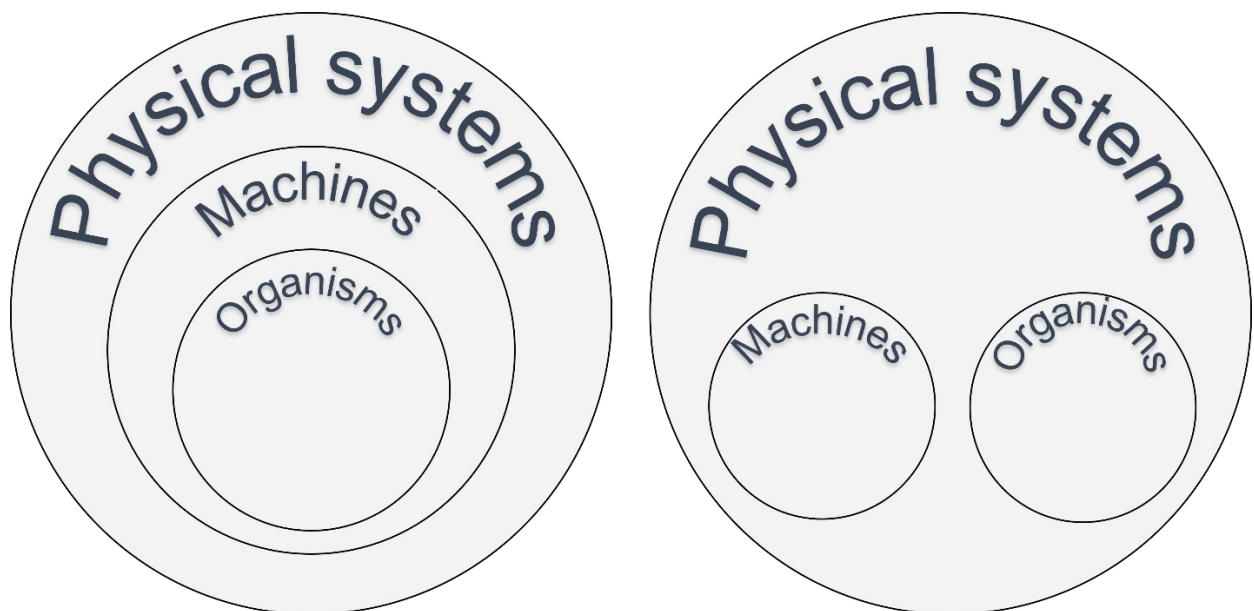

**Fig. 7.** Two perspectives on the ontology of natural systems and machines. **Left:** Organisms, and consequently brains and minds, are special kinds of machines. **Right:** Machines and organisms are distinct classes of physical systems, and consequently, brain and mind are not machines. We develop a modified formulation of the latter perspective.

### 4.2. What are some of the limitations of the computer metaphor?

We have suggested that the computer metaphor derives from the hypothesis of epistemic constructivism and the mechanistic hypothesis of the mind-brain organization and function. This latter hypothesis leads to weaknesses of the computer-metaphor simply because organisms are not machines. This is not a trivial distinction—it is not just that brains are not manufactured or made of aluminum and silicon—rather, assuming that biology is fundamentally mechanical invokes properties that are inconsistent with biology. The mechanical hypothesis incorporated the following tenets; (i) part-function isometry, (ii) decomposable parts, (iii) part-whole independence, (iv) equilibrium structures; and (v) externally prescribed function. The first tenet, part-function isometry, holds that every identifiable part of a mechanical system has a specific function. The second tenet, decomposable parts, holds that the system can be differentiated into neatly separable parts, and via (i) those parts have distinct functions. Features (i) and (ii) together constitute what Robert Rosen called *fractionability*, or the separability of parts with localized functions (Rosen, 1991). The third tenet, part-whole independence, requires that the function of a given element in the system is independent of the organization and processes within the rest of the system. In other words, function is intrinsic to an element or part by virtue of its structure, not by virtue of its context, and so does not change dependent on where it is in the system nor on what the system is being used for. The fourth tenet, equilibrium structures, holds that the mechanisms and their component parts are equilibrium structures. That is, they are stable at thermodynamic equilibrium and their dynamics are driven to minimize deviations from equilibrium. Below, we argue that it is this last tenet that provides the license for all the others and sets the mechanical hypothesis off on the wrong foot entirely for explanations of biological function. The fifth tenet refers to how machines are used by intelligent agents and essentially posits that machines are special kinds of tools whose function is *extrinsically* derived from the intentions of an intelligent agent, rather than being intrinsically determined by the organization of the system.

It has long been noted that living organisms fail to satisfy the above mechanical tenets. Firstly, organisms are not fractionable in the same way machines are (Kondepudi et al., 2017; Rosen, 1991; Von Bertalanffy and Sutherland, 1974). While we may draw boundaries to define physiological structures, their functions are context sensitive and not entirely localized. Consider Rosen's distinction between the wings of a bird and those of an airplane (Rosen, 1991). Flight requires both propulsion and lift, functions that are encapsulated separately in the engine and airfoil (wing) of a plane respectively. However, a bird's wings are both engine and airfoil, such that the functions of propulsion and lift cannot be attributed to distinct anatomical components.

We should expect that the brain is similarly non-fractionable; evidence for this can be found in the debate over the localization of function within the brain. A longstanding result in neuroscience, and a foundational observation in the localization debate, is the remarkably consistent loss of language comprehension in individuals suffering damage to the left temporal hemisphere known as Broca's area (Head, 2014; Luria, 2011). However, research with novel non-invasive imaging techniques has explored the activity of Broca's area, revealing that damage to Broca's area alone is not sufficient to produce the classic aphasic pathology, pointing to a more distributed speech production system (Ardila et al., 2016). On the assumption that semantic linguistic content might thus similarly be localized in Wernicke's area or elsewhere, researchers performed fMRI scans while participants listened to two hours of narrative stories. Responses to different semantic categories of words (e.g., visual, social, emotional, etc.) were widely distributed across the cortex, revealing no localization (Huth et al., 2016). Further, despite well-established evidence of localization like that of Broca and Wernicke, there exist individuals with normal functional capacities who lack the typical associated anatomical structure, such as those whose sense of smell is perfectly operational despite missing their olfactory bulbs (Weiss et al., 2020). Structure and function do not appear to be intrinsically linked in organisms nor in the brain, as typically assumed under the mechanical hypothesis.

Another empirical example of the proposed structure-function relation under the mechanical hypothesis has been the notion of "wiring diagrams" of the brain, especially those advocated by the Human Connectome Project (Seton-Rogers, 2013; Sporns et al., 2005). These "connectomes" may be high-level depictions of the interaction among cortical regions, or more detailed neural level mappings of connectivity. In either case, the essential idea is that we might lay the foundations for understanding function in the brain by rigorously mapping the connections among cortical elements. Connectomes have been mapped in some non-human organisms, including the sea slug *C. elegans* (White et al., 1986), the learning system in the fly *Drosophila* (Takemura et al., 2017), and even for the digestive system of a crab *Cancer borealis* (Bargmann and Marder, 2013). Frustratingly, in all such cases, a complete description of the structural connectivity of neurons does not yield *functional* predictability of the network's activity (Bargmann and Marder, 2013; Cobb, 2020). Further, standard functions can derive from different structures, as distinct networks of neurons can separately produce the same swimming behaviors in two species of nudibranch (Sakurai and Katz, 2017). While such an approach may shed the stronger assumption of functional localization by seeking to understand distributed networks, it implicitly retains the assumption of structure-function isometry by seeking to explain the network's function in terms of connectivity.

The functions of physiological elements (however construed) are not insulated from context as required by tenet three, part-whole independence. Outside of the brain e.g., muscular activity displays different functions for different contexts. When one has their arm at the side of their body, activating the pectoralis major will *adduct* the arm, bringing it towards the front of the body. However, while the arm is raised perpendicular to the body, activating the pectoralis major will *abduct* the arm, lifting it away from the body (Turvey et al., 1982). The point here is that the function of particular muscles (i.e., component parts) depends on the larger context of body position (i.e., the whole). Such context-conditioned variability is widespread within anatomy, and we should expect that the brain is no exception. e.g., the dynamics of a neuron are affected by volumetric interactions of neurotransmitters with neuromodulators (Nusbaum et al., 2017), as well as the individual history of a given nerve (Gina et al., 1994). Some researchers have pointed to the developmental trajectories of the brain to highlight its context-sensitivity. Some neural systems exhibit redundant functions that support flexible compensation under varying contexts (i.e., damage or loss function), as well as systemic reorganization of networks to provide different functions at different developmental periods (Johnson, 2017). In some cases, damage to localized cortical regions has consequences for the activity of distal areas (Carrera and Tononi, 2014), pointing to the contextual operation of cortical regions. Further, there is evidence that the same brain-region may be recruited for varying tasks, suggesting that function may be more directly related to the coordination dynamics of multiple regions rather than individuated networks or structures (Bressler and Kelso, 2001; Ding et al., 2000).

The remarkable plasticity of the brain should be considered a form of context-sensitivity, in that the activity and structure of physiological elements must be reorganized due to perturbations. The system must then be "sensitive" (in a general sense) to perturbations and respond in an appropriate way to maintain functionality. Machines have limited flexibility to respond to perturbations because they are equilibrium structures. No doubt clever engineers can design sophisticated systems that do respond to mechanical breakage; a run-flat tire on a BMW is a crude example of a redundancy accomplishing this. However, we argue that the type of stability associated with the proper function of a machine is categorically distinct from that of an organism. The stability of machines depends on the parts being near equilibrium. While of course most machines operate outside of equilibrium, depending on energy flows to drive work within the system (e.g., the gas in your car or the battery in your computer), these irreversible energy flows *destabilize* the system and cause it to wear and breakdown[8]. When the energy flow stops in a machine, it returns to an equilibrium state and maintains its structure. Contrast this with organisms, whose structural stability *derives from* the flow of energy. Further, if energy flows are cut off for an organism, structural and functional stability degrades. Recent advances in cellular biology provide radical evidence that even the components of cells, typically assumed to be rigid building blocks of biological systems, are actually in constant flux, with stability maintained through continuous fluxes of matter and energy (Nicholson, 2019).

From a physics perspective, this distinction is due to different kinds of dynamical stability; machines operate on near-equilibrium or "energy-well" stability (Bickhard, 2009), while organisms exhibit far-from-equilibrium stability. The stability of organisms is more like that of a driven dynamical system which settles on a dynamical mode. Notice that when an equilibrium system is perturbed, it will "adapt" by returning to equilibrium, or a state of rest—not the operational state of the system. When a far-from-equilibrium system is perturbed, it responds by returning to the previous self-organized state, or by assuming a new self-organized state (Nicolis, 1989). The restoration of function demonstrated by brains is more like the return to a dynamical state in a non-equilibrium system than the return to equilibrium of a mechanical one. The brain appears to maintain this functional stability through at least two means, *redundancies and plasticity*.

*Redundancies* are effectively duplicated systems that participate in some functions so that if one is damaged, the other can maintain function (Johnson, 2017). e.g., although only a side-effect of a more traditional lesion-based localization of function study, Newsome and Pare (1988) discovered that parallel visual networks in the occipetal lobe appear to take over functions lost due to lesion. In some stroke patients suffering damage to motor networks on one hemisphere, motor networks on the unaffected hemisphere demonstrate changes in activity that appear to compensate for the functional losses during recovery (Bajaj et al., 2016; Liu et al., 2015). Redundancies in brains and other far-from-equilibrium systems should be distinguished from those found in machines. In a machine a redundancy must be designed, requiring that the nature of the breakdown must be anticipated so that an appropriate failsafe can take over. In machines it is reasonable to expect that engineers could make such predictions, but the dynamic nature of the biological world seems to forbid anticipating all (or even enough of) the possible failures and appropriate solutions in advance. Further, a mechanical redundancy typically serves only a failsafe function, and are essentially "offline" or nonfunctional absent any perturbations. In the brain it appears that the systems that function as redundancies are active networks prior to that perturbation and are either adapting their function or being incorporated into additional processes, rather than lying dormant and being brought "online" as needed.

---

8  One might think of a machine as "using" energy to operate. However, it is more accurate to describe a machine as a complex set of constraints on an energy flow. A car, e.g., does not "use" energy to move; rather, the car is a sophisticated set of constraints on the conversion of chemical to mechanical energy. We as intelligent agents "use" the machine to direct energy fluxes in a functional way. An organism may similarly be cast as a system that constrains and directs energy fluxes but note that the constraints (system organization) *emerge due to the non-equilibrium fluxes*. Thus, we can distinguish machines from organisms in terms of equilibrium or non-equilibrium constraints respectively.

*Plasticity* is the broader dynamic change in the structure and function of nervous tissue (Fuchs and Flügge, 2014). In some animals, learned behaviors can be disrupted temporarily if the relevant tissue is damaged but will be recovered if allowed the time for other neural structures to modify their activity and compensate for the loss (Otchy et al., 2015). e.g., plastic neural network changes outside the occipital lobe support functional recovery from optic neuritis (Werring et al., 2000). Such observations suggest that functional stability in brains does not depend only on structural stability, as redundant networks may take over function or plastic changes to the structure may restore function. This kind of flexibility is intrinsic to the physics of far-from-equilibrium systems, but requires special handling from the perspective of equilibrium structures or machines. Specifically, from the perspective of equilibrium structures, the machine would need to be prepared in advance to respond appropriately to a particular disruption.

The final distinction is between the origin of functions within machines and organisms (or biological systems more broadly, inclusive of the brain). While we usually think of machines as having a purpose—most are devised with particular end in mind—that purpose is not required to explain how it operates. Rather, its activity can be explained simply by the causal chain of local interactions among constituent elements. Further, that purpose is always externally derived. A machine's purpose is not intrinsic to it but derives from how it is used by an agent. An object is a hammer when used to pound nails, and a paperweight when it keeps papers on a table. A computer is a sophisticated encoding device for representing vast sums of information when used by an intelligent agent as such, and merely a warm perch for a cat. We hold that the function of organs or activities alternatively derives from the organization of the system embedded in the environment. Function is not prescribed prior to development (as it is in the blueprints of a machine) but emerges through self-organizing dissipative processes interacting with environmental context (Bickhard, 2009; Christensen and Bickhard, 2002). The structure and behaviors of organisms derive not from pre-ordained instructions but from morphogenic dynamic processes that are sensitive to the developmental context (Adolph and Robinson, 2013; Edelman, 1992; Lewis, 2000; Thelen and Gunnar, 1993).

The discussion has argued that machines and organisms are fundamentally different types of physical systems and warrant different explanatory frameworks. We suggest that the mechanical hypothesis is deeply ingrained in contemporary approaches to brain science, especially in the quests for functional localization and wiring diagrams. We contrasted near- and far-from-equilibrium stabilities to suggest that the brain's operation is best understood in terms of the latter. In the following section, we argue that the brain, and biology more generally, should be recognized as members of the class of dissipative structures rather than machines. Because dissipative structures have a different set of intrinsic properties than machines, they offer a means of explaining many biological properties in physical terms.

## 4.3. What metaphor should replace the computer metaphor?

In line with the arguments above, we seek a framework for understanding the mind and brain that does not assume a mechanical basis of biological (and mental) function. The class of natural systems encompassing organisms should be non-fractionable, demonstrate non-equilibrium dynamical stability, and have intrinsic functions. We propose that the proper place for mind and brain is within the class of dissipative structures (**Fig. 8**). Dissipative structures are spatio-temporal organizations that emerge in systems that are held far from equilibrium (e.g., via the continual input of energy or matter). Canonical examples include Benard convections in fluids subject to thermal gradients or the striking patterns emergent in chemical oscillators such as the Belousov-Zhabotinsky reaction (Nicolis, 1989). Organisms are themselves dissipative structures—self-organized systems driven by flows of energy and matter—and there has been a rich history of applying the theory of dissipative structures to biological phenomena, especially pattern-formation and rhythmic processes (Goldbeter, 2018, 2017; Kugler et al., 1980; Kugler and Turvey, 1987). More recently, our group has investigated how some foundational aspects of biological behavior are explained through dissipative structure theory. Perhaps most strikingly, dissipative structures offer a theoretically based account of end-directedness, much like that observed in biology.

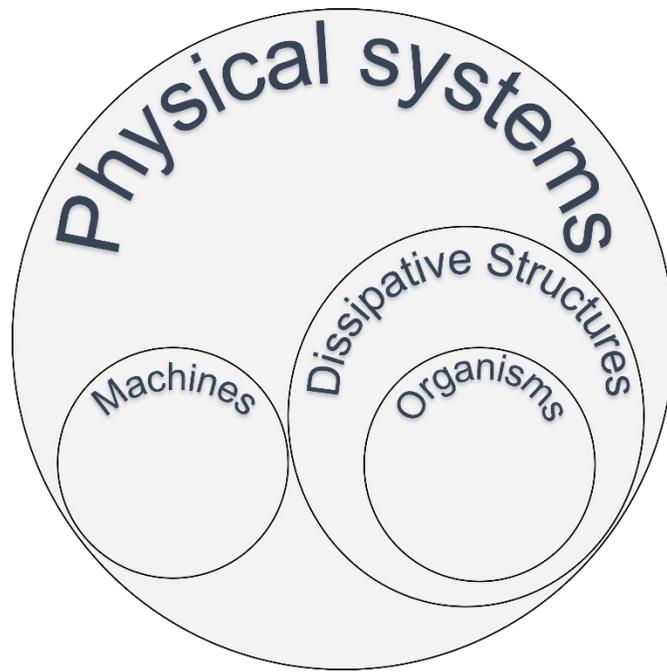

**Fig. 8.** A third, and our preferred, ontology of natural and artificial systems. Not only are machines and organisms ontologically distinct, organisms are particular types of a more general physical system, dissipative structures.

End-directedness, here, refers to a system preferring a particular end-state, such that multiple paths can be taken to that state. Biological systems famously demonstrate end-directedness in their pursuit of resources, shelter, mates, etc. Nonliving physical systems are sometimes thought not to display end-directedness, and this distinction is even offered as a potential dividing line between animate and inanimate systems (Jacobs, 1986; Swenson, 1999; Villalobos and Ward, 2015). However, nonliving physical systems are end-directed in well-established ways. e.g., consider a box of gas that is isolated from all flows of energy and matter. This simple system will go to its equilibrium state. Exactly how it will get there (i.e., what particles will collide and when) is not known, but the end-state is assured by the second law of thermodynamics. Our box of gas provides an excellent example of end-directedness, but the particular end to which it evolves, equilibrium, is of little use for living systems. Any system at equilibrium will need an applied external force to do anything at all and, worse yet, could not be alive. Since organisms are both alive and exhibit self-initiated growth, motility, and other processes, it seems clear that being directed to an equilibrium state is a literal dead end theoretically.

Importantly, dissipative structures are also end-directed, but towards a very different end state. Dissipative structures appear to be end-directed to increase the rate at which they produce entropy (Endres, 2017; Kondepudi et al., 2015; Martyushev and Seleznev, 2006; Swenson and Turvey, 1991). While a complete thermodynamic explanation of this phenomenon is a topic of current debate in physics, the phenomenon itself is firmly established empirically (Chung et al., 2022, 2017; De Bari et al., 2019; Endres, 2017; Kondepudi et al., 2015; Vallino and

Huber, 2018). The implications of dissipative structures being end-directed towards states with greater flow of energy are quite profound, even in very simple systems, such as the one we discuss below. The implications for more complex dissipative structures, such as those in the biological domain, have not yet been explored in any detail. For our purposes herein, the end-directedness of dissipative structures provides a theoretically grounded explanation for how biology can have an intrinsic function. Next, we review work with a simple dissipative structure that illustrates some of the implications of being end-directed towards states the increase the rate of entropy production. By working with a minimal dissipative structure, we can show how end-directedness yields complex behavior in a system with very few degrees of freedom and relatively tractable physics.

One such system is what we call our electrical dissipative structure (EDS) composed of metal beads in a dish with a shallow bath of oil (Joseph and Hübler, 2005; **Fig. 9**). A source electrode is positioned above the dish, separated by an approximately 5-cm air-gap. A metal rings surrounds the beads in the dish. A grounding electrode is connected to this metal ring. Charges are sprayed out from the source electrode, accumulate on the oil and the metal beads, and are conducted through the metal ring to ground. The system is maintained out of equilibrium by this flux of electrical charges. The beads become charged dipoles and are attracted to the grounding ring. After some time, the beads will tend to self-organize into strings of beads called "trees" that branch out from the grounding metal ring. These trees serve as pathways for the conduction of charge and move about, flexing and swaying as well as translating along the interior edge of the metal ring.

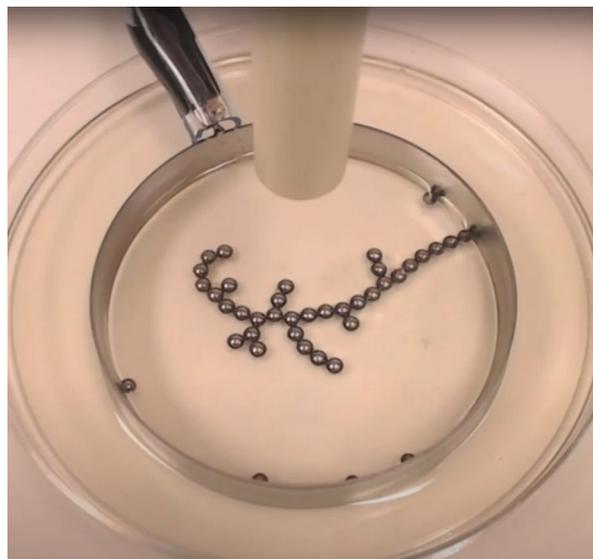

**Fig. 9.** The EDS with a tree structure. A source electrode (white tube) above the dish supplies electrical charges. The metal ring in the dish is connected to the grounding electrode. Trees will tend to sway and flex and translate along with the interior of the metal ring.

An essential construct for understanding the dynamics of this system is the rate of entropy production (REP). In far-from-equilibrium systems, irreversible processes that drive changes in the system also produce thermodynamic entropy[9], a physical quantity much like energy with units of J*K-1. In the EDS, the REP is directly proportional to the electrical current flowing through the ground electrode. In subsequent discussions of the EDS, REP and electrical current may be interchanged. Experimentation has repeatedly revealed that the system self-selects for morphologies and dynamics that maximize the rate of entropy production (Davis et al., 2016; De Bari et al., 2021, 2020, 2019; Dixon et al., 2016; Kondepudi et al., 2017). e.g., the initial formation of tree-structures is always accompanied by a sharp increase in the REP, as the tree becomes an effective pathway for electrical current. Further, the system will tend to evolve into tree structures from initially separate beads but will not evolve from tree structures to individuated beads. The self-selected structured state and the higher REP state are thus coincident in the system. Interestingly, tree structures producing greater REP are also more resilient to perturbation. Thus, the self-selection of higher REP states is a form of self-maintenance, demonstrating a self-organized "preference" for stability. In addition to self-selection of morphological states, the trees' dynamics demonstrate a tendency to maximize the REP. Through a combination of experiments and simulation we have come to understand that the tree follows gradients of increasing electrical charge-density, thus moving continuously into regions of the dish that increase the current and REP. This property enables a variety of interesting dynamics, from oscillations and synchronization of multiple trees (De Bari et al., 2020, 2019) to the coordination of dynamics among trees (Davis et al., 2016; De Bari et al., 2021).

Critically, as described above, we consider this self-selection of dynamics that maximize the REP as a rudimentary form of *end-directedness* (akin to that seen in biology). Here end-directedness may be construed as the time-evolution of a far-from-equilibrium system constrained by the optimization of a physical quantity. Moreover, because maximization of the REP also increases stability, the trees are end-directed to maintain themselves. Structural self-maintenance has often been taken as a necessary condition for the emergence of life, as in the concept of autopoiesis (Varela et al., 1974). Bickhard and colleagues (Bickhard, 2009; Christensen and Bickhard, 2002) have suggested that this self-maintenance is a foundational form of normativity, with certain states being intrinsically "good" or "bad" for the system determined by their impact on stability, and for our purposes on the REP. Certain behaviors and morphologies thus have *intrinsic functions and purposes*, derived from this far-from-equilibrium stability and optimization of the REP.

As has been remarked many times before, machines cannot be said to have intrinsic functions or purposes, a crucial limitation of the computer metaphor for explaining biological function (organisms, we assert, necessarily have intrinsic functions). Dissipative structures on the other hand offer a principled basis for the emergence of function, grounded in self-stabilizing processes and concurrently the optimization of a physical quantity (while syntactically distinguished here, these two should not be construed as separate processes in the EDS; optimizing the REP stabilizes the system and vice versa). In the EDS, end-directedness has been identified with the maximization of the REP. Many other far-from-equilibrium systems appear to demonstrate a tendency to maximize the REP (Endres, 2017; Swenson and Turvey, 1991), and much discussion has been devoted to the possibility that maximizing the rate of entropy production (MEP) is a general principle of far-from-equilibrium systems (Martyushev and Seleznev, 2014, 2006). Despite this, the current argument does not hinge on MEP in particular, as other variational principles could satisfy the same role. Further, if MEP is a general principle of far-from-equilibrium systems, we do not intend to advocate that *all* purpose and function in biology is simply to maximize the REP. Such a principle may be foundational for the emergence of living systems but need not proscribe the possibility of the emergence of other goals and purposes that should be considered just as real and valid for understanding a system's activity. We speculate that MEP or some other variational principle may be the foundational source of end-directedness in the system and that all other goals ultimately derived from it, often in non-obvious ways. Even in our very simple systems, we find end-directed behaviors that appear to be contrary to MEP, e.g., moving from an energy-rich region into energy-poor region (De Bari et al., 2019). Even in such cases, these paradoxical behaviors can be shown to be explained by MEP playing out over the current constraints on the system (De Bari et al., 2019).

---

[9] This is to be contrasted with statistical entropy, such as Boltzmann or Shannon variants. The two concepts overlap in near-equilibrium systems, but in general, and especially far-from-equilibrium, are distinct quantities.

We also distinguished machines and organisms on account of their different kinds of stability, near- and far-from-equilibrium stability. Dissipative structures display this latter kind of far-from-equilibrium self-stabilization. Nicolis (1989) cleverly distinguishes between stability of mechanical and far-from-equilibrium systems by comparing an idealized (i.e., frictionless) simple pendulum to a chemical oscillator. In each case, the system's stable state is an oscillatory mode with a certain frequency and amplitude. If the pendulum is perturbed (pushed) it will assume a new mode with a different amplitude and frequency. If the chemical oscillator is perturbed such as by locally increasing the temperature, the system will relax back to the original limit cycle (i.e., frequency and amplitude) once the temperature gradient is dissipated. In further contrast, if a pendulum is subject to irreversible entropy-producing processes (here, friction) the oscillatory dynamic will decay. Alternatively, the limit cycle of a chemical oscillator is *driven by* entropy-producing processes (here, chemical reactions). The chemical oscillator will decay if irreversible processes are cut-off. The stability of organisms is closer to the far-from-equilibrium stability of a chemical oscillator than a simple pendulum, more like a dissipative structure than a machine.

Of course, clever engineers can create machines that demonstrate dynamical stability, as in the steam-powered Watt governor, which displays classic fixed-point dynamics much like far-from-equilibrium systems. Both a Watt governor (or any similar machine) and a dissipative structure will display such dynamical stability due to physical constraints on energy flow. The crucial difference is that in a machine, those constraints are equilibrium structures that originate through a process entirely independent of the energy flux that powers the machine. Further, those constraints are degraded by the flow of energy. Alternatively, in dissipative structures, those constraints emerge due to energy flow and will degrade if that energy flow ceases. e.g., the convective cycles in a Benard cell constrain the motion of the individual particles (Nicolis, 1989), and the dissipative flow of particles drives those cycles; if the flow stops, the constraints dissolve. The constraints in machines are equilibrium structures, while those of dissipative structures and biological systems are more like stabilized processes (Nicholson, 2019) that require continual throughput of energy and matter.

The EDS exhibits remarkable self-stabilizing properties and adaptation to perturbations. When tree-structures are mechanically perturbed (e.g., by manually breaking them apart with a rod) they will spontaneously re-form into stable structures. We have referred to this capability as *self-healing*, analogous to the healing capabilities of organisms. In this self-healing process, the system very often produces different structures post-perturbation while producing the same level of current (Kondepudi et al., 2017), not unlike the plastic changes in the brain. As an analogue of redundancies, multiple tree structures can demonstrate coordinated activity when coupled through a shared pool of electrical charges on the oil-surface (Davis et al., 2016; De Bari et al., 2021, 2020), and their joint activity tends to maximize the REP. When perturbed such that the REP decreases, the trees will adjust their dyad-level (Davis et al., 2016) or individual (De Bari et al., 2021) dynamics in a way that restores the REP. In the previous section we distinguished between mechanical and far-from-equilibrium redundancies. Mechanical redundancies do not serve a function pre-perturbation, while in brains there is evidence that already-functional networks adapt their activity to compensate for the perturbation. In the EDS the redundant systems are active dissipative structures, each functioning (i.e., foraging for electrical charges) pre-perturbation, changing their dynamics after the perturbation. Thus while both machines and far-from-equilibrium systems can demonstrate redundancies, they are accomplished in substantively distinct ways. These properties also neatly illustrate how the EDS does not readily satisfy fractionability, as function and structure may change independently and in contextually sensitive ways.

## 4.4. What empirical findings support your preferred alternative metaphor?

Principles of self-organization and nonlinear dynamics have been applied to the study of biological behavior and the nervous system, especially through the framework of *coordination dynamics* (Bressler and Kelso, 2001; Kugler et al., 1980; Kugler and Turvey, 1987; Tognoli et al., 2020). Coordination dynamics investigates how multiple constituents with intrinsic dynamics (typically oscillatory) interact with one another. Critically, the approach aims at identifying coordinative regimes defined by a *collective variable* that captures the global dynamics of the ensemble of constituents. Given the focus on oscillators as constituents, a common collective variable is the *relative phase* of two (or more) oscillators. Coupled oscillators tend to exhibit two primary coordination dynamics, in-phase (relative phase = 0°) and anti-phase (relative phase = 180°), with a typical preference for in-phase dynamics. These coordination patterns have been observed in both intrapersonal (Haken et al., 1985; Kelso, 1984) and interpersonal (Richardson et al., 2007; Schmidt et al., 1990) biological contexts, as well as in the dynamics of coupled structures in the EDS (De Bari et al., 2020). We have found evidence that, in the EDS, the stability of these modes is directly related to which is most functionally adaptive (i.e., produces the most entropy) given system parameters (De Bari et al., 2020). The kernel insight is that, in order to explain the remarkably adept coordination of inordinately many physiological degrees of freedom (nervous, muscular, or otherwise), biology capitalizes on the self-organizing processes native to far-from-equilibrium systems (Kugler et al., 1980; Kugler and Turvey, 1987). The activity of the nervous system involves many oscillatory processes appropriate for the tools of coordination dynamics, from the neural-level cyclic activity to the collective oscillatory dynamics of central pattern generators (Fred, 1980; Marder and Bucher, 2001).

More recently, coordination dynamics has been applied directly to the nervous system by investigating the relative activity of cortical regions (Alderson et al., 2020; Bressler and Kelso, 2001; Kelso, 1995). The approach involves recording cortical activity while subjects are engaged in various cognitive tasks and addressing the degree to which the collective activity of multiple brain regions is associated with changes in cognitive activity. In classic fashion, the collective activity is assessed in dynamical terms, including the convergence of relative phase and frequency of oscillatory events across multiple recording sites. Critically, the authors advocate that this coordinated activity is characterized by *metastable* dynamics. A metastable dynamic may be crudely cast as an approximation of a fixed point or limit cycle dynamic; the system tends toward the dynamic but does not converge on that dynamic entirely. Because of this, the system demonstrates more flexibility, transitioning between metastable states under varying circumstances without being "captured" by those states. This flexibility appears to enable several properties in line with thinking of the brain as a dissipative structure. When monkeys were engaged in a visual task, changes in cognitive activity (i.e., anticipatory vs. active processing) were accompanied by changes in the coordinated activity of multiple sites (Bressler and Kelso, 2001; Ding et al., 2000). Similar results were observed in human subjects (Alderson et al., 2020). The prevailing interpretation is that the adaptive recruitment of multiple regions into a coordinated dynamic is the same kind of self-organizing process that underwrites coordination of action in living and non-living dissipative systems.

To summarize, the computer metaphor of mind and brain has two primary threads: the hypothesis of mental operation, which we call epistemic constructivism, and the proposition of mental ontology in the mechanistic hypothesis. Our evaluation of the computer metaphor of mind and brain has focused on the underwriting mechanical hypothesis and highlighting disparities between the nature of machines and biological systems. There have been many thoroughgoing critiques of epistemic constructivism, as detailed at the outset. Without revisiting these critiques, we argue that our preferred framework, dissipative structures, is consistent with these perspectives and has even been a starting point for these scientific approaches to behavior and mind (Bickhard, 2009; Kugler et al., 1980). We have investigated the epistemic properties of the EDS and other dissipative structures, drawing comparisons to the tenets of Ecological Psychology (Chung et al., 2022; De Bari et al., 2020). Further, we point out that even some "representation-hungry" phenomena like end-directedness (Kondepudi et al., 2015) and anticipation (Dixon et al., 2021), typically assumed to require the tools of epistemic constructivism, can be reasonably attributed to the EDS.

We identified four components of the mechanical hypothesis and pointed to their limited application in the brain and biology. We reviewed how, with respect to each tenet, the brain behaves more like a dissipative structure than a machine. Some have identified similar criteria for machines, especially properties related to the fractionability, and used such conclusions to highlight similarities between biology and mechanics (Bongard and Levin, 2021). While their compelling analysis does highlight the ever-increasing sophistication of machines and their integration into biological processes, it does not consider the fundamental far-from-equilibrium properties of living systems central to their operation. The properties of non-equilibrium stability and intrinsic function remain, in our opinion, clear demarcations between machines and organisms.

While we focused our attention on one specific dissipative structure, we identified five components of the mechanical hypothesis and pointed to their limited application in the brain and biology. We reviewed how, with respect to each tenet, the brain behaves more like a dissipative structure than a machine. Our suggestion is not that the EDS itself constitutes a metaphor for the brain, but rather that the brain will minimally have the properties we find in our simple EDS by virtue of being a dissipative structure. These properties, such as adaptivity, self-healing, and energy-seeking, are quite mysterious if one starts with equilibrium structures or mechanical systems but are a natural consequence of the end-directedness of dissipative structures. It will, of course, be important to establish the details of how thermodynamic forces and flows are manifested in the complex dissipative structure that is the brain. It seems likely that the multi-scale nature of the brain, along with its complex chemistry, will be central to understanding how it functions as a dissipative structure. It may not necessarily be the case that the brain self-stabilizes in order to maximize the REP. Rather we are suggesting that brain activity and plasticity can only be understood with reference to far-from-equilibrium self-organization, and not with a mechanical metaphor. The new metaphor must be in the physics of dissipative structures, not equilibrium structures.

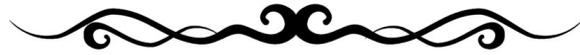

## 5. Complexity: Understanding brains and minds on their own terms—Luis H. Favela[10]

To comprehend brains and minds as computers is to explain their activities and organization in terms of information processing, which is essentially the systematic manipulation of representations. This understanding has served as the fundamental guiding commitment of research in artificial intelligence, cognitive psychology, neuroscience, and related disciplines. While computational terminology has been applied metaphorically to brains and minds, it has certainly been applied in literal senses as well. The Human Brain Project's SpiNNaker computer is presented as an example of the inability of research committed to the computer metaphor (and literalism) to facilitate understanding of brains and minds. Long-term progress requires shifting to investigative frameworks that approach brains and minds on their own terms, and not via inappropriate metaphors. Complexity science is such a framework, in which brains and minds are understood via core features of complex systems: emergence, nonlinearity, self-organization, and universality. Extended Haken-Kelso-Bunz models are presented as empirically-supported research exhibiting a fruitful complexity science approach to brains and minds across spatial and temporal scales of activity and organization. The result is a clear case for the effectiveness of investigating brains and minds on their own terms as complex systems.

### 5.1. What do we understand by the computer metaphor of the mind and brain?

At its most basic, to understand brains and minds as "computers" is to treat their workings in terms of information processing. It can be helpful to define the kind of processing carried out by brains and minds by way of twentieth-century psychologist Ulric Neisser's (1967) explications of the two core elements of cognition, which contributed substantially to the conceptual and theoretical foundations of cognitive psychology. First, cognition involves *constructive processing*, which refers to the reconstruction or transformation of stimuli (i.e., from within or outside the system) into units for the cognitive system to act on. In more contemporary terms, these units are commonly called *representations*. Second, cognition involves *information processing*, which refers to the ways those units (i.e., representations) are manipulated in order to produce behaviors, perceptions, reasoning, and so on. In more contemporary terms, these manipulations are commonly called *computations*. Of course, this is a simple description of Neisser's characterization of cognition, one that glosses over details that can be significant at times, such as the idea that the two processes are not always distinct—that is, constructive processing can be a form of information processing and vice versa. For now, such differences are not important.

---

10   Correspondence: luis.favela@ucf.edu (L. Favela).

What is important now is that conceptualizing brains and minds as essentially involving the manipulation (i.e., computation) of units (i.e., representations) lined up in a number of meaningful ways with strategies utilized throughout the early development of artificial intelligence (AI) research (i.e., 1950s through 1990s; Wooldridge, 2021). The modern form of AI is commonly understood as originating with Alan Turing's pioneering work on the theoretical foundations of computers (Turing, 1950b), which was further developed and formalized by researchers such as Alonzo Church and Herbert Simon (Boden, 2006; Buchanan, 2005). The predominant guide to research following those early years (~1950s) was the *symbolic AI* strategy, which was committed to the claim that AI ought to be based on human minds (e.g., McCarthy et al., 2006). Accordingly, AI research should aim to identify the various kinds of mental processes (e.g., reasoning) and formalize them (e.g., logic and rules) in a manner implementable in AI systems. The formalisms appealed to by AI researchers were based on those developed by Turing and colleagues, namely, rule-based syntax that operated over semantically-imbued symbols, such as 1s and 0s. It is clear how the core commitments of the symbolic AI strategy line up with Neisser's approach: brains/minds and AI systems manipulate (e.g., operate over) units (e.g., symbols) in some sort of systematic way (e.g., follow rules).

While the symbolic AI approach gets much of the attention in discussions of the history of AI (e.g., "symbolic AI" is synonymous with Good Old-Fashioned Artificial Intelligence or GOFAI; e.g., Boden, 2014; Haugeland, 1985) and much of the criticism (e.g., Dreyfus, 1992; Harnad, 1990), a different neurophysiologically-based strategy ran in parallel (e.g., McCulloch and Pitts, 1943; see Wooldridge, 2021). The *neural networks strategy*, much like the symbolic AI strategy, was committed to the guiding principle that AI ought to be based on those systems we already know are intelligent, specifically, humans. However, unlike the more abstract symbolic AI strategy, the neural networks strategy held to a more concrete commitment of basing the *architecture* of AI systems on the physiology of the organ purported to be the seat of human intelligence, namely, the brain. To be more specific, AI architecture should be based on the brain's structure, which is comprised of more basic elements, i.e., neurons and their connections. While research programs in line with a generally symbolic AI approach continue to this day ("Cyc technology overview," 2022), it is clear that the general neural networks strategy has prevailed as the predominant guiding approach to AI. This is made evident in large part due to the dominance of machine learning and deep neural networks; though, such approaches are not without deep challenges (e.g., Carlson et al., 2018; Hosseini et al., 2020). Like the symbolic AI strategy, the neural networks strategy also lines up with the Neisser characterization of cognition as information processing: brains/minds and artificial neural networks manipulate (e.g., firing rates) units (e.g., network nodes) in some sort of systematic way (e.g., encoding and decoding).

Given this (very) brief historical overview, we can return with better understanding to the point that started this section. From cognitive psychology (e.g., Neisser, 1967), to the symbolic (e.g., McCarthy et al., 2006) and neural networks (e.g., McCulloch and Pitts, 1943) strategies guiding much AI research, information processing has been central to the investigation and explanation of brains and minds. In that sense, describing brains and minds as "computers" is not metaphorical. Strictly speaking, a metaphor is a figure of speech whereby a word is applied to something else in an analogous sense, such that the utilized word's meaning is not *literally* true of that which it has been applied to ("metaphor, *n*.," 2021). So, if one were to say, "Ana has a heart of gold," it would not be literally true that the organ in Ana's chest that pumps blood is made of a material with atomic number 79. In the same way, for the term "computer" to be applied to brains and minds in a metaphorical sense would be to say that the latter do not literally do what computers do, namely, process information in the particular ways that computers do. On the contrary, cognitive psychologists and AI researchers have *literally* meant that brains and minds are computational devices. That is to say, brains and minds are information processing systems in the *same way* that computers are: they systematically operate over representations. This is not to say that all cognitive psychologists and AI researchers believe that brains and minds are literally computers. Certainly, there are cognitive psychologists and AI researchers who explicitly believe that the relationship of brains and minds with computers is metaphorical. Additionally, there are those whose belief that brains and minds are computers is implicit as revealed by their research frameworks (e.g., the concepts and explanations used) and the works they produce (e.g., journal publications). The same holds true of the neurosciences. Specifically, several neuroscience subdisciplines (e.g., cognitive neuroscience, computational neuroscience, and sensory neuroscience) treat the brain and mind as literally being computers in the senses described above; and this belief is held explicitly and implicitly by various neuroscientists.

### 5.2. What are some of the limitations of the computer metaphor?

The previous section argued that across research in AI, cognitive psychology, and several neuroscience subdisciplines, brains and minds are typically understood as being computers—or carrying out "computations"—in a literal and not metaphorical sense. There is an enormous literature in each of those disciplines alone that engage with the limitations of understanding brains and minds as being computers—both in the literal and metaphorical sense. In cognitive psychology (as well as other subdisciplines of psychology and cognitive science), there is a decades-long tradition of anticomputational and antirepresentational approaches stemming from considerations of mind (inclusive of cognition) as being necessarily embodied (Chemero, 2009; Gibson, 1979). In AI research, there is also a long tradition of identifying fundamental shortcomings of, e.g., developing body-less systems with intelligence primarily based upon abstract knowledge, such as symbolic logic (Brooks, 1991; Dreyfus, 1992). More recently, the neurosciences are exhibiting increasing criticisms of explaining brain structure and function in traditional computational terms, such as defining neural activity in terms of coding (e.g., Brette, 2019), explaining intelligent behavior via a central controller (e.g., Buzsáki, 2019), and parsing the mind's capacities along neatly delineated folk psychology-defined cognitive ontologies (e.g., Poldrack and Yarkoni, 2016).

For a bit of background, consider one of the large recent government-funded projects established to advance understanding of brains and related phenomena, the Europe-based Human Brain Project (HBP). Beginning in 2013, the HBP is a ten-year, €1.019 billion budget project that aims to, "tame brain complexity … [by] building a research infrastructure to help advance neuroscience, medicine, computing and brain-inspired technologies - EBRAINS … to create lasting research platforms that benefit the wider community" ("Overview," 2021). EBRAINS is an infrastructure for sharing brain data for the purposes of modeling and simulating neuroanatomy and various kinds of brain activity ("About EBRAINS," 2021). As evidenced across the official HBP website, it is undeniable that the brain and related phenomena are treated in computational and representational terms. Cognition, e.g., is explicitly defined as such (e.g., "representations are the basis for higher cognitive processes") and is meant to be explained via "'deep learning neuronal network[s]" ("Understanding cognition," 2021).

At eight years in, HBP leadership published a list of the project's six most impressive achievements (Sahakian et al., 2021). These include a human brain atlas visual data tool, touch-based telerobot hand, neuro-inspired computer, and being cited in 1,497 peer-reviewed journal articles. There should be no doubt that much of this research is impressive, particularly when put into various contexts, such as the potential for advancing robotic limbs to improve the lives of people who have had amputations. However, it is far from clear whether any of these achievements have illuminated our understanding of brains and minds in a significant way.

Take e.g. the "neuro-inspired computer" mentioned above: Spiking Neural Network Architecture or SpiNNaker (Furber et al., 2014). SpiNNaker is a massively parallel, multicore computing system that is comprised of +57,000 nodes that contain +1 million ARM9 cores and seven terabytes of RAM ("Architectural overview. SpiNNaker Project," 2021). With that many core processors, the amount of power required to run SpiNNaker is staggering. Consider that a MacBook Pro laptop with ten cores runs on an average 100 watts (W) ("MacBook Pro (16-inch, 2021) - Technical Specifications," 2022). If SpiNNaker's ARM9 cores are at least as energy efficient as a MacBook Pro's cores are, then—not including the energy needed to power cooling systems—SpiNNaker would consume 100,000,000 W, or 100 megawatts (MW). For the sake of comparison, the version of IBM's Watson that won first place in the game show Jeopardy! consumed ~85,000 W (Kelly III and Hamm, 2013). Now, consider that a human brain utilizes ~20 W (Balasubramanian, 2021). It can be tough to appreciate what is learned about the actual structure or function of brains and/or minds from a computer that is purported to be "inspired by the connectivity characteristics of the mammalian brain" (Furber et al., 2014, p. 652) and yet consumes about 5,000,000 times more energy than a human brain. Moreover, though SpiNNaker consumes so much more resources than a mammalian brain does, it does not have a fraction of the latter's capabilities.

SpiNNaker is an illustrative example of the limitations of attempting to understand brains and minds as computer. First, it is unclear that appealing to networks as the core defining feature of mammalian brain architecture is what makes them appropriately understood via the computer metaphor and, in turn, why the SpiNNaker computer is accurately described as brain-like. Yes, like brains, SpiNNaker sports a massively parallel network architecture with information transferred via spike-like activity (Furber et al., 2014). However, it is unconvincing to primarily appeal to features such as parallelism and networks to motivate the claim that SpiNNaker's computational architecture is *like* the mammalian brain's purported similar computational structure. It is, however, evident that *many* systems share those features, which include, but are not limited to, gene regulatory systems, Hollywood actors who have worked on the same movies, sexual partners, and the United States' airline system (Barabási and Bonabeau, 2003). Consequently, SpiNNaker does not provide further understanding of brains and minds on their own terms if the manner by which their relationship is metaphorical is had by so many other similar systems that could have been described in the same ways, such as having a parallel network architecture.

Second, given that SpiNNaker is literally a computer, if it is supposed to implement a brain-like architecture, then it is implausible to believe that brains are literally computers as well. Consider again the incredible contrast between the power requirements of brains and SpiNNaker; specifically, a 5,000,000 times difference. As a starting point, it seems intuitively plausible that if two systems have comparable architectures (e.g., parallel networks) aimed at producing equivalent capabilities (e.g., control a limb), then their energy requirements would be similar as well. An obvious reply is that the scale of those systems matters when considering energy requirements; that is, a human brain is the fraction of the size of SpiNNaker's infrastructure, thus it uses a fraction of the energy. With that said, it is obvious that the same ratio of power use (human brain ~20 W: SpiNNaker 100,000,000 W) does not also hold for the capabilities of those systems. What the human brain can do with ~20 W is many orders of magnitude beyond what SpiNNaker can do with 5,000,000 times more power. This fact makes the following claim likely to be true: SpiNNaker is literally a computer (i.e., it has a computer's structures and functions) and brains are not computers in any literal sense (i.e., they do not have a computational structure underlying its functions).

Thus, while SpiNNaker is an impressive computer in many ways, it will likely produce a limited range of understanding that is derived from highly idealized models and simulations. Putting the point bluntly: No matter how much computational processing power is achieved by SpiNNaker, additional understanding of the real nature of brains and minds will not be achieved. To be clear, the point is not that powerful computers cannot process models and simulations that can contribute to valuable research, such as descriptions of neurophysiology and robotics. The point is that attempts at illumination via comparison will be extremely limited because brains/minds and computers have radically different organizational principles underlying their capabilities. Consequently, trying to reduce the real nature of brains/minds to being equivalent to that of computers, or trying to base computers on inaccurate and false conceptions of brains/minds, is to promote confusion and impede progress. In this way, continuing to adhere to computer metaphors (or literalisms) when investigating brains and minds would be tantamount to guiding cardiovascular research via the "gold metaphor" of hearts.

### 5.3. What metaphor should replace the computer metaphor?

As can be gathered from the previous sections, I do not think the computer metaphor—and especially *computer literalism*!—is facilitating progress towards understanding the real nature of brains and minds. That is not to say that computer terminology should be banned from all discussions of brains and minds. Certainly, figures of speech such as, "my brain is processing that information now," "that neuron represents his grandmother's face," and "the core processor is the brain of my MacBook Pro," can be acceptable in certain contexts. Nevertheless, when figures of speech shift from being just metaphors for more informal purposes to literal comparisons to inform and guide empirical brain research, then concepts such as "information processing" and "representations" are inappropriate. In the case of brains and minds, what is not needed are more metaphors that facilitate false understanding of the natural phenomena under investigation. What is needed are concepts that convey the real nature of the phenomena on their own terms.

As the SpiNNaker example shows, there must be something radically different about the physical implementation of brains and minds than computers. Information processing (i.e., computations operating over representations) does not seem to be the theory that can provide real understanding of brains/minds and computers; especially pertaining to the structures that instantiate their capabilities. Case in point, relative to computers, mammalian brains consume a minuscule amount of power and neural activity occurs at slower timescales, all while supporting a wider range of possible sophisticated behaviors (e.g., eating sushi, juggling chainsaws, writing a novel, and so on). So, if information processing (i.e., computer metaphor) cannot provide understanding of brains and minds on their own terms, then what can?

The language of *complexity science* has the potential to replace the computer metaphor of brains and minds. Complexity science is already making inroads in the cognitive, neural, and psychological sciences by contributing to productive research programs, which includes inspiring hypothesis generation, providing data analysis methods, and theories to inform explanations. In what remains of this section, I provide a brief summary of complexity science. In the next section, I conclude by providing an illustrative example of the fruitful application of complexity science to the study of brains and minds. By the end, it will be demonstrated that complexity is a suitable replacement for the computer metaphor.

As a starting point, *complexity science* can be generally understood as the interdisciplinary investigation of complex systems, where *complexity* refers to certain kinds of order exhibited by systems due to their many interacting parts at a range of spatial and temporal scales (Allen, 2001; Érdi, 2008; Mitchell, 2009). Common examples of the kinds of interactions and order include chaos (Bar-Yam, 2016), circular causality (Érdi, 2008), multiscale interactions (Bishop and Silberstein, 2006), and phase transitions (Van Orden and Stephen, 2012). At first pass, a review of the complexity science literature could lead somebody to conclude that the apparently wide variety of concepts, each with varying definitions, means there is nothing more to "complexity" than a family resemblance of arbitrary investigator-selected phenomena.

As argued elsewhere (Favela, 2020a), the current state of complexity science should not be surprising as it is very much in the early years of development (cf. Kuhn, 1962). Still, that is not reason alone to disregard the fact that complexity refers to something very real in the world (on this point, also see Nicolis and Nicolis, 2007; Solomon and Shir, 2003). Moreover, complexity science stems from a traceable set of disciplines—with their own tried and tested concepts, methods, and theories—which contribute to its foundations as it develops into a mature science in its own right. These include, but are not limited to, artificial life, evolutionary biology, and Gestalt psychology (cf. Goldstein, 1999). When it comes to applying complexity science to the study of brains and minds, three disciplines have arguably had the most significant influence.

First, is *systems theory*, which includes the concepts and theories of cybernetics and general systems theory (Wiener, 1948). A key lesson from here is the identification of irreducible system-level activity—i.e., emergent—as the target of investigations and assessing the roles of feedback and the multiscale contributions of components. The second major contributor is *nonlinear dynamical systems theory*. Dynamical systems theory (DST) employs mathematical tools, such as differential equations and phase space plots, to evaluate systems that change over time (for more detailed overview, see Favela, 2020b, 2021). Whereas DST studies linear systems (outputs are proportional to inputs) nonlinear DST (NDST) employs many of those same tools to study phenomena displaying outputs that are exponentially or multiplicatively related to inputs. Accordingly, NDST is suited to study more exotic phenomena such as those exhibiting phase transitions or unexpected qualitative shifts. Such phase transitions are demonstrated by water shifting among its solid, liquid, and gaseous states; coordination among an individual's limbs; and spontaneous organization of groups, such as schools of fish becoming a bait ball to defend against predators. One of the most fascinating contributions NDST has made to complexity science, is fractal geometry (Mandelbrot, 1967). Fractals are self-similar structures, where the pattern at one spatial or temporal scale is duplicated at other scales. Fractals can be perfectly self-similar, such as geometric fractals like the Sierpinski triangle (**Fig. 10A**). Fractals can also be statistically self-similar, which means only particular features are repeated at different scales and to certain degrees. Natural fractals with statistically self-similar structures abound, such as broccoli, clouds, coastlines, and coral (**Fig. 10B**). Since the 1990s, there is increasing evidence of statistically self-similar fractals across a wide range of physiological phenomena, including structures like bronchial tubes and retinal veins, but also processes like heart beats and spontaneous neuron activity (**Fig. 10C**).

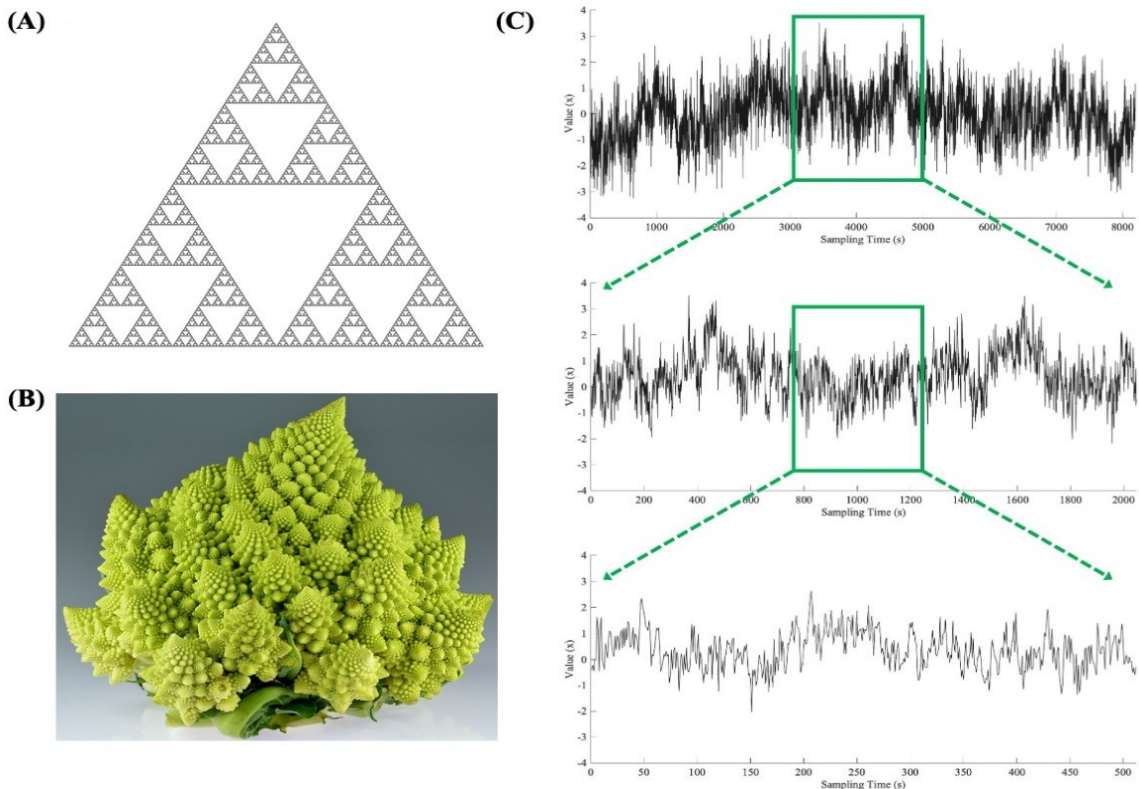

**Figure 10.** Fractals. (A) The self-similarity of fractals can be perfect, as depicted by the Sierpinski triangle. Modified and reprinted with permission from Wikipedia (2013; CC BY-SA 3.0). (B) Fractals can be statistically self-similar, such as fractals found in nature, e.g., Romanesco broccoli. Reprinted with permission from Wikipedia (2021; CC BY-SA 4.0). (C) Physiological activity can also be statistically self-similar. Here, synthetic time series generated from detrended fluctuation analysis (Favela et al., 2016) of data recorded from spontaneous activity of single neurons in the rat main olfactory bulb (Stakic et al., 2011). Statistical self-similarity shown in windows based on power of two—(top) 8,192 seconds,

(middle) 2,048 seconds, and (bottom) 1,024 seconds—with overall temporal trends repeated within each window of time.

The third major contributor to complexity science to highlight is *synergetics*. Synergetics is itself an interdisciplinary field that is often applied to the investigation of systems with many interacting components (Haken, 2007). One of the primary contributions synergetics makes to complexity science are concepts and tools for studying self-organization. In the context of the brain and behavioral sciences, self-organization refers to the spontaneous formation of patterns and structures in nonequilibrium open systems that exchange energy, information, and matter with their environments (Kelso, 1995, 2021a, 2021b). A core aim of a synergetics-based approach is to identify laws that govern a range of phenomena. One way this is accomplished is by means of a modeling strategy guided by order and control parameters (Haken, 1988). An order parameter is the system-level (i.e., collective) variable that captures the macroscopic state of the system. Control parameters are the variables that guide system dynamics. An example of a lawful governing model of coordination is the Haken-Kelso-Bunz (HKB) model (equation 1; Haken, Kelso, & Bunz, 1985), which was originally applied to the study of bimanual coordination:

$$\dot{\phi} = -a\sin\phi - 2b\sin 2\phi \qquad (1)$$

Here, the order parameter ($\phi$) in eq. 1 is defined as the relative phase state of interacting parts and processes, and the control parameters (*a,b*) are the contributors to the state of the order parameter. The model was originally applied to the study of bimanual coordination, where one person moved the pointer finger of each hand up and down. The order parameter is the state of those two fingers, such as in or out of phase with each other. The control parameters are the fingers, quantified via their frequency of movement. The HKB model (and its various modifications) can be said to be "lawful" in that, e.g., it identifies consistent features of coordination across a wide range of phenomena, such as synaptic terminals in astrocytes, limb coordination within and between individual people, and many more (Kelso, 2021b).

When reviewing the relevant literatures and integrating the contributions of systems theory, NDST, and synergetics, a number of concepts come to the forefront as being core to complexity science: emergence, nonlinearity, self-organization, and universality. Even though most of these have been discussed above, they are all worth summarizing here. First, *emergence* refers to irreducible system-level activity and organization (see systems theory). Second, *nonlinearity* refers to exponential and multiplicative interactions (see NDST). Third, *self-organization* refers to spontaneous pattern and structure formation without the need for a central controller or pre-specified instructions (see synergetics). Finally, and not yet discussed, is *universality*. In the simplest terms, universality refers to the fact that nature exhibits recurring patterns of activity and structural organization in vastly different substrates and contexts. Fractals are one such universal activity and organization found in nature. The universality of fractals is both qualitative and quantitative. Qualitatively speaking, coral, lungs, neurons, and trees all seem upon viewing to have rather similar branching patterns: larger "branches" that split into smaller branches, which split into smaller branches, and so on. The universality of fractals is not limited to such qualitative "seemings" alone. Many phenomena exhibiting fractal patterns are also quantitatively similar. Since Benoit Mandelbrot first developed the fractal dimension to quantify the coastline of Britain (Mandelbrot, 1967), other mathematical tools have been developed to assess fractals—and other forms of self-similarity, such as scale-freeness—with increasing sophistication. Examples of such tools include box counting, detrended fluctuation analysis, multifractal analysis, and wavelets.

From the background of systems theory, NDST, and synergetics, complexity science offers a rich investigative framework for understanding brains (and possibly minds) in terms of the key characteristics of emergence, nonlinearity, self-organization, and universality. Furthermore, each of those concepts have sophisticated qualitative and quantitative tools to empirically record and evaluate them. In the final section, I present some empirical findings that support this investigative framework.

## 5.4. What empirical findings support your preferred alternative metaphor?

In concluding, one set of empirical findings are presented as supporting the claim that complexity is the proper replacement for the computer metaphor of brains and minds, and that complexity science is the suitable investigative framework. The findings stem from the extended HKB models. Since the HKB model was first introduced (Haken et al., 1985; see previous section for discussion), it has undergone various extensions and been incorporated into other models in order to capture a wider range of phenomena (for review see Kelso, 2021a). Throughout its modifications, the core principle remains, namely, the regular features of systems that facilitate lawful coupling (or coordination) of interacting components. In Kelso's own words, the HKB model can be understood as lawful in that:

> *The HKB law of coordination takes the form of an equation that expresses how patterns of coordination defined by informationally meaningful collective variables/order parameters evolve and change due to nonlinear interactions between parts and processes.*
>
> — Kelso (2021b; p. 317)

One of the first extensions of the HKB model was the incorporation oscillatory dynamics via the parameter $\Delta\omega$, which incorporates the oscillator frequency of components (eq. 2 in Fuchs et al., 1996).

$$\dot{\phi} = \Delta\omega - a\sin\phi - 2b\sin 2\phi \qquad (2)$$

The HKB model has since been incorporated into various models (**Fig. 11**) that can account for an even wider range of behavior from the scale of brains (e.g., neuronal oscillations) to the behavioral and cognitive (e.g., learning, memory, and perception; for references see (Kelso, 2021b).

The HKB model and its extensions are a paradigmatic instance of successfully applying complexity science to the study of brains and minds. First, it is continuous with other successful sciences, especially NDST and synergetics. Second, while it is a model—and thereby is an idealization that, strictly speaking, introduces falsehoods and simplifications—it has a wealth of empirical data to support its various versions. Third, and, again, though it is a model, it explains its target phenomena on their own terms. Within a complexity science framework, those terms are emergence, nonlinearity, self-organization, and universality. Those systems that HKB models are empirically supported in being applied to exhibit *emergence* in that their activities and structures are irreducible to its constituent parts. The systems also exhibit a wide range of *nonlinearities*, such as those that occur via phase transitions and symmetry breaking. Additionally, the systems are *self-organized*, meaning their behaviors, no matter how intricate, occur without a controller or preprogrammed rules. Specifically, the activities and organization result (emerge) from the (nonlinear) interactions among components. Finally, the HKB models exhibit *universality*. The kind of coupling and coordination captured by the HKB model is widespread in nature. When extended or integrated with other models—such as the Kuramoto model of oscillatory behavior—even larger classes of phenomena are explained, including various kinds of neuronal dynamics, human-machine interactions, and multiagent coordination.

The HKB model is but one example of the empirical fruitfulness of complexity science. Part of the interesting work of doing complexity science is identifying and applying those laws of nature and universal classes of activity and organization (e.g., fractals and self-organized criticality; Bak et al., 1988; Jensen, 1998; Plenz and Niebur, 2014). Most significantly for the current context, achievements like the HKB model demonstrate how complexity science aims at investigating, explaining, and understanding phenomena on their own terms. HKB is a law of coupling and coordination that emerged from doing the empirical work of investigating certain behaviors, namely, bimanual finger movements. The dynamics of finger movements did not need to be first defined as "coupling" in order for the HKB model to be explanatory. On the other hand, the computer metaphor is typically applied to phenomena already defined as being computers, which results in a sort of circular justification. None of this is to say that complexity science is the perfect and final science for studying brains and minds. But it does have a lot to offer conceptually, methodologically, and theoretically—certainly more than the computer metaphor.

|   | Oscillator Level | Collective Level |
|---|---|---|
| 1 | $\ddot{\chi} + (\alpha\chi^2 + \beta\dot{\chi}^2 - \gamma)\dot{\chi} + \Omega_\chi^2 \chi = (a_{xy} + 2b_{xy}(\chi - y)^2)(\dot{\chi} - \dot{y})$ <br> $\ddot{y} + (\alpha y^2 + \beta\dot{y}^2 - \gamma)\dot{y} + \Omega_y^2 y = (a_{yx} + 2b_{yx}(y - x)^2)(\dot{y} - \dot{x})$ | $\dot{\phi} = -a \sin\phi - 2b - \sin 2\phi$ |
| 2 | Symmetry Breaking | $\dot{\phi} = \delta\omega - a \sin\phi - 2b - \sin 2\phi$ |
| 3 | $\ddot{\chi} + \tau\Omega(\chi^4 + \chi^2 - 1)\dot{\chi} +$ <br> $\Omega^2 \left( \chi - \rho - \theta \left( \frac{\dot{\chi}}{\tau\Omega} - \chi + \frac{\chi^3}{3} + \frac{\chi^5}{5} - I \right) \right) = (a_{xy} + 2b_{yx}(\chi - y)^2)(\dot{\chi} - \dot{y})$ | Discrete to Continuous Regimes |
| 4 | $\ddot{\chi} + (\alpha\chi^2 + \beta\dot{\chi}^2 - \gamma)\dot{\chi} + \Omega_\chi^2 \chi = (a_{xy} + 2b_{xy}(\chi - y)^2)(\dot{\chi} - \dot{y})$ <br> $\dot{\Omega}_\chi = \epsilon_\chi \dot{\chi} y$ <br> $\dot{\Omega}_y = \epsilon_y \dot{y} x$ <br> $\ddot{y} + (\alpha y^2 + \beta\dot{y}^2 - \gamma)\dot{y} + \Omega_y^2 y = (a_{yx} + 2b_{yx}(y - x)^2)(\dot{y} - \dot{x})$ | Adapting to Stable States |
| 5 | Phase Pattern Bias | $\dot{\phi} = \delta\omega - a \sin\phi - 2b \sin 2\phi + c \sin(\psi - \phi)$ |
| 6 | $\dot{\varphi}_i = \Omega_i - \sum_{j=1}^N a_{ij} \sin(\varphi_i - \varphi_j) - \sum_{j=1}^N 2b_{ij} \sin 2(\varphi_i - \varphi_j)$ | Scalable Multiagent System |

**Figure 11.** Extensions and incorporations of the Haken-Kelso-Bunz (HKB) model. (1) Original HKB model (dashed lines) for capturing the collective scale (right) and definition of parameters for oscillatory behaviors (left). The parameters in blue ($\Omega_x$ and $\Omega_y$) are intrinsic frequencies of oscillators. The parameters in orange ($a$ and $2b$) capture the coupling phases. (2) Incorporating metastable states via symmetry breaking captured by green parameter ($\delta\omega$). (3) Modifying oscillator dynamics to allow for regimes from discrete to continuous states via three lavender parameters ($\tau$, $\rho$, and $\theta$). Such extended HKB models capture dynamics of human-machine interactions. (4) Extended HKB applied to partners with adapting oscillatory dynamics. Oscillator parameters in blue ($\Omega$) are coupled via adaptation rate purple parameter ($\epsilon$). (5) Another model of human-machine dyads. Here, red parameters refer to bias ($\psi$) of various strength ($c$). (6) HKB model scalable for coupling among multiple agents utilizing same orange coupling phase parameters in original HKB model. For additional explication of the models and supporting empirical evidence, see Tognoli et al. (2020). Modified with permission from Tognoli et al. (2020).

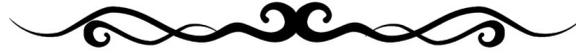

# 6. Radical embodied computation: Emergence of meaning through the reproduction of similarity by analogy—Fred Hasselman[11]

The computer metaphor of mind and brain is unviable as an explanatory vehicle for the complex adaptive behavior of living systems. I present anomalies in the empirical record of the neurosciences that expose profound problems with the assumption that the brain performs machine-like computations. As an alternative, I suggest that evolved agents' adaptive coordination of behavior is based on a massive redundancy of reality instead of massive modularity of mind. A research program of Radical Embodied Computation based on contemporary theories of physical information and natural computation takes the order generating process of the reproduction of similarity by analogy as its main topic of inquiry. Finally, I provide empirical evidence for the ability of all evolved agents, including those without a nervous system, to exploit the massive redundancy of reality to coordinate their behavior.

## 6.1. What do we understand by the computer metaphor of the mind and brain?

In order to be able to operate as a skilled agent in a dynamically changing environment, an organism will have to coordinate its behavior adaptively and intelligently relative to the constraints of its internal structure and the demands of the task at hand (Bruineberg and Rietveld, 2019). The computer metaphor of the mind and brain of skilled agents comes in many different shapes and forms but generally describes a discrete input-output machine that performs computations relatively independent of its immediate environment. Most accounts share the following three ingredients: (i) a virtual-physical dualism; (ii) unauthorized information theory[12], and (iii) a component-dominant causal ontology (Newton's curse).

    The first ingredient, the virtual-physical dualism, refers to a mind-body dualism in which the mind of a skilled agent is considered a virtual machine that is emulated on the evolved hardware of its physical body. More generally stated, a divide is introduced between the physical medium and its behavior, which is interpreted as the result of some form of computation. The dualism is expressed most explicitly in the classical Computational Theory of Mind:

> *Computer theorists [...] often speak of identities of virtual architecture. Roughly, you establish the virtual architecture of a machine by specifying which sets of instructions can constitute its programs [...] the*

---

11  Correspondence: fred.hasselman@ru.nl (F. Hasselman).
12  "Unauthorized," due to an anecdote related by Walter Freeman on the inaugural meeting of the Society for Complex Systems in Cognitive Science (Amsterdam, 2009), in which he recalled a lab-visit by Claude Shannon, the progenitor of classical information theory, who expressed severe concerns about the applicability of his theory for describing information processing in biological systems.

> *present question is why anything except virtual architecture should be of any interest to the psychologist.*
>
> — Fodor (1983, p. 33)

How exactly the physical and virtual architectures interact is often omitted or considered irrelevant (e.g., nonreductive physicalism, Fodor, 1974). It is common to posit only a minimalist correspondence principle that assumes correlates must exist between the states of the virtual and physical machines:

> *… symbol structures [… ] are assumed to correspond to real physical structures in the brain and the combinatorial structure of a representation is supposed to have a counterpart in structural relations among physical properties of the brain.*
>
> — Fodor and Pylyshyn (1988, p. 13)

Virtual-physical dualism can be found at many different scales of observation (**Fig. 12**), e.g., the Neuron Doctrine, posited by Ramón y Cajal in 1888 (cf. López-Muñoz et al., 2006) considers the neuron to be (i) a fundamental structural and functional unit of the nervous system, (ii) an input-output machine with discrete states, (iii) in which input from dendrites is processed by the cell body and sent as output to the axon. The same virtual architecture is applied to larger assemblies of neurons:

> *I refer to 'path theory', which states, roughly, that the functions of the central nervous system are controlled by chains of neurons laid down as a path so as to conduct the impulse to its appropriate end-organ: that the paths are strictly constant, thus accounting for the fact that the reflexes and reactions are largely constant: that learning consists of the opening up of new paths and that memory consists of the retraversing of some old path by another impulse.*
>
> — Ashly (1931, p. 148)

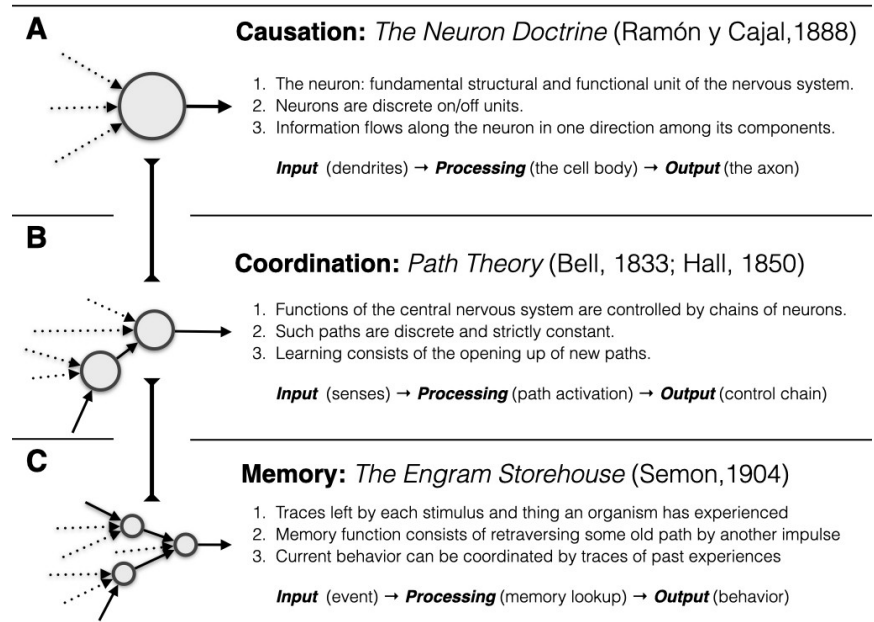

**Fig. 12.** Virtual machines posited to exist at different levels of organization of the nervous system: (A) the neuron; (B) the path; (C) the engram.

In 1904, Richard Semon (cf. Lashley, 1950) applied the same architecture to even larger scales when he suggested the engram to be the unique physiological trace left in the central nervous system by each stimulus and thing that an organism had experienced or learned (Bruce, 2001; McConnell, 1968). The brain could be considered a storehouse for engrams (Gibson, 1966), a biological machine responsible for processing, storing, and retrieving **information** to coordinate behavior. Although these machine models of biological computation were posited over a century ago, they remain omnipresent in contemporary literature:

> *Memories are presumably stored in subgroups of neurons that are activated in response to a given conjunction of sensory inputs.*
>
> — Reijmers et al. (2007, p. 1230)

> *The activity of individual neurons during learning is likely to determine their recruitment into the memory trace.*
>
> — Nonaka et al. (2014, p. 9309)

The second ingredient concerns the conflation of the formal concept of information (a quantity that resolves uncertainty about the configuration of an information source) with meaningful or **semantic information** (similarities between configurations of different information sources). Classical, algorithmic, and quantum information theory, explicitly exclude it as part of their explanatory domain (Desurvire, 2009, p. 38). As Shannon (1948) lucidly explained:

> *The fundamental problem of communication is that of reproducing at one point either exactly or approximately a message selected at another point. Frequently the messages have meaning; that is, they refer to or are correlated according to some system with certain physical or*

> *conceptual entities. These semantic aspects of communication are irrelevant to the engineering problem.*
>
> — Shannon (1948, p. 378)

The semantic aspects of a message are irrelevant for reproducing it, if they were relevant for message reproduction, a universal theory of information and communication would not be possible.

The third ingredient concerns the assumed relationship between parts and wholes in explanations based on the computer metaphor, which is best described as Newton's Curse:

> *[...] conceptualizing causal primacy in terms of a reduction of wholes to parts, where the wholes are causally impotent epiphenomena, i.e., merely aggregates of microphysical constituents.*
>
> — Leeuwen (2009, p. 38)

The idea is that, like a machine, the phenomena of the body and mind can be explained as a composition of independently operating parts whose functions essentially "add up" to generate behavior at the level of the whole. As we go from neuron to brain, the functions of the physical substrate essentially stack up like lego blocks as information is passed from one scale of the organization to the next. This move leads to a science of human behavior and cognition in which research efforts focus on identifying functionally independent components (component-dominant dynamics) instead of the interactions between them (interaction-dominant dynamics).

To summarize, the computer metaphor of body and mind as presented here is a general metaphor of machine-like computations performed by the biological subsystems of the body, most prominently the neurons in the brain. The metaphor can be found in scientific theorizing dating back to the very conception of the cognitive, behavioral, and neurosciences and has led to an interpretative view of computation, the brain can be thought of as performing logical operations, but the focus of research should be on understanding how minds emerge from the physical characteristics of the complex biological systems nested inside our bodies (Boyle, 1994, p. 452). Before such a change can happen, the cognitive, behavioral, and neurosciences will have to acknowledge the anomalies that exist in the empirical record to even the most general assumptions underlying the physical realization of the virtual computing machines.

### 6.2. What are some of the limitations of the computer metaphor?

No matter how minimal a correspondence principle between the virtual and the physical is defined, all scientific claims based on the computer metaphor assume neurons in the brain perform computations and that these computations are somehow causally entailed in bringing about coordinated, **adaptive behavior**. **Fig. 13** summarizes a selection of cases that can be said to be anomalous to that claim. When MRI scans of the brain show a large black hole inside the skull of a patient, indicative of a liquid occupying 50 –75% of the volume that typically contains vast amounts of interconnected neurons, anyone would be surprised to learn the patient is an otherwise healthy 44-years-old French civil servant, married, with children (Feuillet et al., 2007). In China, a 24-year-old woman, married with a daughter, went to a hospital because of persisting nausea and was found to be the 9th recorded case of Cerebellar agenesis: her cerebellum was missing completely (Yu et al., 2015). Due to Rasmussen syndrome, a a 3-years-old Dutch girl underwent surgery to remove her language dominant hemisphere. This chronic focal encephalitis had caused a severe regression of language skills, but at age seven, except for slight spasticity of the left arm and leg, she is living an everyday life and is fully bilingual in Turkish and Dutch (Borgstein and Grootendorst, 2002).

| | |
|---|---|
| **K 50-75% of brain tissue is missing**<br>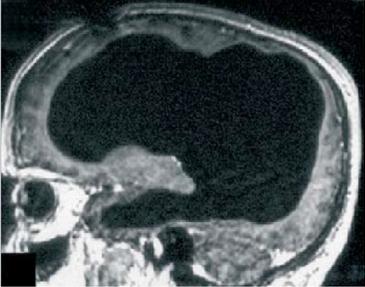<br>Feuillet et al. (2007) | Age: 44<br>Gender: Male<br>Residence: France<br>Cause: Childhood hydrocephalus<br>Discovery: Hospital due to mild weakness in left leg<br>Health: Complaints went away after shunt was placed<br>Social/Cognitive Skills: Married with children, Civil servant |
| **Cerebellum is missing**<br>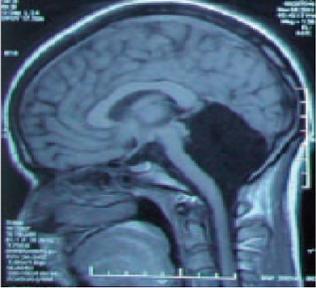<br>Yu et al. (2015) | Age: 24<br>Gender: Female<br>Residence: China<br>Cause: Cerebellar Agenesis<br>Discovery: Hospital due to persistent nausea<br>Health: Mild to moderate disfluencies in motor coordination<br>Social/Cognitive Skills: Married with children |
| **Hemisphere is missing**<br>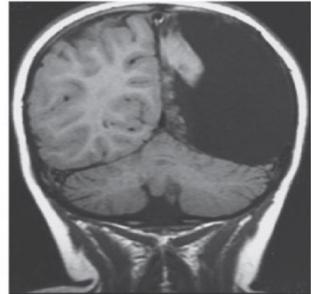<br>Borgstein and Grootendorst (2002) | Age: 7<br>Gender: Female<br>Residence: Netherlands<br>Cause: Hemispherectomy of language dominant hemisphere at age 3<br>Discovery: Severe Language impairment due to Rasmussen Syndrome<br>Health: Slight spasticity of left arm & leg at age 7<br>Social/Cognitive Skills: Bilingual Turkish/Dutch, Typical development |
| **Hemisphere is missing**<br>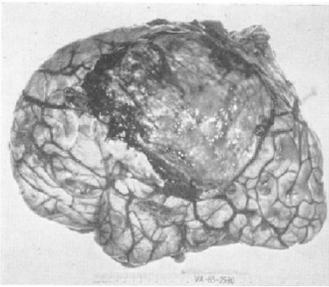<br>Smith (1966) | Age: 47<br>Gender: Male<br>Residence: United States<br>Cause: Hemispherectomy of language dominant hemisphere at age 47<br>Discovery: Mild Weakness in right arm<br>Health: Post-operative impaired speech production<br>Social/Cognitive Skills: Pre/Post-operative PIQ = 107/104, Married with children |

**Fig. 13.** Face the brains that should not be: Examples of anomalies in the empirical record to the computer metaphor of mind and brain (see text for details). All figures reproduced with permission from publishers of the respective articles.

Majorek (2012) provides an exhaustive review of research articles and case reports that represent more than 100 individuals who live rather ordinary lives given their out-of-the-ordinary brains. The vast majority experience relatively minor discomforts due to their condition. To be more specific about the extent of problems these cases pose for the idea that the brain is a central processing unit, consider the loss of computational power due to loss of brain tissue: the cerebellum makes up 10% of total brain mass but contains 80% of all the neurons in the brain, the cortex comprises 80% of brain mass, but only 20% of neurons (Azevedo et al., 2009). This composition is due to the distribution of non-neurons (e.g., glial cells) to neurons, which is about 3.7 in the cortex, but only 0.2 in the cerebellum. In the case of the woman with cerebellar agenesis, only 20% of the neurons available in a typical brain were sufficient for this individual to behave as a skilled agent in a dynamically changing environment. It was estimated that the brain tissue (neurons and glial cells) of the French civil servant was reduced by 50-75%. A hemispherectomy would remove "only" 10% of the neurons in a typical brain but almost four times as many non-neuronal cells. Notwithstanding, most contemporary cognitive neuroscience is based on studies of the activity of neurons in the neocortex, which is reduced in mass by 50% due to the surgery. Vining et al. (1997) studied the burden of illness in 58 children who had undergone hemispherectomy due to various kinds of debilitating afflictions of the brain and, remarkably, found that most children were better off with half a brain:

> *We are awed by the apparent retention of memory after removal of half of the brain, either half, and by the retention of the child's personality and sense of humor. Yet we look forward to the time when there are less mutilating approaches to these problems. Until then it seems that half of a brain is less burdensome to these children than a whole brain where one side is badly misfiring.*
> 
> — Vining et al. (1997, p. 170)

The authors report that these children leave the hospital on average ten days after the surgery. How is that possible if the brain is responsible for all our experiences of the world and our actions upon it?

What is often pointed out as an explanation is the extreme plasticity of the brain in young children, which would allow the remaining part of the brain to take over functions from the half that was removed. Aside from the biologically implausible mechanisms that need to be invoked for such an explanation to be valid (cf. Majorek, 2012), the scientific literature reveals that the remarkable recovery from hemispherectomy also occurs in much older patients. Consider the case of E.C., a 47-year-old right-handed, right-eyed patient who had his left (language) dominant cerebral cortex removed (Smith, 1966). E.C. had a pre-operative performance I.Q. (WAIS) of 108. Seven months after his dominant hemisphere was removed, his performance I.Q. was 104. He scored 85 out of 112 items correct on a verbal comprehension test. One would expect that removing a hemisphere storing many decades of unique traces of experienced events would scale to a much larger effect on I.Q. and cognitive ability.

The virtual machines hypothesized to exist at the level of the neuron and neuronal assemblies (e.g., paths, traces, the engram) do not escape clashes with a much more complex and dynamic reality. The validity of the storehouse or database metaphor of the brain has been criticized for many years (Wolpaw, 2002), but it was Lashley (1950) who, after a lifetime spent searching for the engram, declared no such structures could be found anywhere in the central nervous system. What is striking about Lashley's objections is that he suggested memory traces could not be unique or constant, localizable, that there are multiple representations of the same experience, and these representations are not reducible to the physical parts of the brain but to dynamically changing relations between those parts. Induced loss of brain tissue in the associative areas of animals does not cause amnesias but difficulties with abstraction and generalization. Wherever the engram may hide, its true nature must be a "multiplicity of interactions that can only be inferred from the final results of their activity" (Lashley, 1950, p. 27).

At the cell level, anomalies to the three postulates of the Neuron Doctrine have also been found. The idea that the neuron is the fundamental unit of the nervous system is challenged by the different types of glial cells that, in total, make up half of the mass of the brain and perform a wide range of functions such as forming communication networks, control over capillary blood-flow, neuron discharge and the pruning of connections between neurons. One astrocyte cell can control thousands of neurons (Miller and Gauthier, 2007; Takatsuru et al., 2014). The idea that the neuron is an independent discrete unit that is triggered by input and generates output is challenged by the fact that activation of neuronal pathways traced from the moment a stimulus is presented to the senses is not possible beyond the neurons in the primary sensory cortices, after which any association between input and output is lost (Freeman, 2008). A further complication is the finding that individual spiking series, as well as large scale neuronal avalanches, display scaling behavior that conforms to the statistics of selforganized critical processes, not to linear arrangements of deterministic oscillators with characteristic/fundamental frequencies (Chialvo, 2004; Rubinov et al., 2011). Finally, the postulated locus of impulse generation and the direction of flow is incorrect, due to the observation of antidromic spiking; an impulse can travel "upstream" in the opposite direction. Moreover, an impulse can be (spontaneously) generated anywhere (Bukalo et al., 2013).

To summarize, in addition to the empirical findings that are inconsistent with the models displayed in Fig. 12, the main problem exposed by the anomalous cases in Fig. 13 is twofold. (i) Drastic reduction of the number of neurons in the brain does not impact the ability of individuals to lead relatively ordinary lives; individuals with extreme atypical brain development can complete an education, maintain a job, and support a family. (ii) The brain cannot be considered a central processing unit composed of functional modules and storage capacity for experienced events, the removal of an entire hemisphere does not lead to the expected impairment of cognitive functions and memory loss if it were to fulfill those functions.

## 6.3. What metaphor should replace the computer metaphor?

The persistent problem for any theory seeking to explain intelligent behavior by a complex adaptive system that appears to coordinate its behavior based on previously experienced events, is explaining the existence of thermodynamically improbable order representing highly contextualized meaningful information about particular facts of those experienced events. The anomalous nervous systems summarized in Fig. 13 suggest there cannot be a collection of snapshots of the past imprinted onto the biological substrate. Alternative perspectives to the computer metaphor indeed seek to dispense with internal information storage and processing by "offloading" to relationships between the structure of the evolved body and the environment it evolved in. Examples are the theory of direct perception (Gibson, 1979); the ontology of affordances and effectivities (Turvey et al., 1981) and more recently *Radical Embodied Cognition* (Chemero, 2009), *Physical Intelligence* (Turvey and Carello, 2012) and *Embodied, Embedded, Extended, Enactive (4E) Cognition* (Newen et al., 2018). I conclude from the anomalies discussed in the previous section that the alternative perspectives are not radical enough in their conception of the opportunities for coordination of behavior presented by the nested structure of reality. Instead of a massive modularity of mind (Carruthers, 2006), there is a massive redundancy of reality. In what follows I will argue that coordinative structures can be realized through the recognition and reproduction of multi-scale redundancies and that **physical systems** that do so, engage in the encoding and decoding of meaningful information.

I suggest *Radical Embodied Computation* as a research program that can generate viable alternatives to theories based on the computer metaphor. Radically embodied computational theories depart from three core assumptions, the first is that a massively redundant reality exists that is composed of many nested spatial and temporal scales on which physical processes interact by exchanging energy, matter and information. A second assumption is that the state configuration of physical systems can be described in terms of physical information and their behavior as the result of natural computations (Fredkin, 1990; Hopfield, 1994; Wheeler, 1990; Wolfram, 2002). Informational Realism (cf. Floridi, 2014, p. 59) and Digital Physics can be said to "[…] regard the physical world as made of information, with energy and matter as incidentals" (Bekenstein, 2003, p. 59). When systems self-organize from one state configuration to another, the amount of information they represent changes, because different sets of degrees of freedom become available, while others are fixed. This can be described as information processing, or **embodied computation** (Flack, 2017; Kondepudi et al., 2017; MacLennan, 2012; Polani et al., 2007). The third assumption is that all behavior of living systems concerns the *reproduction of similarity by analogy*. In fact, any structure or process that generates or persists thermodynamically improbable order in the universe, at some level of analysis, reproduces redundancies as a self-organizing dynamic pattern over time, or as an emergent structural pattern through a spatial configuration, or both. In complex systems this can be observed as the emergence of long-range temporal correlations and (multi-)fractal geometry in system observables (Gilden et al., 1995; Goldberger et al., 2002; Kelty-Stephen et al., 2013; Wijnants et al., 2012). The order generating phenomena at the level of the behavior of an organism are hypothesized to belong to the same domain as is described by the physical, chemical and biological principles and laws suggested to be responsible for the emergence of complexity in the universe, e.g. in the statistical physics of **self-replication** (England, 2013) and assembly theory (Marshall et al., 2021). The idea is that living systems are able to recognize, couple, interact, or resonate with the multi-scale invariant patterns presented by their internal and external environments. What is generally meant by terms such as recognition, coupling, multiscale interaction and resonance is often formally equivalent to the dynamic patterns produced by multivariate (coupled) dynamic system models (cf. Broer, 2012).

As an example, consider a well-known system from organic chemistry in which two physical information sources (nucleotides and amino acids) are connected through a coding structure, transfer RNA. The relationship between RNA and DNA is apparent, however:

> *[…] the conversion of the information in RNA into protein represents a translation of the information into another language that uses quite different symbols. Moreover, […], this translation cannot be accounted for by a direct one-to-one correspondence between a nucleotide in RNA and an amino acid in protein.*
>
> — Alberts et al. (2002, p. 334)

Transfer RNA is an analogy that gives meaning to the information structures by translating their similarities, i.e., by recognizing or signaling the invariant structure that exists between them, but cannot do so directly:

> *Two separately identifiable patterns are related by analogy if the existence and frequency of the one is correlated with the existence and frequency of the other in the absence of direct forces between the two patterns that could cause the correlation. That is, correspondence between codon and analogon came about, and is maintained, by reproduction of an initial random event.*
>
> — Walker (1983, p. 809).

The transfer RNA molecule can be said to connect the world of nucleotides to the world of amino acids (Barbieri, 2003, P. 98), it represents the transmission of a message "correlated according to some system with certain physical or conceptual entities" (Shannon, 1948). The tRNA translates redundancies that were captured and encoded millions of years ago, but it takes just 20 seconds to several minutes to reproduce those patterns in the present (Alberts et al., 2002). The communication system as a whole is the embodiment of contextual meaning, but the semantic aspects of the message are still irrelevant for successful transmission. The translation leads to the identification of mutual information between the different information sources, which reflects the emergence of meaning (Kolchinsky and Wolpert, 2018).

The next step is to extend this idea to explaining "higher cognition" without machine-like storage and processing of meaningful information, the "Representation-Hungry Challenge" (Kiverstein and Rietveld, 2018). This concerns the problem which Bruinenberg et al. (2019) define as the coordination of behavior based on "aspects of the sociomaterial environment that are not sensorily present". Coordination of behavior that is *not* considered representation-hungry, concerns the perceptual coupling of an evolved agent to the regularities and invariants of its ecological niche, which Bruinenberg et al. (2019) define as lawful ecological information. This perceptual coupling represents the offloading of functions to ecological principles and laws (Petrusz and Turvey, 2010). Bruinenberg et al. (2019) describe general ecological information as regularities of the ecological niche of the kind that if X occurs, it is likely that Y will occur as well (i.e., statistical regularities). Coordination of behavior based on statistical regularities could indeed be a way to account for higher cognition. However, I argue that from the perspective of radical embodied computation, this distinction is unnecessary and, in both cases, refers to the decoding of meaningful information by an order generating process.

The fact that organisms can perceptually couple with their environment is due to the evolutionary process of speciation by natural selection, which occurs on time scales at which the identities of genomes fluctuate. The coordination of behavior based on lawful ecological information is an uncontroversial evolutionary account of how specific knowledge about the world ended up represented by the physical configuration of an organism. The only difference between lower and higher cognition, or, lawful and general information, seems to be the time it takes for meaning to become encoded by the body. In fact, the evolutionary events that caused fish to have gills, birds to have wings and humans to walk upright, are also no longer sensorily present in the sociomaterial environment, but their after-effects certainly are! The events in a family history that caused an individual to be born at a particular geographical location, in a specific sociocultural environment, in a specific moment in time, are no longer present, but they do play a role in explaining why the behavior of an individual is considered psychopathological or not (Olthof et al., 2022). The thermodynamically improbable order generated by living systems through their structure and behavior represents an accumulation of broken symmetries of the past (cf. Hopfield, 1994), a living record of meaningful information.

To partly resolve the "Representation-Hungry Challenge," I suggest to distinguish between immediate causal entailment and entailment that is mediative for explaining the behavior of skilled agents (**Fig. 14**). The former refers to causes for the current state of affairs whose effects are immediate (laws of physics, genotype, sociocultural environment, personality), the latter refers to the causes that can be traced as mediators in the realization of the current state of affairs (age, time of day, quality of sleep last night, current emotional state). The scale divide is an effect horizon, its main purpose is to serve as an explanatory vehicle to indicate there is a structure that is permissive of the behavior in the present, and a structure that is causative. The permissive structure appears as a set of constants painted on the horizon, representing boundary conditions for the causative structure. The causative structure lies within the horizon and is the set of efficient causes of the behavior under scrutiny in the present. This interaction-dominant causal ontology is consistent with theories about causation and direction of coupling in complex systems in which there is no privileged scale at which causality resides. In general, fast processes evolving at shorter time scales can be said to be responsible for bottom-up, aggregate effects, whereas processes at slower scales have a top-down effect, setting boundary conditions for the faster processes (coarse graining as downward causation, (Flack, 2017; biological relativity, Noble et al., 2019).

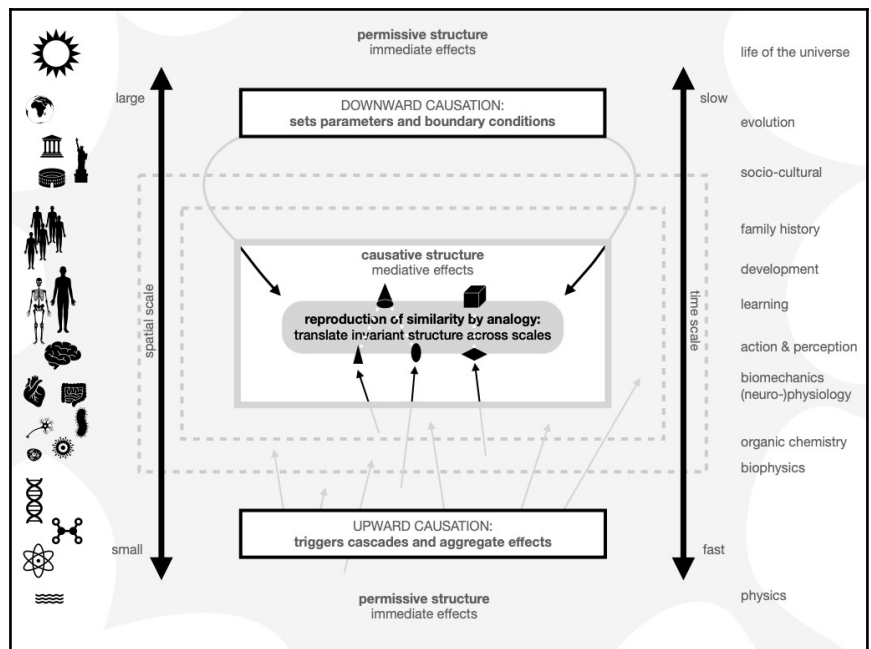

**Fig. 14.** The massive redundancy of reality provides sufficient structure for evolved agents to coordinate their behavior as is if it were based on a unique internalized interaction history. The grey boxes represent different effect horizons that can be placed at an arbitrary divide between spatial and temporal scales to separate immediate and mediative effects into permissive and causative structures, respectively. In addition, this illustration shows that upward causation involves structures and processes that have cascading or aggregate effects relative to higher scales. Downward causation involves structures and processes that set parameters and boundary conditions for lower scales. Meaningful information emerges whenever an analogy reproduces similarities between structural or dynamical patterns that exist on different scales.

The horizon also can separate spatial scales, to explain phenomena such as swarming behavior, the (temporary) collective coordination of behavior between many individual agents, it seems obvious that system boundaries should at times be able to extend beyond the individual or its component parts (Brush et al., 2022). Note that for temporal scales, the relationship between the relative order of the scale and their classification as immediate versus mediative may be more consistent than for spatial scales. However, what should be considered fast or slow, small or large, is arbitrary, unless there is knowledge about naturally occurring divides (e.g., adiabatic separation, Hopfield, 1994). Quantum physical phenomena are permissive of phenomena observable at larger spatial scales. In most circumstances it will not be sensible to declare quantum phenomena as the efficient causes for the macro scale structure of biological systems, but the macro states could not manifest without the immediate entailment of quantum phenomena. The same holds for speciation by natural selection, it is not wrong, but also not very informative, to suggest the adaptive behavior of an individual organism in the present was caused by an ancient event in its evolutionary history, but the genome does not provides a structure that is permissive of such behavior.

## 6.4. What empirical findings support your preferred alternative metaphor?

In order to produce viable alternatives to the computer metaphor, the perspective of radical embodied computation should of course be supported by empirical evidence. I will address two questions: (i) Do skilled agents indeed reproduce multi-scale invariant structure of the internal and external environment to coordinate their behavior? (ii) Can the brain function as the redundancy reproduction system?

The most direct evidence supporting the first question is the phenomenon of Complexity Matching in which synchronization between systems occurs at the level of complex dynamical patterns. Evidence of selective matching of dynamical behavior to scaling exponents in different observables measured simultaneously throughout the body suggests that such a complex multiscale coupling relationship between physiological and psychological processes could actually exist (Coey et al., 2018; Rigoli et al., 2014). Complexity matching has also been reported for dyadic interactions, e.g., interpersonal coordination of coupled movements (Almurad et al., 2017; Marmelat and Delignières, 2012) and overt behavior during joint problem solving (Abney et al., 2014). Speech perception experiments have shown that human participants make use of the multifractal structure of the signal to classify speech sounds (Hasselman, 2015; Ward and Kelty-Stephen, 2018). The Complexity Matching or Complexity Control hypothesis of human behavior is comparable to principles for optimal and maximal information transport between complex systems such as posited by fluctuation-dissipation theorems and non-equilibrium thermodynamics. The general idea is to regard self-affine structure as a complex resonance frequency (e.g. the $1/f$ resonance hypothesis, Aquino et al., 2011; chaotic resonance, Freeman et al., 2001; complex stochastic resonance Kelty-Stephen and Dixon, 2013). The multifractal spectrum is a code that exposes the existence of redundancies between different scales, the ability to match, resonate, synchronize or couple to this code implies the reproduction of similarity by analogy.

To answer the second question, consider studies on (congenitally) decorticated animals and humans. Reviewing studies on the behavior OF decorticated rats (cf. Whishaw, 1990), Merker (Merker, 2007) concluded it would be difficult to distinguish them from animals with a cortex, they can engage in social behavior (grooming, mating) and do not perform worse on learning tasks. In a case report of a congenitally decorticated rat identified as R222, the authors concluded:

> *Long before encephalization there was centralization of function in the upper brainstem for sensory integration, learning, memory, motivation, and organization and expression of complex behaviors [...] R222 lived a long "normal" life by defaulting to brain organization that has sustained and propagated vertebrate life since inception.*
>
> — Ferris et al. (2019, p. 9)

Based on many behavioral experiments and neuro imaging of R222, the authors were able to corroborate a conjecture by Merker (2007) that the basic minimum brain structure required for a "normal" life is everything that evolved before the neocortex. The neocortex is not even required for consciousness. Shewmon et al. (1999) report 4 cases of children with total to near-total absence of a cerebral cortex who possessed many "higher" cognitive skills, such as distinguishing familiar from unfamiliar people and environments; social interaction; musical preferences; appropriate affective responses, and associative learning. The authors suggest these children had a mind without a brain because their caregivers continued to interact with them as autonomous individuals, providing a safe, stable, predictable environment.

Most evolved organism do not have a neocortex, in fact, most do not have a nervous system. However, even the behavior of plants can be described as retrospective, prospective, flexible, i.e., apparently coordinated by previously experienced events (Carello et al., 2012). What could be the function of the neocortex? The nervous system of invertebrates is rather simple and has been found to contribute variability, or, *adaptive indeterminacy* to their behavior, rather than exert executive goal-directed control (cf. Maye et al., 2007). Perhaps the role of the neocortex is to make the behavior of evolved agents less determined by the highly predictable structure of our massively redundant reality. Perhaps it is there to introduce some randomness and novelty to our adaptive behavior.

## 7. Tunnel vision, tunnel action, tunnel mind: Just get out—Fred Keijzer[13]

The computer metaphor invites views on mental, neural, and behavioral processes built around the *input-output relations* between an inner and an outer domain usually cast in terms of information processing. This metaphor also operates in ways that make the *material constitution* and context of these processes and domains less relevant. There are two problems here. First, the metaphor suggests that we know more about these processes and domains than we do. The metaphor also shields this unwarranted confidence from the life sciences' broader empirical context that provides examples and conceptual frameworks that bear critically on much work within the cognitive, neural, and behavioral sciences. In both ways, the computer metaphor limits the range of conceptual and empirical options to make further progress. By discarding the computer metaphor and positioning the various cognitive sciences within the general life science domain, new views on minds, brains, and behavior become possible that have a closer fit to the other sciences. The early evolution of the nervous will be used as a showcase that provides new approaches to understanding cognitive and experiential phenomena.

---
[13] Correspondence: f.a.keijzer@rug.nl (F. Keijzer).

## 7.1. What do we understand by the computer metaphor of the mind and brain?

The computer metaphor has taken many different forms. Here, I address two issues: First, this metaphor invites an interpretation of mental and neural processes that hardly diverges from a dualistic view of the mind that goes back hundreds of years. The central innovation is that the idea of an *immaterial mind* is now recast in terms of a *material brain* acting as a computational device running mental processes. Second, this metaphor operates in ways that make the *material constitution and bodily context* of mental processes and their interactions less relevant. Though the brain is obviously a material system, its present status as the modern material version of the mind—a **mind-brain**—comes from acting as a computational system (e.g., Piccinini and Scarantino (2010), among others many others). Its physical constitution is relevant as far as it provides a medium for computational processes.

Both aspects fit a long-standing way of thinking that separates the (human) mind from both the body and the rest of the world. The computer metaphor invites a *computational dualism* that remains conceptually close to Cartesian dualism—a separate inner space of reasoning that accesses a separate external world only through perception and action. The interpretation of the brain as a computational medium invites a **brain-body dualism** in turn that sets the brain apart from the rest of the body.

All in all, the computer metaphor offers a conservative view of the mind that remains close to views predating modern science, leaving alone the current views of the natural world. I want to stress two specific issues for their impact on the cognitive sciences.

### *The central role of input-output relations*

The concept of computation is tied to input-output functions, and computing involves executing such functions, calculating the output states given specific input states. Initially, computers were humans, often grouped and organized in large office environments, who performed useful calculations by hand. This task was later delegated to digital computers that performed this function faster and more reliable. The computer metaphor for human thinking derives from this context. Mental processes could be interpreted as computational processes that are at heart well understood and do not involve a spooky immaterial substance. The latter was a major scientific improvement that enabled what came to be *The Mind's New Science* (Gardner, 1987). The various cognitive sciences that we know today all go back to this shift in perspective, which enabled the nonreductive scientific study of the mind in a form that remained close to its pre-existing conception.

Input-output relations play a central role here. There are various related issues. First, while there is ample flexibility in how the mind-brain might implement computational processes, thm being *computational* is tied to articulated inputs and outputs as these enable us to specify these processes as algorithms, rules, or just "computation." As it has been formulated:

> *All that the rules need to do is specify what relationship obtains between the strings of digits constituting the inputs and the strings of digits constituting the outputs.*
>
> — Piccinini and Scarantino (2010, p. 238)

Having definite inputs and outputs is usually not problematical in the context of computation. They more or less define the problems or functions addressed. However, for the computer metaphor, the situation is less clear as it is cast as a way to explain all—or at least large chunks—of (human) behavior. Here it becomes difficult to establish in a nonarbitrary way what human beings' overall inputs and outputs are. Usually, they are selected by researchers in convenient and relevant ways for a particular task at hand. But specifying the human input and output in a more general way is not a self-evident issue.

The human task most central to the computer metaphor, computing, is a highly specialized one, which (some) humans can learn, given sufficient training and context. Computing is not the obvious core of human thought and action, which might be more in the line of manual and social skills or fun things like dancing. Recasting the environment that we inhabit and the things we do in terms of "strings of digits" is a formidable challenge. As long as we focus on vision, the input definition may look complex but achievable, but we rely on much more sensory devices on and in our bodies. Obviously, there are the standard five senses, but here tactile sensing also includes temperature and nociception. Internally, we have a wide range of proprioceptive devices, including the vestibular apparatus, but also a wide sensitivity to visceral stimuli. The concept "input" requires more explication than it often receives. For the action side, the situation is similar as basically everything we do—even remaining motionless—requires the full musculoskeletal system and its ongoing coordination even when the focal action consists of entering a single button press, e.g., pressing the 'x' on a keyboard.

Concepts like input and output impose a restricted access between the external world and an inner mental domain. To provide a metaphor for the computer, input and output are like tunnels connecting two different worlds, while the inner mind becomes a tunnel of its own, the one where information is processed and sent onwards. Like the classic notion of mind, the computational mind-brain remains separate from the physical world, being merely linked to them via their input (sensory) and output (action) connections. The brain as the material realizer obviously is part of the world, but is treated as being sufficiently isolated from the body and the world to maintain the classic view of mind.

The computer metaphor nicely fits the rational, reasoning mind that dominated large parts of historical philosophy and twentieth-century analytic philosophy (Hooijmans and Keijzer, 2007). This rational mind made us special and kept us separate from the rest of the (living) natural world. Now, hundred and sixty-plus years after Darwin's *Origin*, it still keeps our mind conceptually isolated from the rest of the (living) natural world: a world of reason versus one of causes, connected by perception and action.

### *Backgrounding the material constitution of the cognitive system and processes*

The computer metaphor with its emphasis on input-output relations stresses the functional and computational organization of the mind-brain and provides a framework that specifies which aspects of the human body are relevant. Choosing appropriate and relatively abstract inputs and output definitions that are useful for a given task or setting is one method by means of which the details of the physical body can be set aside. For the internal computational processing part, a similar abstract characterization is sought. Piccinini and Scarantino (2010) define computation in this context as any process the function of which is to manipulate medium-independent vehicles according to a rule defined over the vehicles. They also describe "medium-independence:"

> *a medium independent vehicle is such that all that matters for its processing are the differences between the values of the different portions of the vehicle along a relevant dimension (as opposed to more specific physical properties, such as the vehicle's material composition)*
>
> — Piccinini and Scarantino (2010, p. 238)

The computational interpretation determines which parts of the physical system actually constitute the computational vehicles and which do not. Importantly, the computational interpretation, including the notions of input and output, comes first, while the actual physical system—the living and acting organism—acts subserviently as an implementation or realizer for the computational processes.

Thus, under this metaphor, cognitive processes and the mind itself are thought to exist as a set of abstract processes and/or functions that are not necessarily tied to a brain, even when, in practice, they are. As said, the computer metaphor fits nicely with classic dualism and the idea of the mind as a separate space of thought and reason.

As a result, a living and acting organism is not considered cognitively relevant in its own right but as the carrier of an inner set of computational processes that together constitutes the organism's mind. The remainder of the organism can be left out of consideration as long as the overall functionality remains intact, notably perceptual processes and motility. Also, the human mind provides the yard-stick for what we take the mind to be, which further limits the physical systems deemed relevant from a cognitive science perspective. Given the computer metaphor, it is self-evident that, when it comes to cognition, the brain is an essential part of the body while, e.g., our intestines are not. Interest and relevance depend on the presence of the mind, not on the human body itself.

The focus on the brain comes with an additional and important characteristic that is widespread within the cognitive sciences: *brain-body dualism*. The computer metaphor zooms in on the brain—its neurons and action potentials constituting crucial computational mechanisms—and possibly other parts of the nervous system. Maybe the senses and muscle system are taken in as central components of an organism's embodiment. However, other parts like glands, liver, stomach, intestines or immune system are not taken on board by the metaphor. The result is a new form of dualism between the brain and the rest of the body, which seems not, or much less, relevant for the mind.

Brain-body dualism makes perfect sense as long as it remains implicit and in the background. Once it becomes an explicit characteristic, it becomes an awkward implication, as will become clear below.

## 7.2. What are some of the limitations of the computer metaphor?

The computer metaphor gave birth to the current cognitive sciences and is still shaping our thinking about potential scientific approaches for the set of phenomena we currently designate with terms like mind and brain. I discuss three ways the metaphor hinders conceptual and empirical advances within the cognitive sciences.

### *Conceptual conservatism and false confidence*

In his classic book *Progress and Its Problems*, philosopher of science Larry Laudan (1978) argued for the central importance of empirical and **conceptual problems** in the sciences. In his view, both historians and philosophers of science tended to underestimate the central role of conceptual problems when it came to scientific progress. Laudan also argued that the conceptual issues of scientific theories are not always clearly distinguished from non-scientific conceptual implications. He discusses *worldview difficulties* that arise when a particular scientific theory is incompatible or difficult to reconcile with a body of widely accepted but prima facie non-scientific beliefs (1978, p. 61). Laudan treats these worldview difficulties as intra-scientific difficulties except that the inconsistency or lack of reciprocal support is between a scientific theory or claim and some "extra-scientific beliefs" deriving from areas like metaphysics, logic, ethics, and theology. The well-known Libet experiments and their presumed implications for the notion of free will provide an example where such worldview difficulties play(ed) a major role. Much of the critical response to Libet's original studies and claims was motivated by the aim to counter the problems they raised—or seemed to raise—for our continuing acceptance of us having free will (Libet, 1999; Schurger et al., 2012).

The computer metaphor is also tied to worldview difficulties but then in a reversed way. Instead of being inconsistent with important "extra-scientific beliefs," there is a very close fit with long-standing views of the mind deeply embedded in our extra-scientific views of our own subjective experience (epistemological, metaphysical, phenomenal). Notwithstanding scientific developments that stressed physiology and behavioral studies, the rise of the computer metaphor reinvigorated the idea of an inner domain of reasons, reinterpreted as a computational device. Once the metaphor was in place, the notion—and the topic—of mind as a domain of logic, reasons, and language became a scientifically credible target that could be studied as a computational system in relative independence of underlying brain processes and their biological context.

The connection with neuroscience changed over time when the brain itself became reinterpreted as a computational device with the rise of neural networks and parallel computing. Now, after several decades of the brain, the brain is often treated as a device that has morphed into the material substance constituting the mind. From an initial opposition, the new interpretation became one of mutual reinforcement where the classic inner mind idea became tied to the human brain, strengthening itself in a new way.

Having a classic philosophical concept figuring as the central target of a modern scientific domain need not be problematical. However, we should be wary about its impact on otherwise reasonable attempts to reorient that same domain based on ongoing progress, both empirical and conceptual. The concept of mind is a powerful and influential conceptual structure that has a major influence on how the problem space of the cognitive sciences is conceived: an inner domain of reasons connected with inputs and outputs to an independently existing external world. This conceptualization is so deeply established and generally accepted that it is difficult to see it as a conceptualization that may be questioned, as classic phenomenology indeed does (Zahavi, 2003). Diverging from this conceptualization will be harder to accept than staying in its wake. Such divergences will raise conceptual problems in ways that have a very long history in science, such as, e.g., the difficulties related to Ptolemaic mathematical astronomy given the acceptance of an Aristotelian cosmology during the later Middle Ages (Lindberg, 2010). The worries about Libet's experiments and their potential impact on the reality of free will also illustrate the forces at play.

I worry that the computer metaphor reinforces the ongoing acceptance of a classic interpretation of the mind, although now with a different operating system.

In this way, the computer metaphor can easily convey confidence in its soundness by reinforcing long-standing and largely extra-scientific views, while the conceptual and empirical status of the sciences inspired by the computer metaphor may not warrant such confidence.

### *Body loss*

When I travel by train, I find it enticing to watch disembarking passengers walk away across the platform at intermediate stops. The motions and rhythms involved in walking are so different between individuals. People move fast, slow, energetic, slumped, tired, jerky, proud, elegant, awkward, stiff, and in many other ways. These differences derive from physical variations in the state of one's tendons, joints and muscles, but also differences in height, body mass, fatigue, individual walking habits, as the shoes that are worn. In addition, moods and other psychological states have their own effects. This fascinating kaleidoscope of different walking patterns and rhythms shows how (human) bodies have lives of their own that are intrinsically coupled to inner neural processes. Thelen and Smith (1996) made it a central case in their classical studies on how movement development in babies depends on an ongoing dynamical coupling between neural processes and the fast-growing baby body. These movements are also not merely physical but, as we all know from our own experience, also involve the pleasure that simple movements like walking often bring, leave alone the joy of dancing, playing, stretching, running fast, or feeling the sand of a beach beneath your bare feet. Our moving bodies are a significant and direct source of feelings and, in that sense, an integral part of mental phenomena in their own right.

All these movements, rhythms, and feelings of our living bodies disappear when we bring the computer metaphor into play. The metaphor abstracts the "inessential" parts away and focuses on what is deemed relevant to the mind-brain and its connections to the objective world. Our active, sensuous, and sometimes hurting bodies that constitute our fleshy presence in the world are dismantled and analyzed to tease out the part they play as input channels, output channels, and a connecting mental mechanism situated in our heads.

The motivation for this major excision of bodies from cognitive science derives from the differentiation between the mind-brain and the remainder of the body, which is just the body and not part of the mind. This differentiation between the merely physical and the mental domain is deeply engrained in our culture more generally and pervades the cognitive sciences. e.g., the classic Turing Test builds on the idea that the body is not necessary for thinking. We are also familiar with thought experiments involving isolated brains or other computational devices hooked up to the world and still functioning as minds. These ideas seem self-evident and to make sense. The computer metaphor perfectly fits such hypothetical cases that—supposedly—show the living body to be inessential.

Nowadays, the relevance of embodiment and the environment is widely acknowledged and given attention (see, e.g., Newen et al. (2018) for an overview of this extensive literature). However, as long as the computer metaphor and the inner mind-brain remain central ingredients, embodiment easily becomes an elaborate input-output channel around the central mind-brain that *need not be a full biological body* as robotic devices may count as well (Degenaar and O'Regan, 2017; Hooijmans and Keijzer, 2007). Overall, it is a widely shared view that the body can be artificial and *need not* be biological as far as it is relevant for cognitive science.

Interestingly, this position makes sense from a conceptual standpoint, but empirically it is difficult to defend. Conceptually, it is hard to defend that living systems provide necessary and sufficient conditions for cognitive systems and that, therefore, the bodies of cognitive systems must be *biological*. However, from an empirical perspective, such an argument is beside the point. The observation is simply that, as a matter of fact, in all naturally occurring cases, the mind-brain's embodiment is biological. Humans are organisms with everything that implies. The humans stepping across the station's platform are biological entities as they stride rhythmically to their various destinations. Humans—including their living bodies—are the natural instances of the cognitive sciences. Difficulties only arise if one starts to think in terms of a mind-body opposition and the empirical bodily context is set aside.

Sheets-Johnstone (2009) formulates her critique both strong and clear, "embodiment" amounts to a lexical band-aid on the more than 350-year-old wound inflicted by the Cartesian split of mind and body, which for her refers clearly to our living and animated bodies.

There are clear empirical reasons why the cognitive sciences should focus on full walking, thinking, and sensuous living bodies. The computer metaphor suggests cutting away the largest part of our bodies from consideration as being inessential to our thinking and feeling. This leaves our physical mind-brain, accompanied by abstract input and output boxes to connect it to the world. We must be critical about this excision. First, how can we be so sure—given our current limited state of knowledge—that we need this erasure to make headway with the study of mental phenomena? So far, we have no clue how a computational device would ever come to have the kinds of experiences that humans experience in an ongoing way. So, second, how can we defend this excision when it leads us to an abstraction that excludes the sensuousness of our natural bodies and the wide range of ways in which it immerses us, cognitively and experientially, within the world? Finally, where and how could we ever make such a cut? Our bodies themselves are the central interaction device that lets us partake in the world. The very idea of a potentially clear cut between mind and body derives from a dualistic conceptualization of the mind, but a cut that involves the mind-brain, itself a body part that evolved in conjunction with the rest of the body, becomes a very messy and bloody operation.

### *Empirical isolation*

As discussed elsewhere (Keijzer, 2021), backgrounding the material living systems that constitute naturally occurring cognitive systems discourages the conceptual integration of the cognitive sciences within the wider range of natural sciences, particularly the life sciences. The connection between the cognitive and natural sciences is often cast as *naturalization*: cognitive and experiential phenomena as we know them must be understood in ways that are *compatible* with the natural sciences but not necessarily derived from them. Another option is to stress a plain *naturalism* that aims to develop an account of cognitive and experiential phenomena that interprets these phenomena as being fully derived from the context provided by the natural and, most relevantly, the life sciences. Instead of naturalizing the mind, this scientific project tries to conceptualize what cognitive and experiential phenomena are by understanding how they arise—evolutionary and organizationally—from existing natural processes. How this latter project could take shape remains an open issue, but there are several relevant points in this context.

The computer metaphor focuses on IO functions, and backgrounding of the material substrate—the living body—tends to limit the range of empirical fields plausibly relevant for the study of mental phenomena. Given the metaphor's background in fields like computation, formal systems, and logic, the role of the biological background of cognitive and experiential phenomena is not self-evidently a central concern. In addition, this restraint dovetails nicely with long-standing tendencies in psychology to maintain some distance with strong reductive approaches such as biological theories.

In cases where a field like neuroscience is taken on board, the focus remains primarily on human cognition and experience and subsequently on the human central nervous system and its operation. In addition, a limited number of animal model systems help clarify specific scientific questions relevant to the human case. Kandel's work on *Aplysia* in relation to LTP and the study of memory is an example of the latter (Kandel and Schwartz, 1982). Finally, the IO focus on perceptual stimuli and behavioral output as general functions, irrespective of the organism or agent involved, again reinforces thinking about the cognitive domain in a more general way that need not be tied too closely to a wide range of biological fields and organisms.

The situation changes radically when the cognitive sciences are positioned within the domain of the life sciences. Working from the basic observation that natural instances of mental phenomena are exhibited by organisms, it can be assumed that cognition and experience are labels for phenomena that arose and belong within this general context. Once such an approach is taken, general considerations and theories from biology provide a clear background from which mental phenomena can be understood and studied.

This turn to the life sciences depends not only on the need to avoid losing the human body from cognitive consideration. There is currently a wealth of new empirical work that targets a *very wide array of nonhuman cases* of what tends to be called minimal or basal forms of cognition (Baluška and Levin, 2016; Levin et al., 2021; Lyon et al., 2021).[14] These cases consist, e.g., of decision making by slime molds, plant signaling that attracts predators that will eat the insect attacking the plant, and the organization and operation of basic nervous systems. While one may resist the use of the word "cognitive" for this wide range of phenomena, it is becoming increasingly clear that the phenomena that are cognitively relevant and interesting are widespread and arguably universally present in all life forms.

In this context, the words "biology" and "biological" do not so much refer to the various details of the neural systems that underly cognitive processes or even the human body, but to the wider biological domain in which cognitive phenomena have evolved exist in a wide variety. In this case, the question of what nervous systems are and how they operate become open issues in need of further research and conceptual work.

A good example of what such a program could look like at a wide conceptual level is provided by Godfrey-Smith's recent work on the constitution and evolution of animal minds and experience (2020, 2016a, 2016b). Building on Hoffmann (2012), Godfrey-Smith (2016a) sketches how metabolic processes root in processes at the nanoscale, where spontaneous motion maintains a continuous "molecular storm" that drives and shapes molecular events in ways very different from the push and pull of events at our bodily scale. From here on, evolution comes into play, including the origins of nervous systems and questions about the evolution of animals and eventually experience. In this approach, issues about the molecular constitution of life, the origins of animals and other groups, developmental and physiological processes, and the operation of nervous systems are all brought together in a broad overarching framework in which human cognition and experience acquire a context. Godfrey-Smith focuses here on the broader picture and explores potential ways of redrawing fundamental issues such as the hard problem. However, more specialized targets, such as, e.g., Cisek's (2019) proposal to link the evolution of the functional architecture of the mammalian (human) brain to expanding capacities of behavioral feedback control, also relies on such a broader biological and evolutionary background.

The wider biological context invites attention to basically all organisms' behavioral and perceptual processes, providing a testing ground for developing theories and explanations that encompass this wide range and provide a common ground in which the eccentricities of the human case can be highlighted.

Building on this wide biological context does not involve a critique of computational approaches or techniques. On the contrary, this wider context provides new domains and testing grounds for such techniques. The animal body itself requires computational models to understand its development and maintenance (Manicka and Levin, 2019). From this perspective, the still influential idea of a division between automatic and simple processes maintaining the body in contrast to the complex information processing occurring in the brain must be discarded.

---

14  A recent overview of these developments can be found in the double special issue of *Philosophical Transactions B*, Volume 376, Issues 1820 and 1821, "Part I: Conceptual tools and the view from the single cell" and "Part II: Basal cognition: Multicellularity, neurons and the cognitive lens" edited by Lyon et al. (2021).

To conclude, the computer metaphor tends to limit and isolate what we consider to be the relevant set of empirical phenomena for the cognitive sciences. This creates difficulties in accepting the relevance of our bodies for cognition and experience. It also keeps a distance from nonhuman cases of cognitively relevant phenomena that range from bacteria to plants, fungi, and animals.

## 7.3. What metaphor should replace the computer metaphor?

Instead of presenting a new metaphor, I suggest pushing more strongly towards a conceptually wider and more empirically inclusive methodological change. This would also include a more open and *wider target than the mind and brain*. Both concepts take humans as the paradigm case and, in doing so, limit the breadth of a systematic empirical and conceptual study of, what I would like to call, *cognitive and experiential phenomena*. The latter phrase is more open and inclusive than "mind" and not hindered by chestnut problems whether animal species so-and-so does *really* have a mind or are really conscious. In this way, all organisms are treated as valid, plausibly cognitive and potentially conscious cases.

Instead of "brain," using "nervous systems" (plural) would also be more inclusive and more open to a fuller integration of their operation as a part of whole animal bodies. As the many nonhuman cases discussed within the field of basal cognition (Levin et al., 2021), the presence of a nervous system is not a condition for cognitively interesting behavior, and decision-making as these phenomena also occur in organisms without one. This, in turn, raises important questions as to why nervous systems initially evolved and what their functions might be (Arendt, 2021; Keijzer et al., 2013). This is an important empirical question that requires serious consideration of possible answers rather than assuming that there can be only one, and we already know what it is.

Pamela Lyon's **biogenic approach** provides an important and useful methodological proposal for articulating a wider empirical focus. This approach states that the cognitive sciences can better start from general biological principles than by taking the specific human condition as its basic exemplar (Lyon, 2006; see also Lyon et al., 2021; Lyon and Caporael, 2016). As such biogenic principles, she names continuity, control, interaction, normativity, memory, selectivity, valency, and others. The aim is explicitly not to reduce cognitive phenomena to basic molecular mechanisms but to trace such phenomena "in the flesh" and understand better what they amount to, including such things as the normative assessments of both the organisms' state and their environment.

A biogenic approach stresses the need to focus on a wide range of organisms to investigate what cognition is and what it does, something, I would add, that would also apply to experiential phenomena. This may imply a call to turn away from the central focus of the cognitive sciences: human cognition and consciousness. But this is not a necessary implication at all. Instead, the point is to position human cognition and experience in a much more comprehensive range of natural examples. Some of these cases will be similar yet different, others could provide more basic forms of cognitive and experiential phenomena, and still, other cases may be radically different and will tell us about alternative forms that such phenomena might take. The key issue here is to *add to the pool of relevant phenomena and data*, not to exclude the human case that remains as important as it ever was.

Adding such a wider evolutionary and organizationally context is already helping to make sense of the cases closer to ourselves and learn more about how the human nervous system operates. A good example is provided by Cisek (2019; this paper), who explicitly uses a phylogenetic approach to develop a conceptual taxonomy of neural mechanisms that map to neurophysiological and anatomical data and that help to improve upon traditional concepts of cognitive science. In a roughly similar way, authors like Feinberg and Mallatt (2016) and Ginsburg and Jablonka (2019) have also developed accounts of consciousness basing themselves on phylogenetic perspectives.

A difficult issue here could be the central role envisioned for living systems at the cost of artificial ones. This could be a major point of debate, but it can be pushed forward for now. Again, the key issue here is to expand the range of relevant phenomena by turning to the many natural cases provided by living systems, cases that have received too little attention until recently. How AI examples fit into this range will become clearer later on when more becomes known about the conceptual and empirical impact of the nonhuman additions.

In whatever way this particular issue will unfold, computational techniques and ideas are crucial tools to develop such a cognitive life science. The main difference lies not with their central role within the cognitive sciences but in being potentially critical about using these tools as basic examples of the cognitive and experiential phenomena that have been present here on earth for billions of years.

## 7.4. What empirical findings support your preferred alternative metaphor?

When discussing the limitations of the computer metaphor, I stressed three points: conceptual conservatism, body loss, and empirical isolation. Here, I sketch ongoing work on the evolution of the first nervous systems, both empirical and conceptual, impacting these three points. At this early evolutionary stage, the applicability of the computer metaphor and the "mind-brain" that it envisions is not the only game in town, nor the most obvious. Various architectural, developmental, and operational interpretations are being developed and under consideration. One of these options is a proposal for a basic behavioral organization of early evolutionary animals that amounts to moving and sensing in ways that diverge in important ways from the computational input-output interpretation. I give a short overview of the field and then say a bit more on the moving and sensing proposal.

### *Early nervous systems*

Proposals about the possible origins of the first nervous systems go back to the 19th century, starting shortly after Darwin launched his theory of evolution. For a long time, development was slow (see, e.g., Mackie (1990) and Lichtneckert and Reichert (2009) for overviews), but in the last 15 years or so, the field has become very active (see, e.g., Moroz (2014) and Arendt (2021) for overviews). I will rely primarily on Arendt's overview, which nicely sets out the conceptual options. I will ignore the empirical evidence here.

As to why nervous systems first evolved, Arendt dismisses the seemingly obvious answer that they enabled more smart behavior (Lyon et al., 2021). Instead, he stresses that nervous systems evolved for information exchange and integration *between* the cells of a multicellular organism. He also differentiates between the **vertical and horizontal transmission of signals**, each tied to different views on early functionality. A focus on the vertical transmission from sensors to effectors —a reflex arc being the most basic form—highlights the cellular level of analysis and focuses on various neural circuits for behavior (**Fig. 15**). In contrast, attention to horizontal transmission comes into its own from a tissue-level perspective. It focuses on the coordination of macroscopic effectors, such as ciliated and contractile surfaces, where large groups of cells must act in unison to have a behavioral effect at the level of the whole organism.

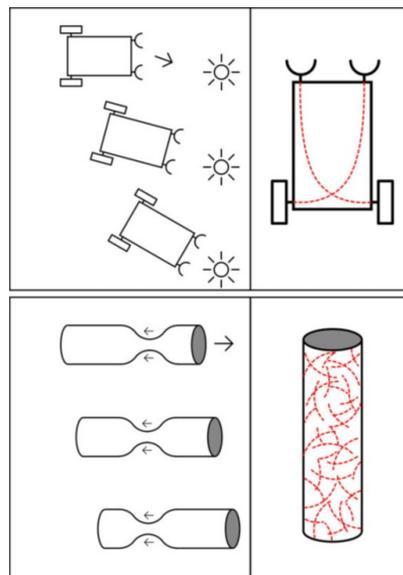

**Fig. 15.** Top: Vertical neural transmission from sensors to effectors, a basic Braitenberg vehicle (left) illustrates the behavior produced by the neural connections between well-placed sensors and effectors (right). Bottom: Horizontal neural transmission across a body surface, where a basic nerve net (right) maintains peristaltic waves of contraction (left).
Reproduced from de Wiljis et al. (2017).

In addition to these two global functional options, there are additional possible variations that also may have impacted the historical set (or sets) of events that have produced the wide variety of nervous systems present in extant animals. Here are some notable possibilities:

- Nervous systems, and neurons, may have evolved independently in more than one evolutionary lineage (Moroz, 2009).
- Extant nervous systems within a specific lineage probably have a chimeric history. Separate nervous systems evolved separately at different locations in the same animal and grew together over time (Arendt et al., 2016).
- Nervous systems may have originated first as a chemical transmission system —like a hormonal system—using neuropeptides to transmit signals to dedicated reception sites (Jékely, 2021). In this case, electrical signaling and neural elongations were a later innovations.
- Neurons may have evolved as a part of an ancestral neuroimmune system (Klimovich and Bosch, 2018).

These options may be combined in various ways and, together with the cell- and tissue-oriented functionalities, provide a rich set of empirical possibilities that still need to be sorted out. These new insights and possibilities of how nervous systems may have functioned at an early stage suggest that the straightforward interpretation of nervous systems as a biological analog of a digital computer is too hasty and requires scrutiny. The entangled relations between chemical and electrical signals and physical forces (also acting as signals), both at and between a wide range of scales, constitute an organization that is not yet well understood.

### *Body central*

In response to the computer metaphor, the option that stresses horizontal transmission and the need to coordinate macroscopic effector surfaces is particularly relevant. A recent version of the basic idea is the so-called **skin brain thesis** (Jékely et al., 2015; Keijzer, 2015; Keijzer et al., 2013; Keijzer and Arnellos, 2017). The skin brain thesis starts from the idea that agency is not a given but an explanandum. Early animals were initially multicellular aggregates that, over evolutionary time became increasingly integrated and acquired forms of agency on the way (**Fig. 16**).

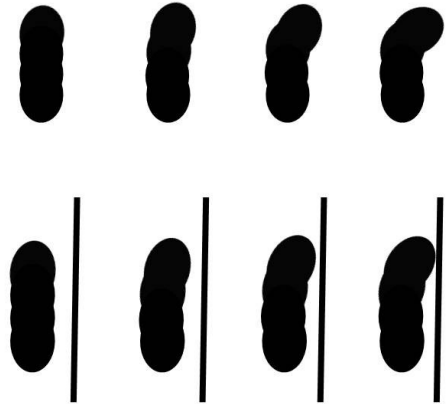

**Fig. 16.** Sensing bodies: a self-induced motion to the right can either proceed (top) or is blocked by an external obstacle (below). The difference in body contraction can be detected without requiring external sensors. Reproduced from Keijzer (2015).

One crucial problem on that road was to find ways to make multicellular aggregates of connected cells move as a single unit. More down to earth, think of it in terms of water-filled balloon-like organisms that can gain evolutionary benefits by evolving into forms that can move about. A fairly easy and cheap solution here is to remain as flat as possible and crawl about by using cilia, cellular extensions that act like small oars and provide a limited propulsive force that allows crawling and swimming. But what made (some) animals so extremely successful was the evolution of cellular (muscle) contraction as a mechanism for motility, enabling much faster and more forceful forms of movement. The skin brain thesis proposes that early nervous systems arose as basic nerve nets that coordinated the patterns of contraction across the animal body, turning the "balloon" into a motile unit.

With this change, the contractile body became a more unified entity, bound together by contractile cells, connective tissues, and a sensitivity to the forces it imposed on the body. Neural signals, both chemical and electrical, had to be integrated with the physical forces at a larger, bodily scale that themselves acted as signals feeding back to the neural and cellular levels. Importantly, the skin brain thesis proposes that contraction-based motility allowed the multicellular animal body to acquire the characteristics of an agent ath the whole-body level, rather than remaining a smart collective of cellular agents. Importantly, this bodily configuration turned the body itself into a sensing device that became sensitive to physical environmental happenings at the scale of the whole body, also tying the constitutive cells together as sensory devices (**Fig. 16**; Keijzer, 2015).

Stressing bodily coordination as a crucial evolutionary step provides a perspective on the relation between animal bodies and their nervous systems that differs importantly from the separate information processing device that the computer metaphor suggests. The skin brain perspective casts nervous systems as an intrinsic part of the body and as being dependent on the body as a functioning entity. The animal body is not a pre-existing platform to which a controlling device becomes attached.

This intrinsic connection will also be the context to make sense of what nervous systems do and how they operate. While it is obvious that nervous systems have grown in many different and often extremely complex new forms, at heart, the neural mechanisms by which muscle and body control are organized remain in place when we sit upright typing away at our keyboards. My hunch here is that this basic evolutionary background principle provides an important understanding of what nervous systems are and do.

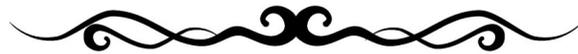

## 8. Resonances in the brain—Vicente Raja[15]

The computer metaphor has constituted a central component of the narrative in cognitive psychology and cognitive neuroscience since roughly the 1960s, although the interpretations of the metaphor are not homogeneous and many of them are not even compatible. However, the general take is that the brain is in the business of processing inputs to deliver appropriate outputs using a roughly computational strategy. Namely, the brain somehow encodes inputs, manipulates them following some algorithms, stores some information extracted from those inputs, recombines some bits of that information, delivers the proper commands for the accurate outputs, and so on. The computer metaphor has offered decades of theoretical progress as well as many computational methods and models within the cognitive sciences. However, it similarly faces longstanding problems that cyclically bring into question the usefulness of the metaphor to guide the scientific approach to fundamental (and not so fundamental) issues in the sciences of the mind. Here I defend an alternative metaphor inspired by the use of the notion of resonance in cognitive science and neuroscience writ large, and more concretely based on the notion of ecological resonance developed in the literature of ecological psychology.

### 8.1. What do we understand by the computer metaphor of the mind and brain?

The computer metaphor is not what it seems to be and this fact makes it difficult to analyze. One may think the computer metaphor is somewhat straightforward. The brain is a computer.[16] That is it. And in some sense, that is literally all the computer metaphor entails. This simplicity is likely one of the main strengths of the metaphor. It does not involve technical terms or sophisticated formalisms, so its meaning is easy to grasp in fairly familiar terms: the brain is a computer that gets some inputs, applies some computing to them, and delivers some outputs. However, this very simplicity is at the same time one of the weaknesses of the computer metaphor. It is so simple that it is helplessly vague. The computer metaphor is pretty much an "everything goes" kind of framework for cognitive science and neuroscience. But why is this the case? Well, this is because it is not clear what "computer" means in the computer metaphor and, depending on what it means, a "computer" may be a very particular device or quite literally any system in the known universe. Before we go on, let us take a closer look at this issue.

---

15  Correspondence: vgalian@uwo.ca (V. Raja).
16  A different formulation, perhaps more accurate, would be something like "the mind-brain system is a computer." This formulation would highlight that the computer metaphor applies to minds, and brains, and their relationship in general. In order to maintain some economy of words, I will use "brain" instead of "mind-brain system," but both formulations can be taken as equivalent with regard to the arguments I will be presenting.

In its original formulation, the computer metaphor was inspired by the digital computer. The cognitive revolution that began in the late 1950s was carried out by a group of scholars from different fields who found a coherent narrative precisely by virtue of their common interest in digital computers (Miller, 2003).[17] The computer metaphor was a natural corollary of this interest in digital computation, and the brain was regarded as that kind of computational system. This is still the most common take on the computer metaphor in most cognitive science textbooks (e.g., Clark, 2014; Thagard, 2005). Whenever one finds a statement like "the brain is the hardware and the mind is the software," the notion of "computer" at play is the "digital computer" one. This formulation of the computer metaphor was significantly productive and gave rise to research related to the nature of the computational states in minds and brains—usually characterized in terms of mental or neural representations (Shea, 2018)—and the computational rules to manipulate them (Milkowski, 2013). It was however a quite restrictive metaphor. Digital computers are a very particular kind of formal system that involves some equally particular forms of processing (e.g., serial processing), syntax, semantics, etc. The restrictive nature of digital computers was the main target of early criticisms of the computer metaphor. The metaphor was regarded as incapable of accommodating central aspects of the workings and the phenomenology of cognitive systems (e.g., Dreyfus, 1972; Turvey, 2018). These criticisms and further advances in the cognitive sciences provided different formulations of the computer metaphor where "computer" did not mean "digital computer" anymore.

Among the other available formulations of the computer metaphor, the one that stands in the most radical contrast to the digital computer is the one that promotes a deflationary interpretation of computation and, therefore, regards it as a kind of universal phenomenon. Put simply, this formulation takes "computer" to be a system that implements a function in a set of finite algorithmic steps. In this sense, it may be said that a "computer" is a universal Turing machine without the need for the von Neumann architecture typical of digital computers. The benefits of interpreting the computer metaphor in this sense is that researchers are less constrained by the particular features of digital computers and have more room to accommodate typical cognitive science questions within the framework. For instance, under this formulation of the computer metaphor, both digital computers and neural networks are computational devices even though they differ in core details such as the nature of the information-processing (serial vs. parallel) or the status of their representations (discrete vs. distributed). This is obviously an improvement in facing criticisms regarding issues like the biological plausibility or the formal scope of computational models. However, these benefits come at a high price: the vagueness or lack of specificity of the computer metaphor. The deflationary definition of "computer" is as general as it gets and applies to a universal domain of systems. Almost all physical, chemical, and biological systems and processes can be described as implementing a function in a set of finite algorithmic steps (Fredkin, 2003; Rozenberg et al., 2012). Indeed, often the same function can be described in terms of different sets of algorithmic steps. The consequences of this fact are deep. First, the computer metaphor does not constrain the kinds of explanation offered in the cognitive sciences. Many computational explanations of the same cognitive event, including incompatible ones, can coexist under the umbrella of the computer metaphor insofar as it is interpreted in the deflationary sense. Additionally, as the deflationary reading of the computer metaphor may be applied to virtually any physical, chemical, and biological system, it really says nothing about the concrete nature of cognitive systems. It allows researchers neither to distinguish the cognitive from the non-cognitive nor to make claims that apply specifically to brains. The computer metaphor becomes in this way more of a desideratum than a research guide.

---

17 The famous meeting at Dartmouth College in September of 1956—which is somewhat mythologically regarded as the beginning of the cognitive revolution—is an example of this fact. The meeting was attended by Noam Chomsky (linguist), Charles Miller (psychologist), Alan Newell (mathematician), and Herbert Simon (political scientist), among others. They belonged to different fields and had diverse interests, but they were united by their interest in digital computation.

There are of course several positions between the Scylla of the digital computer and the Charybdis of the computational universe. Some positions accept the computational metaphor but defend a framework based on *analog* instead of digital computation (Maley, 2021). Other approaches regard computation as a particular kind of mechanism (which is taken to be a more general notion) and claim that this conceptual move has significant effects for the computer metaphor and the explanatory strategy of the cognitive sciences (Piccinini, 2020). Still other approaches take computation to be in the eye of the beholder (Szangolies, 2020) or simply take the notion of computation to be a placeholder for any activity we can measure in the brain. All these alternative interpretations of the computer metaphor show it is more than just one metaphor. It is many metaphors that are not always compatible with each other. Therefore, it is difficult to assess just *one* ensuing research program stemming from it. Actually, it is even difficult to pinpoint a set of common features shared by all of the computer metaphor formulations. But what can we do? Are we doomed due to the ambiguity of the computer metaphor? I would not say so. It is true that providing a concrete definition or a set of sufficient and necessary conditions for the computer metaphor is likely impossible. However, we can list a kind of open-ended family of concepts that usually feature in the explanations provided by paradigms and researchers that work under the umbrella of the computer metaphor. This open-ended family of concepts is the real contribution of the computer metaphor to the cognitive sciences. Let us turn to some of these concepts.

The first concept associated with the computer metaphor is that the brain should be modeled as an *input-output* system. The characterization of the brain in these particular terms might seem fairly innocuous but it indeed has deep implications that will be discussed in the next section. Right now, I just want to highlight that once an input-output computational system is defined, both the input and the output are usually taken for granted and the efforts to explain the computational machinery then focus on whatever happens in between them. In the case of the cognitive sciences, once the brain is regarded as an input-output system, both the sensory input and the behavioral output are taken to be the precedent and the product of the computational activity, respectively. However, neither of them gets cast in computational terms.

Another concept associated with the computer metaphor is *coding*. When the input arrives to the computational system, it gets encoded to be suitable for computational manipulation. This is very easy to illustrate using the example of the digital computer. In a digital computer, the input from the keyboard, for instance, gets encoded as a string of binary code that can be manipulated by the von Neumann architecture.[18] The case of the brain is similar. In a perceptual task, e.g., the sensory input is received and encoded in the nervous activity to then be processed in different regions of the brain. The characterization of the encoding process proposed in the literature is almost as diverse as the computer metaphor formulations—see, e.g., Stone (2012) for the visual system—and even coding has been regarded as a metaphor itself (Brette, 2019). These two facts make the waters of coding a little bit muddy, but its fundamental idea remains the same: metaphorically or not, the input of the cognitive system must be encoded for the system to be able to do computational work on it.

Finally, the computer metaphor leads to the concept of *algorithms* that implement (or at least approximate) cognitive functions. The activity of the brain, when regarded as a computational system, is expected to be suitable for algorithmic explanation. Namely, the way the brain manipulates an input to deliver an output of a cognitive function is expected to be suitable for an explanation based on the sequential application of a limited set of rules. The specifications of these rules may vary significantly and go from explicit computational commands to heuristics or Bayesian statistical inferences, to name a few. But, in the general case, it must be possible to explain the cognitive function in this way.

As noted before, the computer metaphor must be understood as an open-ended family of concepts used to provide explanations within the cognitive sciences. The reason to focus on the concepts of input-output, coding, and algorithm is that they feature a very prominent role in virtually all the flavors of the computer metaphor. There are, of course, other concepts—such as mental/neural representation, filtering, storing, information, and so on—but these three are enough to characterize a general view on the computer metaphor and, importantly, to understand its fundamental limitations to account for cognitive systems.

---

18  This is an openly idealized characterization of the process. There are many levels of description of a digital computer in which we could find a code. I decide to stick to the algorithmic/formal level because it delivers the message without too many technicalities.

## 8.2. What are some of the limitations of the computer metaphor?

Even with all these caveats, it can be said that the computer metaphor describes the brain as a specific kind of system. A system that takes whatever input is available in a task, encodes it, processes it algorithmically, and delivers an output. I take this description to be uncontroversial enough as to be accepted for most practicing cognitive scientists and neuroscientists nowadays. At the same time, it has all the components to understand the main problems of the computational metaphor as a guiding idea in the cognitive sciences. These problems are different but deeply related and I will refer to them as (i) the problem of *relevance* and (ii) the problem of *inference* (see Raja, 2020).

The problem of relevance has encountered several formulations both in the computer and the cognitive sciences. Put simply, the problem of relevance highlights a very fundamental feature of cognitive systems that is difficult to accommodate in computational terms: cognitive systems exhibit sensitivity to and selectivity of some aspects of their environments while ignoring other ones. Additionally, cognitive systems can manipulate the selected aspects of their environment—in the sense of manipulating the input they get from them or in the sense of acting upon them—and are subsequently equally able to be sensitive to and selective of some aspects of the results of that manipulation. In other words, cognitive systems engage with *relevant* aspects of their environment in a context-dependent way. They are even able to generate such relevant aspects *ex novo* through their own actions (Roli et al., 2022). The problem of relevance emerges when a computational algorithm is used in an attempt to implement this relevance-driven behavior of cognitive systems. Both technically and conceptually, computational algorithms seem unable to provide such an implementation.

The technical side of the problem of relevance is exemplified by the infamous *frame problem*, first described by McCarthy and Hayes (1969) and still largely an open question in the cognitive sciences—also somewhat open in computer science as well, although some partial solutions have been already proposed (e.g., Lifschitz, 2015; Shanahan, 1997). In its simplest form, the frame problem describes the issue of knowing which elements of an environment change and which ones remain unchanged when a computational (rule-following) system engages in a specific course of action. In other words, it is the problem of knowing what things are relevantly changed by an action. This version of the frame problem has been claimed to entail an important shortcoming of computational systems in modeling biological and cognitive systems (Danks, 2014; Dennett, 2006; Dreyfus, 1992;

Fodor, 2000b). However, conceptually speaking, the frame problem is only a concrete instantiation of the wider problem of knowing what is relevant (and what is not) in the input the system gets from its environment as the system-environment interactions develop in time. For instance, what aspects of the changing retinal input are the relevant ones to visually know when to stop before hitting an obstacle while running? Object shapes? The optic flow? Textures? All of the above? The brain must know the answer to these questions in order to properly parse and encode the input. This problem is exacerbated by the fact that there is a nonlinear and highly non-lawful relationship between environmental states and the retinal input. This entails that a retinal input may be produced by an almost infinite set of environmental states and leaves what seems to be an ill-posed problem for the brain (see Raja, 2020).

The issue just presented leads to the second problem of the computational metaphor, the problem of inference. This problem highlights the issues that appear when implementing an inference in a computational system and the need for assuming the system already has some prior knowledge to scaffold such an inference. An early example of this problem was identified by William James (1890) when discussing Helmholtz's theory of spatial perception. Helmholtz's theory proposed an unconscious inference that transforms space-less sensations into spatial perceptions. William James then asked: "But how, it may be asked, can association [a form of inference] produce a space-quality not in the things associated? How can we by induction or analogy *infer* what we do not already generically know?" (James, 1890, p. 279; emphasis added). According to William James, spatial perception cannot be taken to be a form of inference because it would require some *a priori* knowledge of space that is, by definition, not present in the stimuli (i.e., the stimulus is space-less by definition). And that is impossible to justify unless one adopts a highly unsatisfactory Kantian stance.

The problem of inference did not go away with the advent of the computer metaphor. For instance, the problem of perceptual inference remains acknowledged nowadays both in the theoretical and the empirical literature (e.g., Chemero, 2009; Dennett, 1978; Turvey, 2018). And this is true even for the most advanced theories and techniques in computational neuroscience and machine learning, such as representation learning or reinforcement learning (see Raja et al., 2021). The only difference between current formulations of the problem of inference and William James's formulation is that inferences are now more explicitly characterized as computational processes entailing a relation between the system's input and the environmental causes of that input. Friston (2005), for instance, claims that "[t]he problem the brain has to contend with is to find a function of the inputs that recognizes the underlying causes;" he then immediately acknowledges that such a functional relation "may not be invertible and that the estimation of causes from input may be fundamentally ill-posed" (p. 820). Similarly, Stone (2012) claims that "[t]he brain is constantly doing its best to find out what in the world is responsible for the image on the retina" (p. 2); he then rapidly highlights the difficulty of the task. Another illustration of the problem of inference's current validity is that proponents of different computational theories and models of cognition feel the need to justify the prior knowledge they must postulate for their theories and models to work; be it an empirical Bayes formulation of a priori beliefs (Friston, 2005, 2003) or an evolutionary explanation of that kind of knowledge (Gallistel, 2020; Gallistel and King, 2010).

Summing up, the models of the brain developed under the umbrella of the computer metaphor share technical and especially theoretical issues that have to do with very fundamental assets of biological brains: (i) the capacity to relate to their environment in terms of relevance and (ii) the capacity to create new knowledge and inferences without the need of *a priori* knowledge. Even the most deflationary accounts of the computer metaphor characterize brains in such a way (i.e., input-output, coding, algorithms, etc.) that these limitations are impossible to overcome. Thus, the main question is whether all metaphors of the brain fall prey to these problems.

## 8.3. What metaphor should replace the computer metaphor?

Let me be very straightforward and say I do not think there is any overarching metaphor available in the cognitive science literature to substitute the computer metaphor. There are of course alternatives to the kind of information-processing, computational-cum-representational framework usually entailed by the computer metaphor (e.g., enactivism, ecological psychology), but these alternatives do not emerge from nor entail an overarching metaphor. This is probably for the good. Another metaphor with the universal scope of the computer metaphor would probably also suffer from being either too particular or too general as to serve as an illustration of cognitive systems. However, the lack of an alternative overarching metaphor does not mean that we cannot use alternative metaphors in particular contexts. And, more concretely, it does not entail that we cannot use metaphors that do not fall prey to the same shortcomings as the computational one. Such is the context in which I want to propose the concept of *resonance*, an undeniably growing concept in cognitive science and neuroscience, as a metaphor for some of the activities of the brain. Then I want to briefly show how the resonance metaphor overcomes the problems of the computer metaphor (Raja, 2021).

Resonance is a widely observed phenomenon in nature (e.g., acoustic resonance, mechanical resonance, and electrical resonance). In its physical definition, resonance occurs when an oscillatory system entrains another system, and the latter oscillates at a greater amplitude at some specific frequencies. In this sense, resonant systems exhibit three fundamental characteristics: filtering, amplification, and synchronization (or coupling). A resonant system amplifies its own activity at specific frequencies (filtering), and this activity depends on its own constitution and the entraining influence of a driven system (synchronization). These three core aspects of resonance are the ones that have led to cognitive science and cognitive neuroscience's growing interest in this concept.

The brain may be metaphorically seen as a resonant system composed of oscillators that gets entrained with a restricted set of inputs—therefore filtering out other inputs—and that amplifies its activity when influenced by those inputs. Such a characterization would be a way to say the brain, composed by neurons (oscillators) is sensitive to selected aspects of its environment (filtering) and maximizes its activity when some of them are present (synchronization and amplification). In this sense, resonance is a good metaphor for at least some interactions between cognitive systems and their environments. Indeed, far from being a radical proposal, we see the concept of resonance playing such a metaphorical role in many research fields within cognitive science and neuroscience. In terms of neurophysiology, resonance has been used to describe single-neuron activity (Kasevich and LaBerge, 2011); and resonance frequency shifts (i.e., subthreshold oscillations in the brain) have been proposed as a mechanism for the complex patterns of coupling between neural networks and their input as well as for the processes of structural and functional coupling between neural networks during learning events (Lau and Zochowski, 2011; Roach et al., 2018; Shtrahman and Zochowski, 2015). Motor resonance has also been used as a metaphor in the case of mirror neurons (Leonetti et al., 2015). Different notions of resonance have also been used in the field of computational neuroscience to describe different activities of artificial neural networks. Some of these concepts are coherence resonance (Yu et al., 2018), network resonance (Helfrich et al., 2019), or stochastic resonance (Ikemoto et al., 2018). More generally, in theoretical neuroscience both dynamical and inferential models of different brain activities have been based on non-linear resonance (e.g., Large, 2008) or adaptive resonance theory (Grossberg, 2013). In psychology, sequential effects have been explained in terms of resonant processes (Gökaydin et al., 2016). And even more generally, in the case of general theoretical frameworks and philosophy, the idea of the resonant brain is not a new one (Grossberg, 2021).

On top of this, Raja (2021, 2019, 2018; see also Raja and Anderson, 2019) has proposed the notion of ecological resonance to explicitly target the problem of perception without assuming the brain is a computational system. Importantly, ecological resonance offers an alternative to computation that overcomes the problems of relevance and inference in perception. Explicitly inspired in the ecological approach to perception and action developed by James J. Gibson (1979, 1966), the notion of ecological resonance aims to explain the role of the brain in the activity of detecting perceptual information. Put simply, a core tenet of ecological psychology is that environmental regularities relevant for behavior (e.g., the distance towards an obstacle as an organism locomotes) are lawfully related to the structural patters of ambient arrays (e.g., the structure of light in the optic array or the structure of sound in the acoustic array). Some aspects of these structural patterns remain *invariant* when the organism moves around its environment and constitute the perceptual information it uses to control its behavior. Within such an ecological framework, the question of interest is: how is the organism able to detect this perceptual information? The answer to this question involves the brain resonating to the perceptual information.

Ecological resonance is defined as the process by which brain dynamics become constrained by the ecological information that is available at, and is already constraining, the organism-environment scale (Raja, 2021, 2019, 2018). As depicted in **Fig. 17**, formally speaking and following the methodological tradition in ecological psychology (Chemero, 2009; Warren, 2006), the organism-environment interactions (O-$E_D$) can be described using dynamical systems theory as function (G) of time where one of the main variables is perceptual information ($\Psi$). As brain dynamics ($N_D$) can also be described in terms of dynamical systems theory, one can represent $N_D$ as resonating to O-$E_D$ by forcing a coupling relationship between perceptual information ($\Psi$) and the relevant $N_D$ variables ($\chi$) such that $\chi = k\Psi$ (where k represents the parameters of the coupling).

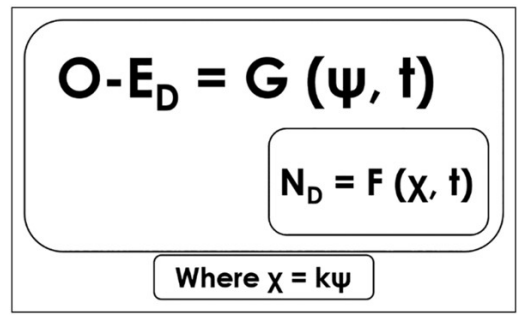

**Fig. 17.** General model of ecological resonance. Reproduced from Raja (2019, p. 409).

This general model of ecological resonance is metaphorical despite being formal. The resonance model does not entail the brain is literally resonating to perceptual information in the physical sense of the term. What it says is that, when trying to explain the way organisms can detect perceptual information, the brain dynamics can be metaphorically understood as resonating to perceptual information. Namely, the brain dynamics filter, synchronize, and amplify the perceptual information that constrains the overall organism-environment dynamics. This entails that if, e.g., you have a dynamical model that accounts for a given behavior (e.g., a baseball player hitting a ball) and that includes a variable of perceptual information (e.g., *tau* ($\tau$) or time-to-contact), you should find that perceptual variable ($\tau$) in the dynamics of the brain activity (e.g., in some relevant parts of your EEG data). If that is the case, it is proper to claim that the brain is resonating to the perceptual information ($\tau$).

Ecological resonance does not fall prey to the problems of relevance and inference in the context of perception and action. First, when brain dynamics are understood in terms of ecological resonance, perception is not characterized in terms of inference. In other words, due to the lawful relationship between elements of the environment and the structural patterns of the ambient arrays, perceptual information is not assumed to be ambiguous and, therefore, the brain is not assumed to need to perform any kind of inference to disambiguate it. In this sense, the brain resembles a self-tuning radio more than it resembles a device performing computational inferences (see Gibson, 1966). And second, the problem of relevance gets diluted once the *input* of the system is better characterized. The computer metaphor ignores the features of the input and assumes a proper description of it is not relevant for the characterization of the computational machine. For this reason, the input to the brain is understood as whatever the researcher defines as an input — from simple inputs to retinal ganglion cells (Sayood, 2018) to clips of the movie Memento (Kauttonen et al., 2018), e.g.. As I have already noted, such a theoretical context makes relevance an issue. On the contrary, ecological resonance is inscribed within the ecological approach to perception and action. One of the main features of this approach is it provides a careful characterization of the input to the brain (perceptual information) in perception-action events. As I have briefly pointed out, the ecological approach offers a description of the organism-environment system in which structural patterns of the ambient arrays fully specify the environment of the organism. To put it in Jamesian terms with regard to spatial perception, the structural patterns of the ambient arrays already contain spatial information. So, as far as space is relevant for an organism (e.g., for survival), its brain does not need decide which part of the input may be used for spatial perception but only to resonate to spatial information. But how is the organism able to know that "that" part of the input (and not others) is *the* spatial information? Once the ecological stance is taken and the availability of perceptual information in the environment is justified, an evolutionary explanation in which perceptual abilities have evolved through the constraints of available information becomes way more compelling (see Warren, 2005).

The resonance metaphor combines all of the just-reviewed metaphorical uses of the resonance notion in the cognitive sciences. As noted at the beginning of this section, it is not an all-encompassing metaphor, but it is applied to particular instances and problems within cognitive science and neuroscience. Many of the different notions of resonance reviewed enjoy well-known empirical support (e.g., motor resonance, adaptive resonance theory, or stochastic resonance), so I will not refer to them anymore. Ecological resonance, however, is a more recent proposal and its empirical support is less well-known within the cognitive sciences. I will focus on it in the last section.

## 8.4. What empirical findings support your preferred alternative metaphor?

Ecological resonance is best supported by the empirical literature on the perceptual information variable known as *tau* or time-to-contact (Lee et al., 2009; Lee and Reddish, 1981). *Tau* is defined as the inverse of the relative rate of change of a closing gap. The closing gap can be described in any domain—e.g., it can be a distance gap, like the one between an approaching ball and the position of a baseball bat, or an energy gap, like the current potential of a neuron and its firing threshold. Given this, *tau* is the variable that specifies the time it will take for the described gap to be completely closed. This is also known as "time-to-contact." In addition to the variable *tau*, *tau*-coupling has been a resource to explain the way two closing gaps are synchronically closed by the activity of an organism (Lee, 1998; Lee et al., 2009). Formally speaking, *tau*-coupling takes the form $\tau_A = k\tau_B$, which is just a particular instantiation of the general ecological resonance form $\chi = k\Psi$ (see the previous section). Thus, *tau*-coupling is formally equivalent to ecological resonance.

This equivalence becomes even more relevant when the literature about *tau* is reviewed. In general, *tau* and *tau*-coupling are important aspects of perceptual information that have been described as constraining organism-environment dynamics in a multitude of situations (for review, see Craig et al., 2000; Craig and Lee, 1999; Lee, 2005; Tan et al., 2009). Importantly, some of these situations involve *tau* and *tau*-coupling both in and between organism-environment dynamics and brain-dynamics. In these cases, *tau*-coupling is an explicit instantiation of the formal model of ecological resonance described in the previous section (e.g., de Rugy et al., 2002; Field and Wann, 2005; Georgopoulos, 2007; Lee et al., 2001; Merchant et al., 2004, 2003a, 2003b, 2001; Merchant and Georgopoulos, 2006; Port et al., 2001; Sun and Frost, 1998; Tan et al., 2009; van der Weel and van der Meer, 2009; Wang and Frost, 1992). Moreover, *tau*-coupling involving both organism-environment and brain dynamics has sometimes been explicitly referred to as an instance of ecological resonance (van der Weel et al., 2019). Thus, the plausibility of ecological resonance has found a good amount of empirical support in the ecological psychology literature.

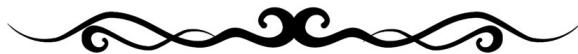

## 9. The brain as a fractal antenna—Jeffrey B. Wagman[19] and Brandon J. Thomas[20]

We argue that all computational processes require data that must be received, represented, and *processed*. The inherent ambiguity in these processes is incompatible with a lawful explanation of psychological phenomena, in particular the successful performance of everyday goal-directed behaviors in organisms all at levels of taxonomic scale. We argue that, as scientific devices, metaphors are best used *a posteriori* to generate testable hypotheses about a well-developed theory. Therefore, we develop a metaphor for mind and brain in the context of the Ecological approach to perception, action, and cognition—an approach in which the successful performance of everyday behavior is a lawful process of detecting and exploiting lawful relations. We propose that in this context, the brain could be understood as a fractal antenna. That is, it could aid in the detecting and exploiting of lawful relations at multiple nested levels, without generating, modifying, or interpreting such lawful relations.

---

19  Correspondence: jbwagma@ilstu.edu (J. B. Wagman).
20  Correspondence: thomasb@uww.edu (B. J. Thomas).

## 9.1. What do we understand by the computer metaphor of the mind and brain?

To us, the computer metaphor of mind and brain is the seemingly simple yet seductive claim that states of mind and brain are (or can be understood as) computational states. In other words, the mind and brain are (or can be understood as) *things that compute*, including—but not limited to—digital computers.

This metaphor (in various guises) has long been part of the attempt to understand the mind and brain. However, it became especially prevalent in the 17th, 18th, and 19th centuries with the attempt to understand all natural processes as *mechanistic processes* by leveraging the explanatory power of Newtonian physics and Euclidean geometry. When applied to processes occurring within the mind and brain, these mechanistic processes took the form of *computation*. Thus, the mind and brain were (or could be understood as) computers because they were *things that computed*—things that performed computations of some kind. This is explicit or implicit in the work of scholars of the brain and mind such as Locke, Leibniz, Descartes, and Helmholtz.

In the early-to-middle 20th century, the specific vehicle for the metaphor became clearer, given the substantial conceptual and technological progress toward developing the modern (programmable) digital computer. Among the landmark steps in this process was Turing's (1936) demonstration that it was possible—in principle—for a programmable digital computing device to solve any well-specified computational problem. This was possible because the device that Turing envisioned—the **Turing machine**—was specifically designed to operate as a **formal system**. In such systems—including but not limited to logic, language, and games like chess—a given goal is achieved by performing a series of system-wide rule-based operations on symbols, creating a local pattern of symbols—a symbol string.

By definition, symbols (and symbol strings) lack inherent meaning but can be interpreted to have a particular meaning in the context of a particular system (e.g., "+" means "to sum" in the context of mathematics and "a move resulting in check" in the context of chess). Symbols are *abstract*—they stand for or refer to other things or processes but not any specific thing or process. Consequently, a given goal is achieved when a sequence of operations successfully generates a symbol string that preserves truth value, regardless of the symbols or symbol strings. e.g., the mathematical expression "1 + 1 = 2" is true regardless of what the numbers themselves refer to. Critically, the abstractness of this process *is the primary strength* of formal systems and hence of digital computers.

Given that formal systems are goal-directed (i.e., truth-preserving), ruled-based, and mechanistic, they seemed to be appropriate models by which to understand states of mind and brain—which also seemed to possess these properties. Filtered through Turing's work—with an assist from Hilbert, Whitehead, and Russell—mind and brain were thought to be (understandable as) computers in the sense that they performed computations of the kind performed by a Turing machine—that is, a sequence of *syntactical operations on abstract symbols*.

Shortly after Turing's breakthrough—and not coincidentally—several technological advances turned the *possibility* of a programmable digital computing device—the digital computer—into a *reality*. In the last quarter of the 20$^{th}$ century, the applicability of such devices to understanding the mind and brain was formalized by Newell and Simon (1976) in the Physical Symbol System Hypothesis. They proposed that a physical symbol system—such as a digital computer—has the *necessary* and *sufficient* means for intelligence (or, more properly, producing intelligent action).

This claim can be interpreted to mean that (i) biological minds and brains (are intelligent only because they) are symbol manipulation devices, and (ii) artificial minds and brains (can be intelligent only if they) are symbol manipulation devices. This is explicit or implicit in the work of scholars of the brain and mind such as Putnam, Marr, and Fodor. Given the shortcomings of the behaviorist approach to language outlined by Chomsky (1959), psychology was a discipline in need of a new model for understanding the abstractness of psychological phenomena. Such a model is what the computer metaphor of mind and brain provided.

But, of course, the digital computer is not the only possible computational model of the mind and brain. Connectionist networks, e.g., take as their inspiration not an abstract formal system such as logic, math, language, or chess but a *concrete neurological system*—the brain itself. Furthermore, in connectionist networks, a given goal is achieved not through system-wide operations performed on abstract symbols but through the *emergence of a system-wide pattern of activity* among the many individual components of a network.

Moreover, there are no explicit symbols in a connectionist network—the components are merely computational relay stations. Each component receives incoming activity from many other components and then relays this activity to other components (or not) based on the relative weightings of the incoming (and outgoing) connections to (and from) those components. Therefore, a series of *local computations* determine how activity propagates through the network, creating the emergence of an interpretable *global pattern of activations* across the network as a whole. In connectionist models, states of mind and brain result from *parallel and distributed processing* (McClelland and Rumelhart, 1989). This is implicit or explicit in the work of scholars of the brain and mind such as McClelleland, Rummelhardt, Churchland, and Pinker.

In the 21$^{st}$ century, computational models of the brain and mind have emerged that—at least to some extent—combine features of both formal and connectionist models. e.g., in predictive processing models, the brain and mind serve as "prediction machines." Furthermore, rather than being *deterministic*, the predictions are *probabilistic*. A given goal is achieved not by rule-based operations on symbols or by the emergence of a particular distributed pattern but rather when prediction errors are sufficiently small. When described this way, the brain and minds operate as *Bayesian inference engines*—they use internal models to make probabilistic predictions about an incoming signal. The incoming signal is compared to a "multilevel cascade" of top-down predictions (Clark, 2013). Any mismatches between the predicted and actual signal are used to modify the model to progressively reduce the error.

Thus, in predictive models, states of mind and brain are neither constructed (symbol strings) nor emerge (as a pattern). Rather they are *hypothesized or reasoned out* under states of uncertainty. In other words, states of mind and brain are "controlled hallucinations." This is implicit or explicit in work by scholars of the brain and mind such as Friston, Gregory, Hinton, Clark, and in the performance of Watson (the question-answering and game-show-winning computer).

### 9.2. What are some of the limitations of the computer metaphor?

As described above, the computer metaphor of mind and brain can take many forms—including but not limited to—formal operations on symbols, parallel distributed processing, and Bayesian prediction. While the specifics of each of these computational models differ, they share a set of requirements and processes common to all computational devices. Namely, (i) all computational processes require data; (ii) that data must be *received* and *represented* in a form that is appropriate for the computational device; and (iii) the represented data are *processed* by some means such that they are rendered interpretable, meaningful, or otherwise more useful to a given user.

In computational models of mind and brain, the data are states of the world (e.g., the sizes, shapes, colors, and distances of surrounding objects) or—perhaps more accurately—the states of the world as *encoded in low-level energy patterns* (e.g., rays of light). These data are received by the sense organs (e.g., points of light on the retina) and are represented in states of mind and brain (i.e., a pattern of brain activity). The represented data is then processed by a set of operations (e.g., symbol manipulation, weightings in a network, or Bayesian statistical inference) such that it is rendered interpretable, meaningful, or otherwise more useful to the organism itself. The result of such processing is the mental experience of states of the world. In other words, the focus in all computational models of mind and brain is on *how states of the world are remade inside the organism* (see Reed, 1996). In particular, the focus is on the *intelligent means* by which this occurs—rule-following, pattern matching, or statistical inference.

Despite the prevalence and promise of computational models of mind and brain, there are many reasons why such models are untenable (Blau and Wagman, 2022; Turvey, 2018). In what follows, we discuss two such reasons—the

**grounding problem** (Von Eckardt, 1995) and the **everyday behavior problem**.

*The grounding problem.* As described above, computer models— of mind and brain or otherwise—require representations in the form of encoded data. The fundamental problem is that—*by definition*—representations bear no necessary relationship to that which they represent. That is—*by design*—representations are entirely *ungrounded*. While this is a primary strength in models of formal systems in which the relationship between representation and that which is represented *does not* matter, it is a primary weakness in models of mind and brain in which *it does*.

Thus, the data received by the sense organs (e.g., the points of light on the retina) and the represented and processed data (i.e., a pattern of brain activity) are *both ambiguously related to states of the world*. The processing performed on these representations aims to bring the mental experience of states of the world into *closer approximation* to the actual states of the world. But even in the best-case scenario, such processes can only create mental states that *just so happen* to be a reasonable facsimile of world states—the so-called "veridical hallucination"—and even then, only when circumstances are just right.

Ambiguity in any process is antithetical to a lawful explanation of that process in which *unambiguous relations* hold across *all possible* (or known) circumstances. Yet ambiguity is a *necessary component* of model of the mind and brain based on the computer metaphor. Moreover, such metaphors neither *reduce to nor emerge from lawful relations*. Consequently, these models drive a wedge between psychological and natural sciences (Turvey, 2013; Wagman, 2010) and create unsolvable mysteries rather than scientifically tractable problems.

*The everyday behavior problem.* Even if we assume computational models are responsible for *a very particular psychological phenomenon*—the mental experience of states of the world—this does not bring us any closer to explaining *a much more general psychological phenomenon*—the successful performance of everyday goal-directed behavior. The fundamental problem is that everyday behaviors are necessarily performed with respect to the world itself, not the mental experience of the world. While the former is physical, objective, external, and law-based, the latter is mental, subjective, internal, and rule-based. Despite these fundamental differences, everyday behaviors must nonetheless be performed *prospectively, flexibly, online, and real-time*.

This challenge is compounded by the fact that in most computational models, the processing results in mental experience of isolated and objective properties that are—at best—*indirectly related to* and—at worst—completely *independent of* performing any behavior (e.g., brightness, color, distance). Moreover, such models are typically mute on the experience of context-dependent and relational properties *directly* related to performing a particular behavior (e.g., reachable, catchable, climbable).

Moreover, performing the movements required of any everyday goal-directed behavior requires coordinating the activity of an inordinate number of anatomical units—the so-called **degrees of freedom problem** (Bernstein, 1967). Therefore, in a *thoroughgoing computational model of mind and brain*, the computational power of the mind and brain would not only be brought to bear in intelligently generating an appropriate mental experience of states of the world from a pattern of brain activity but also in using such mental experience to generate an appropriate *sequence of muscular contractions* resulting in everyday behavior with respect to the states of the world.

Empirical demonstrations that either one of these (i.e., mental experience of states of the world or everyday goal-directed behavior) is possible in an organism without substantial computational power—without a sophisticated brain or *any brain whatsoever*—would weaken the case for a thoroughgoing computational model of mind and brain. Such demonstrations would mean that a computational model of mind and brain is—at best—sufficient but unnecessary.

While it may be challenging, if not impossible, to empirically demonstrate mental experience of states of the world in an organism without a sophisticated brain or any brain whatsoever, it *is* possible to empirically demonstrate prospective, flexible, online, and real-time performance of goal-directed everyday behavior by such organisms (Turvey, 2013; Wagman, 2010). We now provide a brief overview of such empirical evidence.

Worms plug their burrows with leaves to avoid desiccation. But they are *selective* in doing so—both in terms of the size and shape of the leaf they choose and how they choose to grasp and drag that leaf into their burrow. When leaves are not present, they show the *same selectivity* with other materials such as paper strips (see Reed, 1982).

Limpets are aquatic snails that are preyed upon by several other aquatic species. How they respond to predation attempts depends on the relative size and species of the predator. When attacked by a whelk—another kind of aquatic snail— that is slighter larger than the limpet, they fight back. When attacked by a whelk that is much larger than the limpet, they retreat. When attacked by a starfish, they almost always retreat regardless of relative size (Branch, 1979; see also Wagman et al., 2019).

The dodder is a parasitic plant. When no host plants are available, the dodder will grow in no particular direction. If only a low-nutrition host plant is available, the dodder will grow toward the low-nutrition host. If *both* a low- and a high-nutrition host are available, it will grow in the direction of the high-nutrition host. If the dodder is tethered to a very low-nutrition host, it will grow *away* from that host in search of a more nutritious host (Carello et al., 2012; see also Runyon et al., 2006).

Amoeba from the genus *Difflugia* construct shells using 200–300 individual mineral particles. Like the worms, they do so selectively in terms of the size and material composition of each particle. Some species of *Difflugia* use differently sized particles in different places in the process of creating the shell. And some species even create functional "teeth" at the opening of the shell that are used to catch, pry open, and eat microscopic prey (Han et al., 2008; Turvey, 2018).

Such empirical findings create a dilemma for a throughgoing computational model of mind and brain. Either (i) computational models of mind and brain must (somehow) account for mental experience and performance of everyday behavior in *all species— regardless of the sophistication or presence of a brain*, or (ii) computational models of mind and brain can only account for such processes in species with sufficiently sophisticated brains—in which case, *some other kind of model* is necessary to account for such processes in all other species. Neither possibility is desirable, and both are problematic. And in either case, such findings imply that computational models of mind and brain are—at best—sufficient but unnecessary.

### 9.3. What metaphor should replace the computer metaphor?

We do not necessarily support the notion that a metaphor for the mind or brain (or other natural entity or process) is necessary. But, at the same time, good theories are usually amenable to metaphors because the theoretical constructs—the *topic* of the metaphor—often have the required combination of scope and clarity to be likened to something else—the vehicle of the metaphor. And metaphors are useful as (i) scientific devices because the proposed relationship between the two—the *ground* of the metaphor—can generate testable hypotheses about the theoretical constructs and (ii) pedagogical devices because —*by design*— the vehicle is often familiar even if the topic is not.

However, the difficulty comes when what starts as a *metaphorical relation* tacitly becomes an *identity relation*— that is, when the topic of the metaphor eventually becomes identified with (or identical to) the vehicle for the metaphor. In our estimation, this has happened to the computer metaphor for mind and brain. At some point, the computer metaphor of mind and brain underwent a shift from being a *useful means to generate testable hypotheses* about a theory of mind and brain to being *the theory of mind and brain* itself.

We suspect that if the (often implicit) assumptions and implications of the computer metaphor of mind and brain had been seriously considered (Blau and Wagman, 2022; Turvey, 2018), then the metaphor (and subsequent theory) might not have been adopted as widely and unquestionably as it was. Thus, we take the stance that metaphors are best used a posteriori to generate testable hypotheses about a well-developed theory rather than *a priori* to develop (or be) the theory in the first place. Accordingly, we offer an alternative to the computer metaphor of mind and brain by first providing a sketch of a well-developed theory of the successful performance of everyday goal-directed behavior—*the Ecological*

*approach to perceiving, acting, and cognizing*—and then using a metaphor—a **fractal antenna**—to generate testable hypothesis about the role of mind and brain in that process based on that theory.

*The ecological approach to perceiving, acting, and cognizing.* In the ecological approach to perceiving and acting (Gibson, 1979, 1966), as well as psychological phenomena more generally (Blau and Wagman, 2022), the fundamental unit of analysis is the **animal-environment system**—not the animal (or any part of it, including its brain) or the environment in isolation (Turvey, 2018). There are at least two important reasons for this.

The first is that it is at this level that the performance of everyday goal-directed behavior occurs. Specifically, the relationship or fit between a given animal and a given environment (i) determines the set of behaviors that are possible for that animal in that environment—the **affordances** for that animal such as reachable, catchable, climbable; and (ii) supports the performance of any given behavior within this set—reaching, catching, or climbing. The explicit focus on the *prospective, flexible, online, and real-time* performance of everyday behavior rather than on mental experience means that the Ecological approach never has to generalize from the former to the latter. Therefore, it avoids the *everyday behavior problem*—it focuses on how an animal makes its way in the world rather than how a world is made inside of that animal (Reed, 1996).

The second is that the Ecological approach proposes that it is only at this level that the lawful relations that underlie the performance of everyday goal-directed behavior emerge. In particular, the structured energy patterns surrounding a given animal are lawfully (unambiguously, invariantly) structured by the substances and surfaces that surround that animal. Therefore, the structured energy patterns encountered by a given animal at a particular point of observation are informative about that animal's relationship to those substances and surfaces. In other words, those patterns are informative about the set of behaviors that are possible—the affordances—for that animal in that environment. Therefore, performing everyday goal-directed behaviors requires *detecting and exploiting information about affordances—no more and no less* (Wagman et al., 2019). The insistence on lawfulness rather than the acceptance of ambiguity means that the Ecological approach never has to explain how the experiences of the perceiver (just so happen to) relate to the states of the world. *Such experiences are a lawful consequence of detecting information*. Therefore, it avoids the *grounding problem*—it focuses on providing a lawful account of perceiving, acting, and cognizing from the first principles (Turvey and Carello, 2012).

Moreover, given the description of perception as the detection of information —no more and no less—the Ecological approach is never faced with the dilemma of providing different explanations for perceiving, acting, and cognizing depending on the sophistication or presence of a brain or nervous system. In the Ecological approach, perceiving, acting, and cognizing result from *the same lawful processes*

—the detection and exploitation of information—at all levels of the taxonomy. Therefore, the Ecological approach is a theory for all animals—and perhaps for all organisms—serving as a mutually supportive partner to extant theories of evolution (Keijzer, 2015; Turvey, 2013; Wagman et al., 2023).

*Requirements for an Ecological metaphor for brain and nervous system.* It might seem challenging—if not impossible—to develop a metaphor for the role of mind and brain in an approach in which the brain itself is potentially *unnecessary* and *insufficient* (so long as the organism can detect and exploit the structured energy patterns that are informative about affordances). But even if the brain is unnecessary and insufficient, it does not mean that the brain plays *no role whatsoever* in performing everyday behaviors. After all, there is a reason why brains (and nervous systems) emerged and persist in animal species at many taxonomic levels (Fultot et al., 2019; Keijzer, 2015; Keijzer et al., 2013; Swenson and Turvey, 1991).

Developing this metaphor requires an appreciation that, in the Ecological approach, the phenomena of interest occur at the scale of the animal-environment system, and this system is highly nested. In both the animal and the environment, systems are nested within systems, units are nested within units, and processes are nested within processes at all scale levels. While it is the case that the animal-environment system is the scale at which all psychological phenomena emerge, this does not mean that systems, units, and processes are nested within this system (such as the *nervous system or the structured energy arrays*) do not contribute to such emergence. Many different task-specific systems are nested within the animal-environment system (including the nervous system) are flexibly recruited to detect and exploit information in the performance of everyday behaviors (Reed, 1982; Van Orden et al., 2003).

But, of course, any metaphor for the role of the brain and nervous system would need to be consistent with the fundamental principles of the Ecological approach. First, the vehicle for the metaphor would need to be *transparent to the lawful relations in animal-environment systems* (Fultot et al., 2019). It could aid in detecting and exploiting information, but it would not generate, modify, or interpret such information. Second, the vehicle for the metaphor would also need to exhibit the nestedness that is characteristic of units, systems, and processes at (and within) the ecological scale. And third, it would need to be one that could apply across taxonomy levels.

In his most mature description of the Ecological approach, Gibson (1979, p. 246) suggested a possibility that meets all three of these conditions. In particular, he suggested that the process of detecting and exploiting information is one of resonating with that information. If so, perhaps the brain and nervous system can be *understood as resonators*. Importantly, a resonator is sensitive to, reverberates with, and amplifies a pattern of structured energy, making it possible for that energy pattern to influence units, processes, and systems existing at larger or smaller scales.

While a resonator metaphor has recently been used to generate testable hypotheses about neural activity in the **Ecological approach to perception and action** (Anderson, 2014; Raja, 2018; Raja and Anderson, 2019), we wish to develop this metaphor a bit further. We would like to speculate on what *specific kind* of a resonator the brain and nervous might be.

*The brain is a fractal antenna.* In the animal-environment system, systems are nested within systems, units are nested within units, and processes are nested within processes. This nesting includes systems, units, and processes at many different spatial and temporal scales and levels of complexity. Moreover, many (if not all) of these systems, units, and processes have analogous or identical structures across scales. Many (if not all) of these systems, units, and processes exhibit *self-similarity or scale invariance—the defining feature of fractality* (Mandelbrot, 1982). Fractality is an indication that the various micro and macro systems, units, and processes are coordinated across spatial and temporal scales and are poised to flexibly reorganize into a new coordination pattern—if and when conditions demand it (**Fig. 18**).

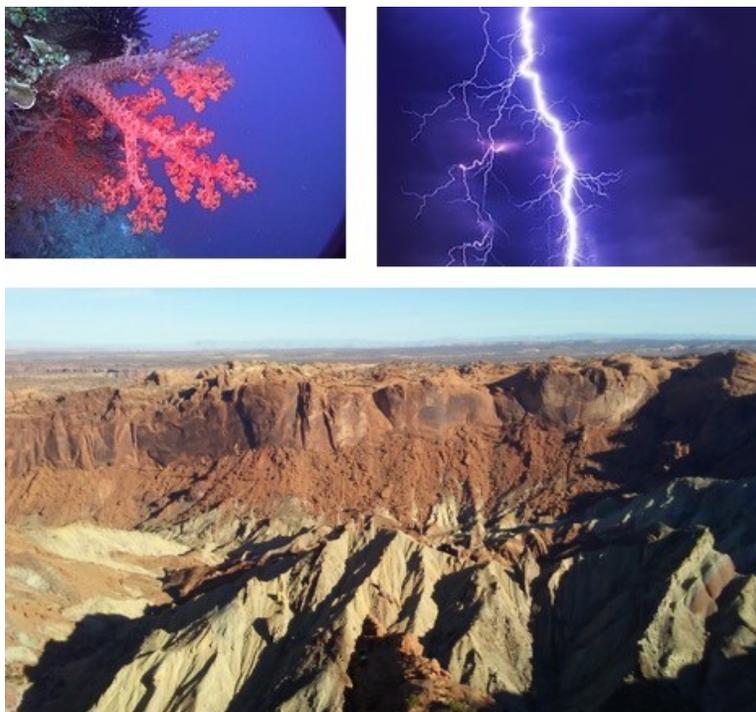

**Fig. 18.** Fractal organization in sea coral (top left), lightning (top right), and geological formations (bottom). In each case, the patterns at all scales all emerge by the same recursive process. Credits: Coral photo from Albert Koch / Wikimedia Commons / Public Domain; Lightning photo from Felix Mittermeir/ Wikimedia Commons/ Public Domain; Geological formation photo from Brandon Thomas.

Whatever kind of resonator the brain and nervous system is, given that it is nested within the animal-environment system, it is likely to exhibit self-similarity or scale invariance. And given that it is recruited in the service of detecting and exploiting information in the animal-environment system, such an organization would allow it to be sensitive to, reverberate with, and amplify information at each of the nested levels of this system (perhaps even simultaneously).

Therefore, we propose that the brain and nervous system can be usefully understood as a fractal or self-complementary antenna (Cohen, 1995). Fractal antennas have a self-similar geometric design that enables them to operate at *multiple nested scales*—that is, they are frequency and bandwidth independent (**Fig. 19**). Consequently, they can simultaneously send and receive transmissions at many different frequencies and bandwidths, simultaneously. This contrasts with a standard antenna that is limited to sending and receiving transmissions only at a particular frequency and bandwidth.

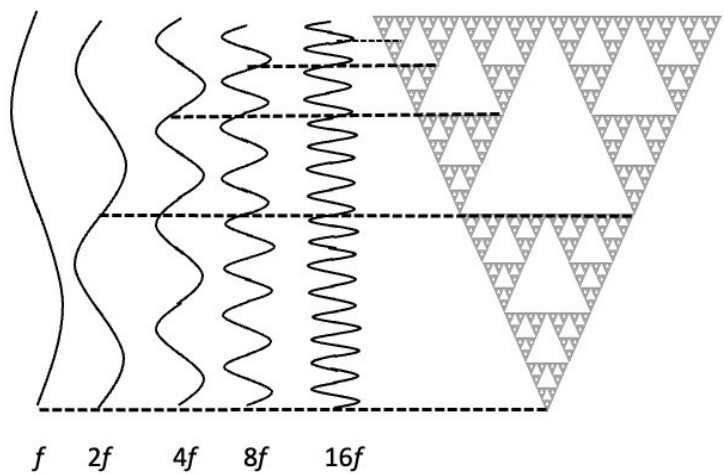

**Fig. 19.** A design for a fractal antenna based on the Sierpiński triangle. The nested triangles are sensitive are capable of resonating to transmissions at multiple frequencies (e.g., 1f, 2f, 4f, 8f, and 16f) simultaneously.

While it is possible to send and receive transmissions without a fractal antenna, the fractal antenna aids in the sending and receiving transmissions at multiple scales, even simultaneously. And the main difference between more sophisticated and less sophisticated fractal antenna is the number of scales over which such signals can be sent and received. In the same way, while it is possible to detect and exploit the information that supports the performance of everyday behavior without a brain, *a brain aids in the ability to do so at multiple nested levels, even simultaneously*. And the main difference between more sophisticated and less sophisticated brains *is the number and scope of the nested levels* (e.g., cultural, socio-technological) over which such information can be detected and exploited.

## 9.4. What empirical findings support your preferred alternative metaphor?

Consistent with the Ecological approach to perceiving, acting, and cognizing, we have proposed that the animal-environment system is the level at which the phenomena of interest occur and the level at which lawful relations that underlie these phenomena emerge. Performing everyday goal-directed behavior requires detecting and exploiting such lawful relations—information about affordances. Furthermore, we proposed that the brain might aid in the ability to do so by resonating to and amplifying such information at multiple scales simultaneously. In this way, the brain may be usefully understood as a fractal antenna.

The empirical findings that support this alternative metaphor are those demonstrate that the detection and exploitation of information in performing everyday goal-directed behaviors exhibit fractality. Fortunately, such empirical evidence abounds. In what follows, we provide a brief overview of a few examples from everyday (and perhaps not so everyday) goal-directed perceiving, acting, and cognizing.

*Fractality in goal-directed perceiving.* Detecting lawful relations in structured energy arrays that are informative about affordances requires actively exploring that structured energy array. To this end, fractal fluctuations appear in the exploratory wielding movements used to perceive properties of hand-held objects, and the degree of fractality predicts the ability to perceive those properties (Stephen et al., 2009). Moreover, (multi)fractal fluctuations are also exhibited in the postural sway movements that occur in perceiving properties of objects that are either held in hand (Mangalam et al., 2020b, 2020a) or attached to the body. And subtle differences in these fractal fluctuations emerge depending on what specific property the person is attempting to perceive (Palatinus et al., 2014). Finally, (multi)fractal fluctuations are also exhibited in the postural sway movements used in the service of visually perceiving the affordances of a surface, and the degree of (multi)fractality predicted whether that surface was perceived to afford a given behavior (Hajnal et al., 2018).

*Fractality in doing.* Finer-grained movements such as those used to manipulate objects are nested within (and in many cases, impossible without) coarser-grained movements such as those used to stabilize posture. To this end, (multi)fractal fluctuations appear in the coarser-grained postural sway movements that support finer-grained eye movement in precision viewing task, and changes in the viewing task change the nature of these fractal fluctuations (Kelty-Stephen et al., 2021). (Multi)fractal fluctuations also appear in the fine-grained dexterous hammering movements used by expert bead craftsmen to remove flakes of a particular size and shape from a larger stone. And the degree of fractality is reduced when less-skilled craftsmen perform this task with unfamiliar raw material (Nonaka and Bril, 2014). Fractal fluctuations also appear in the fine-grained movements of a computer mouse used to control a cursor in a virtual herding task. And the degree of fractality is reduced when the functional relation between mouse and cursor movements is temporarily disrupted (Dotov et al., 2010).

*Fractality in cognizing.* In the Ecological approach, cognizing is continuous with perceiving or acting. Thus, it too requires detecting and exploiting lawful relations in structured energy arrays that are informative about affordances. Accordingly, fractal fluctuations appear in the variability observed in both simple reaction times and word naming times when such tasks are performed over thousands of trials. The fractality is less prominent in word naming, likely due to subtle trial-to-trial differences in this task (i.e., the phonemic properties of the words) not present in the simple reaction time task (Van Orden et al., 2005, 2003). Fractal fluctuations are also present in the finger movements that participants use in the early stages of solving a gear systems problem-solving task. Moreover, there are changes in these fractal fluctuations that occur just before the participants discover a more efficient strategy for solving such problems (Dixon et al., 2012; Stephen et al., 2012, 2009).

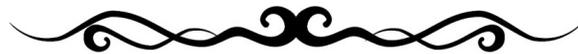

## 10. Discussion

The skepticism over the computer metaphor has created the momentum to search for an alternative metaphor to describe the mind and brain. The above collection of responses confirms that multiple philosophical and scientific investigations support a variety of metaphors to describe the mind and brain. All authors found support for their proposed metaphor in the empirical literature. None of the authors advocated rejecting all metaphors. Instead, their respective positions suggest that we should recognize that metaphors are merely means by which to use what we know to explain that which we do not know—but would like to know. Several contributions refer to (multi)fractal structure/multiplicative cascade as an essential metaphor/construct to understand adaptive behavior. e.g., the fractal antenna is quite similar to the idea of reproduction of similarity by analogy in which the fractal spectrum is the analogy and resonance is the reproduction of similarity. We hope that the foregoing stimulates future discourse about all these new metaphors in studying the mind, brain, and behavior.

### 10.1. Towards a Kuhnian revolution in the study of the mind and brain

Thomas Kuhn's (1962) notion of *scientific revolution* is central to the current discussion. Within a scientific field, a dominating *paradigm* inevitably emerges. In the psychological and neurosciences, the computational paradigm emerged and was fueled by (i) the deficiencies of Behaviorism explanations of language, (ii) the development of Communication Theory to quantify information, and (iii) the advent of digital computers. The paradigm's practices, models, exemplars, and most significant applications are taught to psychology and neuroscience students—e.g., computational explanations have become the mainstay of behavioral, cognitive, and systems neuroscience courses. And, a pattern of *business-as-usual science* follows: these resources are further developed, and the paradigm is expanded beyond its original applications. For instance, computationalism is not only used to explain the mind and behavior but every new development in computer technology (e.g., convolutional neural networks) is also used to explain some other feature of the mind and brain (e.g., visual processing in area V4). Nonetheless, *anomalies*, applications that appear to be intuitively promising for the paradigm but are resistant to absorption, begin to emerge—e.g., findings that neurons cannot implicate symbolic computation, newly discovered roles of glial cells. Anomalies accumulate, and some practitioners begin to see similarities, indicating the beginnings of a new paradigm—e.g., the contributing authors of this article and the complex systems community. Even while some practitioners stick to the still-dominant paradigm and pursue standard research, the authors of the present article identify brain and mind research *crises*, which can serve as catalysts for *paradigm shifts*.

Some argue that the neurosciences have not exhibited revolutions over the past 70 years due to conflicting paradigms, so much as they have demonstrated "revolutionary" technological developments that have altered research aims and practices (e.g., optogenetics; (Bickle, 2016; Bickle et al., 2022). There is no doubt that the creation of new tools is crucial to advancing neuroscience research. The Atwood machine was used to test Newton's second law of motion, while Galileo's inclined plane was used to explore how a body moves under its own weight (Kuhn, 1962). Both tested established beliefs and were thus considered conventional science. Similarly, the advancement of these techniques in neuroscience simply allows for different approaches to questions framed using the old paradigm. "If the tools are good," Dyson (2009) writes, "nature will give a clear answer to a clear question." However, a clear question and answer, as well as an elegant or sophisticated instrument or technique, will not be enough to overcome shortcomings in theory—such as may have occurred in equating the mind and brain with a computer. Indeed, Boyden, one of the developers of optogenetic approaches, has written, "no major paradigm shift in neuroscience has resulted from the use of optogenetic tools …. What optogenetics has done so far is make the study of circuits more tractable" (Boyden, 2015, p. 1200). So, the development of these tools defines normal science—and not the revolutionary science associated with paradigm shifts.

The authors of this article call for an epistemological paradigm shift in psychology and neuroscience. Such a call is not aimed at undermining the ongoing developments in existing techniques, methods, or approaches. Instead, it is to advocate for changes in beliefs, standards, and speculations that will help elucidate those aspects of the mind and brain that have remained out of the computer metaphor's reach. Of course, this new epistemology may call for new techniques, methods, or approaches but may also leverage existing ones. And these proposals may not have any noticeable impact in the short run, as any paradigm shift involves a variety of cultural and contextual factors beyond advocacy itself. The psychological-neuroscience structures in place rest on decades-long investments in the computer metaphor. Nonetheless, many students just beginning their research career or even seasoned researchers might feel that the mind and brain are beyond the computer metaphor's reach. This collection of perspectives might serve as a valuable resource to these individuals, providing fresh grounds for thinking through some of the challenging problems in neuroscience. As noted over 30 years ago, "to deny that the brain is computational is to risk losing your membership in the scientific community" (Searle, 1990a, p. 24). We hope things have changed since then.

**Funding.** P.C. was supported by the Canadian Institutes of Health Research (PJT-166014) and the Natural Sciences and Engineering Research Council of Canada (RGPIN/05245).

**Competing interests.** We declare we have no competing interests.

**Author contributions.** Conceptualization, D.G.K-S. and M. M.; Project administration, M.M.; Supervision, M.M.; Writing – original draft, D.G.K-S., P.E.C., B.D., J.D., L.H.F., F.W.H., F.K., V.R., J.B.W., B.J.T., and M.M.; Writing – review & editing, D.G.K-S. and M. M.